\documentclass[fleqn,usenatbib]{mnras}

\usepackage{newtxtext,newtxmath}
\usepackage[T1]{fontenc}

\DeclareRobustCommand{\VAN}[3]{#2}
\let\VANthebibliography\thebibliography
\def\thebibliography{\DeclareRobustCommand{\VAN}[3]{##3}\VANthebibliography}

\usepackage{graphicx}	% Including figure files
\usepackage{amsmath}	% Advanced maths commands
\usepackage{orcidlink}
\usepackage{booktabs}
\usepackage{comment}
\usepackage{ulem}
\usepackage[dvipsnames]{xcolor}

\definecolor{blazeorange}{rgb}{1.0, 0.4, 0.0}
\definecolor{seagreen}{rgb}{0.18, 0.55, 0.34}
\definecolor{rufous}{rgb}{0.66, 0.11, 0.03}
\definecolor{royalfuchsia}{rgb}{0.79, 0.17, 0.57}
\definecolor{scarlet}{rgb}{1.0, 0.13, 0.0}
\definecolor{royalpurple}{rgb}{0.47, 0.32, 0.66}
\definecolor{darkblue}{rgb}{0, 0, 0.66}
\definecolor{violet}{rgb}{0.5,0.,0.5}
\definecolor{nag}{RGB}{158, 56, 202} 

\title[GRB~250221A]{Evidence of Energy Injection in the Short and Distant GRB~250221A in a High Density Environment}% \textcolor{purple}{This version contains the analysis for the event at z=0.768. For the oldest, please go to "main old.tex"}}

\author[Angulo-Valdez et al.]{Camila Angulo-Valdez\,\orcidlink{0009-0002-6667-3294},$^{1}$\thanks{E-mail: camiangulo@astro.unam.mx (CAV)},
Rosa~L.~Becerra\,\orcidlink{0000-0002-0216-3415},$^{1,2}$\thanks{E-mail: rbecerra@astro.unam.mx (RLB)},
Ramandeep Gill\,\orcidlink{0000-0003-0516-2968},$^{3}$
Noémie~Globus\,\orcidlink{0000-0001-6148-6532},$^{4}$
William H.~Lee\,\orcidlink{0000-0002-2467-5673},$^{1}$\newauthor
Diego~L\'opez-C\'amara\,\orcidlink{0000-0001-9512-4177},$^{6}$
Cassidy~Mihalenko\,\orcidlink{0009-0004-0322-6299},$^{5,17}$
Enrique~Moreno~M\'endez\,\orcidlink{0000-0002-5411-9352},$^{7}$
Roberto~Ricci\,\orcidlink{0000-0003-4631-1528},$^{2}$\newauthor
Karelle~Siellez\,\orcidlink{0000-0002-8568-8523},$^{5}$
Alan~M.~Watson\,\orcidlink{0000-0002-2008-6927},$^{1}$
Muskan Yadav\,\orcidlink{0009-0004-9520-5822},$^{2}$
Yu-Han~Yang\,\orcidlink{ 0000-0003-0691-6688},$^{2}$
Dalya~Akl\,\orcidlink{0009-0006-4358-9929},$^{11,19}$
Sarah~Antier\,\orcidlink{0000-0002-7686-3334},$^{18}$\newauthor
Jean-Luc~Atteia\,\orcidlink{0000-0001-7346-5114},$^{13}$
Stéphane~Basa\,\orcidlink{0000-0002-4291-333X},$^{8,9}$
Nathaniel R. Butler\,\orcidlink{0000-0002-9110-6673},$^{14}$
Simone~Dichiara\,\orcidlink{0000-0001-6849-1270},$^{15}$
Damien~Dornic\,\orcidlink{0000-0001-5729-1468},$^{12}$\newauthor
Jean-Grégoire~Ducoin\,\orcidlink{0009-0008-7341-4825},$^{12}$
Francis Fortin\,\orcidlink{0000-0003-3642-2267},$^{13}$
Leonardo~García-García\,\orcidlink{0000-0001-5125-1043},$^{4}$
Kin~Ocelotl~López\,\orcidlink{0000-0002-9322-6900},$^{1}$\newauthor
Francesco~Magnani\,\orcidlink{},$^{12}$
Brendan O'Connor\,\orcidlink{0000-0002-9700-0036},$^{16}$
Margarita~Pereyra\,\orcidlink{0000-0001-6148-6532},$^{4,20}$
Ny~Avo~Rakotondrainibe\,\orcidlink{0009-0004-0263-7766},$^{8}$\newauthor
Fredd~Sánchez-Álvarez\,\orcidlink{0009-0009-5612-3759},$^{1}$
Benjamin~Schneider\,\orcidlink{0000-0003-4876-7756},$^{8}$
Eleonora~Troja\,\orcidlink{0000-0002-1869-7817},$^{2}$
Antonio de Ugarte Postigo\,\orcidlink{0000-0001-7717-5085}$^{8}$
\\
$^{1}$ Universidad Nacional Aut\'onoma de M\'exico. Instituto de Astronom\'ia. A.P. 70-264, 04510. Ciudad de M\'exico, M\'exico.\\
$^{2}$ Department of Physics, University of Rome - Tor Vergata, via della Ricerca Scientifica 1, 00100 Rome, IT\\
$^{3}$ Instituto de Radioastronomía y Astrof\'isica, Universidad Nacional Aut\'onoma de M\'exico, Antigua Carretera a P\'atzcuaro \# 8701,\\ Ex-Hda. San Jos\'e de la, Huerta, Morelia, Michoac\'an, M\'exico C.P. 58089, M\'exico\\
$^{4}$ Instituto de Astronom{\'\i}a, Universidad Nacional Aut\'onoma de M\'exico, km 107 Carretera Tijuana-Ensenada, 22860 Ensenada, Baja California, México\\
$^{5}$ School of Natural Sciences, Private Bag 37, University of Tasmania, Hobart, 7001 TAS, Australia\\
$^{6}$ Instituto de Ciencias Nucleares, Universidad Nacional Aut\'onoma de M\'exico, Apartado Postal 70-264, 04510 M\'exico, CDMX, Mexico\\
$^{7}$ Facultad de Ciencias, Universidad Nacional Aut\'onoma de M\'exico, Apartado Postal 70-264, 04510 M\'exico, CDMX, Mexico\\
$^{8}$ LAM, Université Aix-Marseille \& CNRS, UMR7326, 38 rue F. Joliot-Curie, 13388 Marseille Cedex 13, France\\
$^{9}$ Aix-Marseille University, CNRS, OSU-Pytheas, France \\
$^{10}$ Observatoire de la C\^ote d’Azur, ARTEMIS, Nice, France\\
$^{11}$ New York University Abu Dhabi, PO Box 129188, Saadiyat Island, Abu
Dhabi, UAE\\
$^{12}$ Aix Marseille University, CNRS, CPPM, Marseille, France\\
$^{13}$ IRAP, Université de Toulouse, CNRS, CNES, UPS, 31401 Toulouse, France\\
$^{14}$ School of Earth and Space Exploration, Arizona State University, Tempe, AZ 85287, USA\\
$^{15}$ Department of Astronomy and Astrophysics, The Pennsylvania State University, 525 Davey Lab, University Park, PA 16802, USA\\
$^{16}$ McWilliams Center for Cosmology and Astrophysics, Department of Physics, Carnegie Mellon University, Pittsburgh, PA 15213, USA\\
$^{17}$ ARC Centre of Excellence for Gravitational Wave Discovery (OzGrav), John St, Hawthorn, VIC 3122, Australia\\
$^{18}$ IJCLAB, Université Paris Saclay, Orsay, France\\
$^{19}$ Center for Astrophysics and Space Science (CASS), New York University Abu Dhabi, Saadiyat Island, PO Box 129188, Abu Dhabi, UAE \\
$^{20}$ Secretar\'ia de Ciencia, Humanidades, Tecnolog\'ia, e Innovaci\'on
}
\date{Accepted January 23, 2026}
\pubyear{\the\year{}}

\begin{document}
\label{firstpage}
\pagerange{\pageref{firstpage}--\pageref{lastpage}}
\maketitle

\begin{abstract}
We present the photometric and spectroscopic analysis of the short-duration GRB~250221A ($T_{90}=1.80\pm0.32$~s), using a data set from the optical facilities COLIBRÍ, the Harlingten 50~cm Telescope, and the Very Large Telescope. We complement these observations with data from the \textit{Neil Gehrels Swift Observatory} and the \textit{Einstein Probe}, as well as radio observations from the Very Large Array. GRB~250221A is among the few short GRBs with direct afterglow spectroscopy, which gives a secure redshift determination of $z=0.768$ and allows the unambiguous identification of the host as a  galaxy with a star-formation rate of $\sim3\,M_\odot\,{\rm yr}^{-1}$. The X-ray and optical light curves up to $T_0+3\times 10^4$~s (where $T_0$ refers to the GRB trigger time) are well described by forward-shock synchrotron emission in the slow-cooling regime within the standard fireball framework. However, at $T_0 \sim 5\times 10^4$~s, both the X-ray and optical bands exhibit an excess over the same interval, which we interpret as evidence of energy injection into a jet with a half-opening angle of $\theta_j=11.5^{\circ}$ through a refreshed shock powered by late central engine activity or a radially stratified ejecta. The burst properties (duration, spectral hardness, peak energy, and location in the Amati plane) all favour a compact binary merger origin. However, our modelling of the afterglow suggests a dense circumburst medium ($n\sim80$~cm$^{-3}$), which is more typical of a collapsar environment. %This tension over the classification of this burst (short-hard vs. long-soft) as inferred from the prompt and afterglow emissions makes GRB~250221A an unusual event and underscores the limitations of duration-based classifications and the importance of multi-wavelength, time-resolved follow-up observations.
\end{abstract}


%%%%%%%%%%%%%%%%%%%%%%%%%%%%%%%%%%%%%%%%%%%%%%%%%%

%%%%%%%%%%%%%%%%% BODY OF PAPER %%%%%%%%%%%%%%%%%%

%%%%%%%%%%%%%%%%%%%%%%%%%%%%%%%%%%%%%%%%%%%%
\section{Introduction}
\label{sec:introduction}
%%%%%%%%%%%%%%%%%%%%%%%%%%%%%%%%%%%%%%%%%%%%
Gamma-ray bursts (GRBs) are the most luminous electromagnetic phenomena observed in the universe \citep{Atteia2017} and are powered by ultra-relativistic jets, with an overall duration dichotomy \citep{Kouveliotou1993} that is  usually associated to two different types of progenitors. Short GRBs (SGRBs) have durations $T_{90}$\footnote{The duration over which the central 90\% of the photons from the GRB are detected.} $\lesssim 2$~seconds and a relatively hard spectrum, characterized through the hardness ratio (HR\footnote{The energy or fluence contained within one band compared to another, where SGRBs usually emit more high-energy radiation than LGRBs.}), and are believed to result from the merger of compact binary systems with at least one neutron star (NS) involved \citep[see, e.g.,][]{Paczynski1986,Paczynski1991,Lee2007,Berger2014,Abbott2017}. On the other hand, long GRBs (LGRBs) are thought to be the result of the core-collapse of a Wolf-Rayet star (with a stripped H and He envelope) whose mass exceeds about $10 M_{\odot}$ and are associated with type Ic supernovae (SNe) \citep[see, e.g.,][]{Woosley1993,MacFadyen1999ApJ,Hjorth2012}.  

However, this duration-based classification has important limitations, which can be seen in several cases, including: hybrid events such as the long GRB\,211211A ($T_{90}=51.37 \pm 0.80$s) \citep{Troja2022,Yang2022,Rastinejad2022} and GRB\,230307A ($T_{90}\sim35$s) \citep{Yang2024,Levan2024}, but with an identified kilonovae (KNe) signature; the short-duration GRB\,200826A ($T_{90}=1.4\pm0.13$s) linked to the collapse of a massive star \citep{Ahumada2021}; and unusual events such as GRB\,210704A likely produced by the coalescence between a neutron star and a white dwarf \citep{Becerra2023}, combining properties exhibited by collapsar and merger-driven bursts. This motivates the use of additional diagnostics, in particular the properties of the afterglow and host environment, to distinguish between progenitors.

Beyond duration and hardness, several studies have therefore explored more physically motivated classification schemes. In particular, the type-I/type-II framework \citep[e.g.,][]{Zhang2007} incorporates host-galaxy characteristics, energetics, and afterglow behaviour to differentiate merger-driven from collapsar-driven events. This broader approach has also been applied by \citet{Tsvetkova2017} when classifying short GRBs, underscoring the need for a multidimensional view of GRB origins.\\
Supernovae and kilonovae provide the most direct evidence of the underlying progenitors. However, they are not detectable at high redshift. 
In such cases, the afterglow and burst environment become the primary diagnostics.
In this context, the multi-frequency afterglow is a powerful probe. While the fireball model \citep{Goodman1986, Paczynski1986, Shemi1990, Rees-Meszaros-92, Meszaros-Rees-93}, captures the overall prompt and afterglow evolution \citep[see][for a review]{Kumar2015,AguiFernandez2023}, many optical light curves show deviations such as flares, plateaus, and rebrightening episodes \citep{Sari-Meszaros-00,Lazzati2002,Petropoulou2020,Dainotti2024,Gendre2025}. The physical origin of these features remains debated, with possible explanations including late-time central engine activity \citep{Li2012}, energy injection \citep{Laskar2015}, refreshed shocks \citep{Vlasis+11,Lamb2019}, or interaction with a structured circumburst medium (CBM). Such signatures are particularly relevant when the progenitor classification is uncertain, since they may reveal information about both the central engine and the environment.

Complementary to the photometric information, spectroscopy is essential for a complete understanding of every event. It provides the only direct means of determining redshift, offers clues about the host galaxy and the environment of the burst, and helps constrain the progenitor system. 
Optical spectroscopy of SGRB host galaxies is particularly challenging because these systems are typically faint and associated with older stellar populations \citep{OConnor2022,Fong2022,Nugent2022}. Nonetheless, when successful, such observations yield valuable information on the global environment in which these transients occur.

Even more powerful, however, is the spectroscopy of the afterglow itself. Unlike host observations, which probe the large-scale properties of the galaxy, afterglow spectroscopy provides a direct snapshot of the line of sight at the moment of the explosion. Optical and near-infrared spectra can reveal absorption features from the host interstellar medium, such as \ion{Mg}{II}, \ion{Fe}{II}, and \ion{Ca}{II}, and occasionally fine-structure lines excited by the intense GRB radiation \citep{Prochaska2007,Vreeswijk2013,Selsing2018}. Such measurements are exceedingly rare, as SGRB afterglows are typically fainter and fade much faster than those of LGRBs, but successful detections deliver uniquely detailed information about the immediate environment of the burst \citep{deUgartePostigo2014,OConnor2022,AguiFernandez2023}.

In this paper, we present an analysis of both the early- and late-time observations of the multi-frequency afterglow of GRB~250221A. 
This is one of the few SGRB afterglows (only the fourth case 
in the sample of over 150 SGRBs discovered by \textit{Swift}) with clear absorption lines in its spectrum, allowing us to securely measure its redshift $z\!\approx\!0.768$ and break the degeneracy with a nearby, unrelated galaxy at $z\!\approx\!0.343$.
Its late-time afterglow shows an unusual rebrightening starting at $\sim$0.6~d post-burst — later and more pronounced than in most previously reported events \cite[see, e.g.,][]{Troja2007,Becerra2019a,Becerra2019b,Pereyra2022,Nardini2011, Kobayashi2003,Li2012,Kann2010} 
— making it an outstanding event for investigating how such features connect to progenitor type and environment. By combining early- and late-time multi-wavelength photometry with spectroscopy of the host candidates, this burst allows us to explore the mechanisms behind rebrightening and assess their implications for GRB classification. We examine its high-energy characteristics during the prompt phase, including duration, isotropic energy, and spectral peak energy, along with multi-wavelength observations of the afterglow in the X-ray, optical, and radio bands extending up to 10~days post-burst. 

The paper is organised as follows. In Section~\ref{sec:observations}, we describe the photometric and spectroscopic data sets, including observations in gamma-rays from \textit{Swift}/BAT, X-rays from \textit{Swift}/XRT and \textit{Einstein Probe}, optical data from COLIBRÍ, the Harlingten 50-cm Telescope, and the Very Large Telescope (VLT), and radio observations from the Very Large Array (VLA). In Section~\ref{sec:hg} we analyse the environment of GRB~250221A and discuss the properties of the potential host galaxy candidates. In Section~\ref{sec:interpretation}, we present the physical interpretation of our data, including the light curve and spectral evolution of GRB~250221A. We discuss our findings in Section~\ref{sec:discussion} and summarise our conclusions in Section~\ref{sec:summary}.

Throughout this work, we adopt a flat $\Lambda$CDM cosmology with BAO with $H_0 = 67.7$~$\mathrm{km\ s^{-1}\ Mpc^{-1}}$ and $\Omega_m = 0.31$ \citep{Planck2020}. 

%%%%%%%%%%%%%%%%%%%%%%%%%%%%%%%%%%%%%%%%%%%%
\section{Observations}
\label{sec:observations}
%%%%%%%%%%%%%%%%%%%%%%%%%%%%%%%%%%%%%%%%%%%%
\subsection{{\itshape High-Energies}}
\label{sec:swift}
%%%%%%%%%%%%%%%%%%%%%%%%%%%%%%%%%%%%%%%%%%%%

The {\itshape Swift}/Burst Alert Telescope (BAT) instrument on the {\itshape Neil Gehrels Swift Observatory} triggered on GRB~250221A $T_0 = $ 2025 February 21 03:34:37 UTC \citep{39396}. Simultaneously, Konus-Wind also triggered the source \citep{39423}. The burst consists of a single short pulse with $T_{90}=1.8\pm0.3$~s in the 15--350~keV band, best fit by a power-law spectrum with photon index $1.43\pm0.23$ \citep{39471}. According to this best fit model, the fluence in the 15--150~keV band is $(3.9\pm0.6)\times10^{-7}$ erg cm$^{-2}$ and the 1-s peak flux is $2.9\pm0.5$ ph cm$^{-2}$ s$^{-1}$.

At the time of the {\itshape Swift}/BAT trigger, all three LIGO–Virgo detectors were operating nominally with ranges of 157, 162, and 52~Mpc. The closest GW candidate (S250221ap, 126~s later) was at $811\pm250$ Mpc with a high FAR, making it incompatible with GRB~250221A.

The {\itshape Swift}/X-ray Telescope (XRT) began observations 79~s after the trigger, detecting a fading X-ray counterpart at RA, Dec (J2000) = 03:57:51.03, $-15$:08:01.3 with an error radius of 1.9~\arcsec at the 90\% confidence level (c. l.; \citealt{39396}). XRT light curves and spectra were obtained from the public online repository \citep{Evans2009} hosted by the UK Swift Science Data Centre\footnote{\url{https://www.swift.ac.uk/}}. 

The X-ray afterglow is observed to decay as a simple power-law with  slope $0.97\pm0.04$ ( 0.3--10~keV), falling below the XRT sensitivity within 1 day.  Subsequent monitoring was carried out with the Fast X-ray Telescope (FXT) aboard  {\itshape Einstein Probe}.
    
Two FXT follow-up observations (PI: Troja) were conducted at $T_0+2.07$~d and $T_0+5.68$~d with total exposures of 5~ks and 4~ks, respectively.
Data were acquired in Full Frame mode with the Thin Filter, and processed using the FXT Data Analysis Software ({\sc fxtdas} v.1.10). 
Aperture photometry of the source was performed using a circular region with a 40\arcsec radius to estimate the source counts and a concentric annular region with radii of 60\arcsec\ and 180\arcsec\ to estimate the background. 
The unabsorbed flux is $\left(2.3_{-0.2}^{+0.3}\right)\times10^{-13}$ erg cm$^{-2}$ and $\left(2.4_{-1.0}^{+1.4}\right) \times10^{-14}$ erg cm$^{-2}$ in the 0.3--10~keV range, respectively (see Table~\ref{tab:observations}).

%%%%%%%%%%%%%%%%%%%%%%%%%%%%%%%%%%%%%%%%%%%%
\subsection{Optical}
\subsubsection{COLIBRÍ}
\label{sec:colibri}
%%%%%%%%%%%%%%%%%%%%%%%%%%%%%%%%%%%%%%%%%%%%
We observed the afterglow of GRB 250221A with the DDRAGO wide-field imager on the COLIBRÍ telescope. COLIBRÍ\footnote{\url{https://www.colibri-obs.org/}} is a Franco-Mexican fast, robotic 1.3~m telescope operated by the Observatorio Astronómico Nacional (OAN) in the Sierra de San Pedro Mártir, Baja California \citep{Basa2022}. DDRAGO is a two-channel imager with the blue channel working in $gri$ and the red in $zy$ \citep{Langarica2024}. The blue channel uses a backside-illuminated, deep-depleted e2v 231--84 CCD in a Spectral Instruments 1110S package. The CCD is $4\mathrm{k} \times 4\mathrm{k}$ with 15~{\micron} pixels. The scale is 0.38~\arcsec/pixel and the field is 25.9~{\arcmin} square. During our observations, the red channel was not available. 

Our observations of GRB~250221A used the $g$, $r$, and $i$ filters, which closely match the SDSS/Pan-STARRS filter system. The color-term coefficients for $g-r$ for transforming to Pan-STARRS DR1 magnitudes are $-0.05$, $+0.00$, and $-0.04$ for $g$, $r$, and $i$, respectively.

The COLIBRÍ control system received an initial {\itshape Swift}/BAT notification of the burst via the GCN system at 2025 February 21 03:35:03 UTC ($T_0 + 26$~seconds). The telescope immediately interrupted its ongoing observations and slewed to the burst location. The first exposure began at 03:35:56 UTC ($T_0+79$ seconds) at an airmass of about 1.7. All exposures were 60~seconds, with a dead time of about 20~seconds between each exposure for read-out and dithering. The final exposure of the first night occurred at 05:11:31 UTC (1.62 hours after the trigger), by which time the airmass had increased to 2.81 and the transparency had decreased noticeably.

The initial exposures were in the $i$ filter (chosen as the reddest filter available, to minimize the possibility of drop-out in $g$ and $r$ for higher-redshift GRBs). The 60-second exposures taken between $T_0+0.30$ and $T_0+1.34$~hours were combined in groups of 5 to increase the signal-to-noise ratio. Towards the end of the first night, starting at $T_0+1.06$ hours, we obtained a sequence of 24 exposures alternating between the $g$, $r$, and $i$ filters. In the following nights, all the observations were only carried out in the $r$ filter.

\begin{figure*}
	\includegraphics[clip, width=0.95\linewidth]{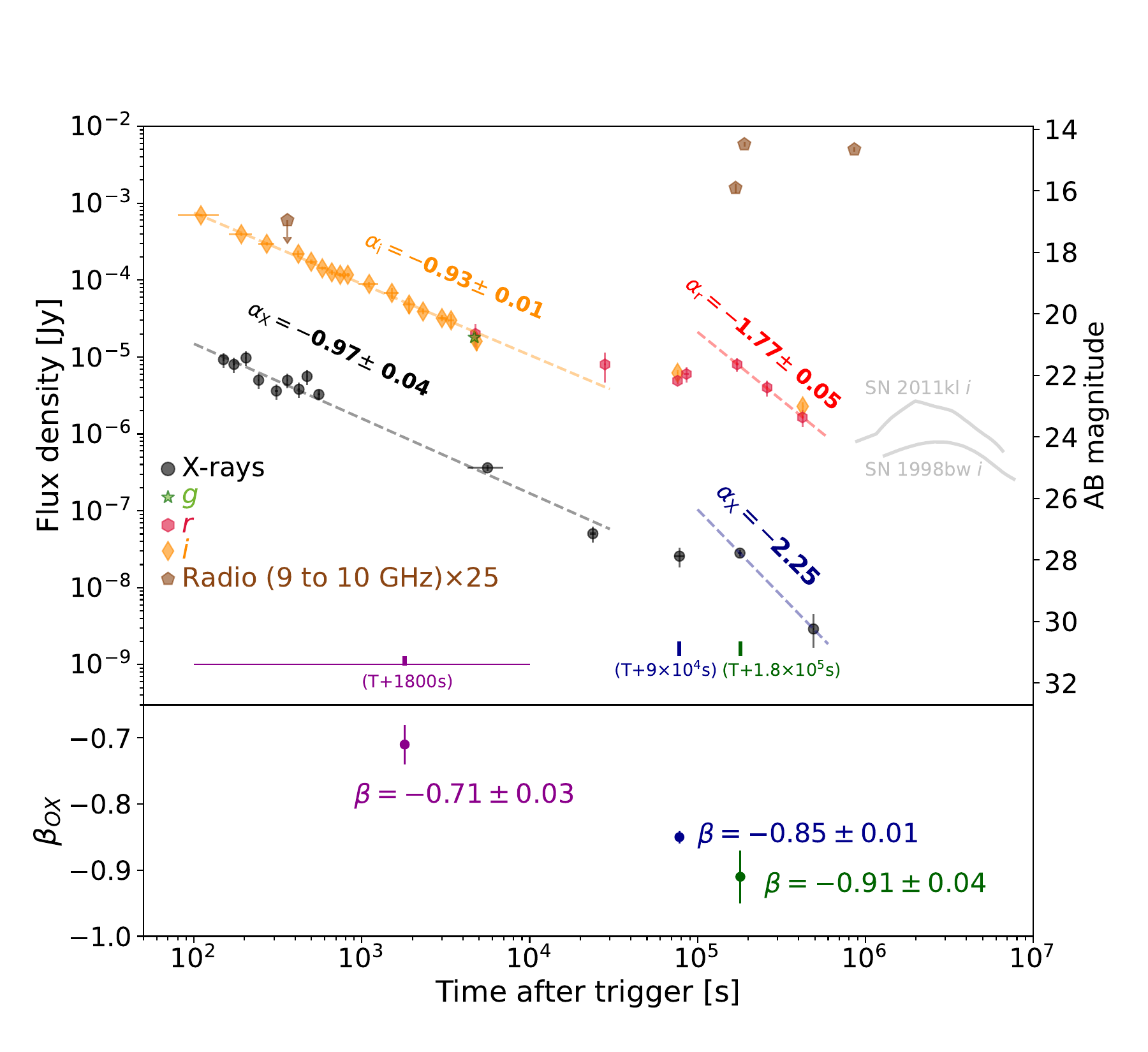}
    \caption{
    {\itshape Top}: X-ray (0.3--10.0~keV, at 1~keV, black circles), optical (yellow diamonds for {\itshape i}, red hexagons for {\itshape r}, green stars for {\itshape g}) and radio (at 9~GHz from ATCA \citep{39501} and 10~GHz from VLA, brown pentagons) observations of GRB~250221A. 
    The dashed lines show the power-law fits to the afterglow for X-rays (fitted before $T \lesssim 2 \times 10^{4}$~s and after $T > 10^{5}$~s) and optical (fitted for $T < 10^{4}$~s for the {\itshape i} band and at $T > 10^{5}$~s for the late {\itshape r} band). {\itshape Bottom}: Optical-to-X-rays spectral indices for three different epochs, the first from $T_0+100$~s to $T_0+10^{4}$~s, with a median arrival time of the X-rays photons at $T_0+1800$~s (see Figure~\ref{fig:sed} and Section~\ref{sec:lcmodel}) (purple), second at $T_0+ 9\times 10^{4}$~s (blue), and third at $T_0+1.8\times 10^{5}$~s (green). We also illustrate the optical light curves of SN1998bw \citep{Galama1998, Clocchiatti2011} and SN2011kl \citep{Greiner2015} (placed at $z = 0.768$) to highlight that the excess is similar in brightness (in the optical) but differs in timescale.
    \label{fig:LC}}
\end{figure*}

We reduced and coadded the images using custom software to carry out sky subtraction, image alignment, and coaddition. Our photometry is included in Table~\ref{tab:observations} and is plotted in Figure~\ref{fig:LC}. 

We identified a bright, uncatalogued source at RA, Dec (J2000) = 03:57:51.07, $-15$:07:59.52, with an uncertainty of 0.5~\arcsec, consistent with the \textit{Swift}/UVOT position \citep{39396}, and reported this in \cite{39397}.

%%%%%%%%%%%%%%%%%%%%%%%%%%%%%%%%%%%%%%%%%%%%
\subsubsection{Harlingten 50-cm Telescope}
\label{sec:harlingten}
%%%%%%%%%%%%%%%%%%%%%%%%%%%%%%%%%%%%%%%%%%%%
We also observed the afterglow of GRB 250221A with the Harlingten 50-cm Telescope of the University of Tasmania Greenhill Observatory. The telescope is equipped with an Apogee Alta U42 CCD with $2048\times2048$ pixels each of $13\,\micron$, giving a field of 27.2 arcmin and a pixel scale of 0.8~\arcsec/pixel, and an Apogee FW50-10S filter wheel with Bessell $B$ and $V$, SDSS $g'$, $r'$, and $i'$, and narrowband filters.

We obtained multiple 60-second exposures in the $r'$ filter. We reduced and coadded the images using custom software to carry out sky subtraction, image alignment, and co-addition. Our final stack contains 960~s of exposure.

%%%%%%%%%%%%%%%%%%%%%%%%%%%%%%%%%%%%%%%%%%%%
\subsubsection{Very Large Telescope}
\label{sec:vlt}
%%%%%%%%%%%%%%%%%%%%%%%%%%%%%%%%%%%%%%%%%%%%
We observed the field of GRB~250221A with the X-shooter spectrograph on ESO’s VLT UT3 (Melipal; 114.27LW.022; PI: Troja). Observations began at $T_0 + 21.4$~hours and were carried out at an average airmass of 1.2 for a total exposure of 4$\times$600~s with seeing of approximately $1.0\arcsec$.

\begin{figure}
	\includegraphics[clip, width=0.9\linewidth]{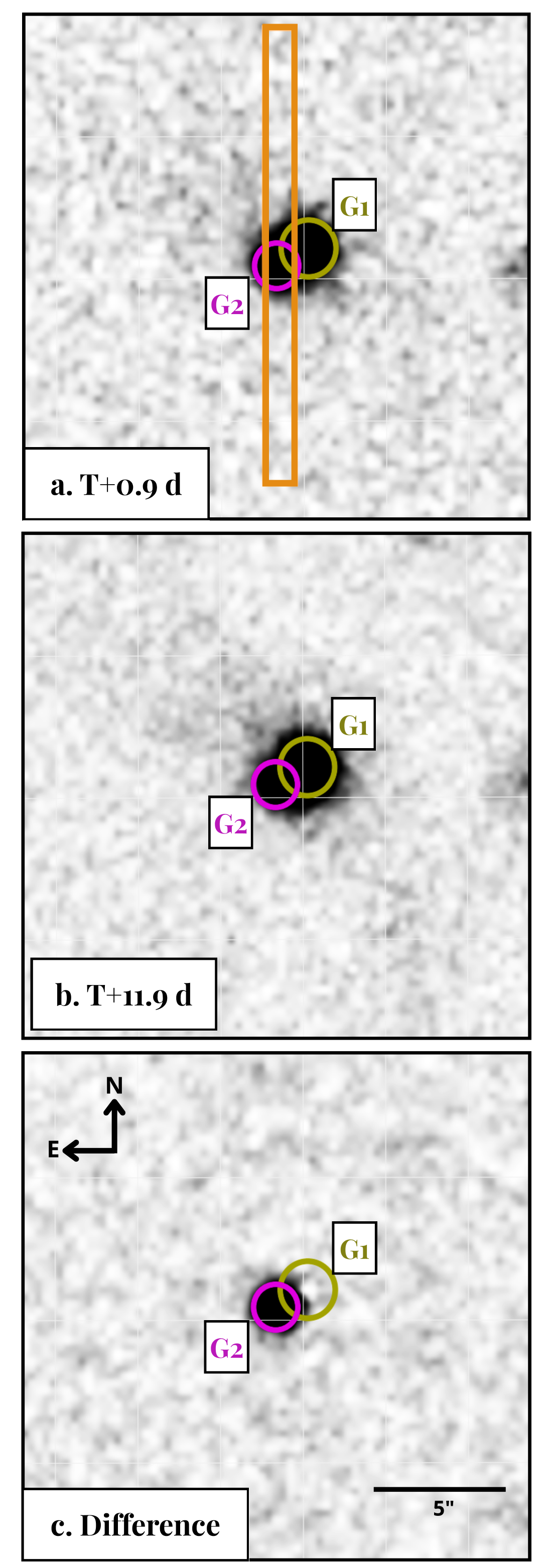}
    \caption{Field of GRB~250221A. We show the R filter image obtained with the VLT/FORS2 after $T_0+0.9$~days (panel a) together with the position of the X-Shooter slit used to acquire the spectrum (orange box)} and $T_0+11.9$~days (panel b). We also show the contribution of the afterglow after image subtraction (panel c). We highlight the presence of two galaxies at $photo-z=0.343\pm0.077$ \citep[G1, green circle; according to the Legacy Survey DR10][]{Dey2019} and at $z=0.768$ (G2, magenta circle). 
    \label{fig:field}
\end{figure}

We used a 1\arcsec\ and 0.9\arcsec\ slit widths in the UBV and VIS/NIR arms respectively, shown by the rectangle in the first panel of Figure~\ref{fig:field}, and a nodding ABAB observing scheme with a nodding throw of 5\arcsec\ to minimize the background in the near-infrared (NIR) arm. 
Exposures from each arm were reduced independently with custom software based on the official {\sc eso} pipeline\footnote{\url{https://www.eso.org/sci/software/pipe_aem_main.html}} \citep{Modigliani2010}. The pipeline performs bias subtraction, flat correction, bad-pixel masking, order tracing and rectification, wavelength calibration, and sky subtraction, followed by flux calibration using the instrument response and standard-star catalogue when available. 

Imaging observations were carried out with the FORS2 instrument on UT1 (Antu) in three epochs (see Table~\ref{tab:observations}; PI: Troja). 
For the first epoch, we observed the target with the 
{\itshape R$\_$SPECIAL}, {\itshape I$\_$BESS}, and {\itshape z$\_$GUNN} filters. Observations began at 00:42:31 UTC on 2025 February 22 ($T+0.89$~days) with an average seeing of $\sim1.3$\arcsec.

The second epoch of observations was conducted on 2025 February 26 ($T_0 + 4.88$~days), starting at 00:30:46 UTC. Images were obtained in the {\itshape R} and {\itshape I} filters, with total exposure times of 1440~s and 1920~s, respectively.
A third epoch was obtained at $T_0+11.86$~days, also in the {\itshape R} and {\itshape I} filters, to facilitate image subtraction and to characterize the host galaxy of GRB~250221A (see Table~\ref{tab:observations}).

Images were processed following standard techniques for CCD data reduction (e.g., bias subtraction, flat-fielding, cosmic-ray rejection), aligned using SCAMP and coadded with SWarp \citep{Bertin2006,Bertin2010}. 

%%%%%%%%%%%%%%%%%%%%%%%%%%%%%%%%%%%%%%%%%%%%
\subsubsection{Other Optical and Infrared Observations}
%%%%%%%%%%%%%%%%%%%%%%%%%%%%%%%%%%%%%%%%%%%%
Optical observations of GRB~250221A were also performed by the Rapid Eye Mount (REM) 60~cm robotic telescope \citep{39406}, 0.36~m telescope in the Al-Khatim Observatory \citep{39410}, 1.6~m Multi-channel Photometric Survey Telescope (Mephisto) \citep{39412}, the Nordic Optical Telescope (NOT) \cite{39413}, GRANDMA/TAROT \citep{39417}, the AZT-20 telescope at the Assy-Turgen Observatory \citep{39422}, the AZT-33IK telescope at the Mondy Observatory \citep{39424}, the 1-meter Sinistro telescope at the Las Cumbres Observatory Global Telescope (LCOGT) \citep{39425}, the 80~cm Tsinghua-Nanshan Optical Telescope (TNOT) \citep{39446}. GRB~250221A was additionally detected in the near IR by REM \citep{39406}.

These observations are reported here for completeness; however, they are not used in the analysis presented in this work, as they do not provide additional constraints on the properties discussed below.

\begin{table*}
    \centering
    \caption{Optical and radio photometry of GRB 250221A \textbf{used in this work}. Optical values are corrected by Galactic extinction.}
    \label{tab:observations}
\begin{tabular}{llrcccll}
\toprule
\toprule
\multicolumn{8}{c}{\textbf{X-rays}}  \\
\toprule
\toprule
Telescope&Instrument & Mid Time [s] & Energy Range & & Exposure Time [s] & Flux density  [$\mu$Jy]& References\\
\midrule
\midrule
{\itshape Swift} & XRT &   149.82 & 0.3--10 & \ldots  &                   & $9.24\pm2.02$       & This work \\
{\itshape Swift} & XRT &   173.06 & 0.3--10 & \ldots  &                   & $8.01\pm1.79$       & This work \\
{\itshape Swift} & XRT &   204.19 & 0.3--10 & \ldots  &                   & $9.72\pm2.08$       & This work \\
{\itshape Swift} & XRT &   242.75 & 0.3--10 & \ldots  &                   & $4.99\pm1.12$       & This work \\
{\itshape Swift} & XRT &   309.46 & 0.3--10 & \ldots  &                   & $3.61\pm0.81$       & This work \\
{\itshape Swift} & XRT &   360.39 & 0.3--10 & \ldots  &                   & $4.98\pm1.09$       & This work \\
{\itshape Swift} & XRT &   422.20 & 0.3--10 & \ldots  &                   & $3.80\pm0.84$       & This work \\
{\itshape Swift} & XRT &   471.64 & 0.3--10 & \ldots  &                   & $5.58\pm1.26$       & This work \\
{\itshape Swift} & XRT &   554.48 & 0.3--10 & \ldots  &                   & $3.26\pm0.53$       & This work \\
{\itshape Swift} & XRT &  5607.76 & 0.3--10 & \ldots  &                   & $0.36\pm0.04$       & This work \\
{\itshape Swift} & XRT & 23784.89 & 0.3--10 & \ldots  &                   & $0.05\pm0.01$       & This work \\
{\itshape Swift} & XRT & 78038.44 & 0.3--10 & \ldots  &                   & $0.03\pm0.01$       & This work \\ 
{\itshape EP}    & FXT &178848.0  & 0.3--10 & \ldots  & 5000              &$(2.82_{-0.27}^{+0.29})\times10^{-2}$& This work\\
{\itshape EP}    &FXT  &490752.0  & 0.3--10& \ldots  & 4000              &$(2.91_{-0.13}^{+0.16})\times10^{-2}$& This work\\
\toprule
\toprule
\multicolumn{8}{c}{\textbf{Optical}}  \\
\toprule
\toprule
Telescope&Instrument & Mid Time [s] & Filter & AB Magnitude & Exposure Time [s] & Flux density [$\mu$Jy] & References\\
\midrule
\midrule
COLIBRÍ & DDRAGO &  108.0   & i & 16.80 $\pm$ 0.04 & 60   & $634.08 \pm 21.35$ & This work \\
COLIBRÍ & DDRAGO &  180.0   & i & 17.41 $\pm$ 0.05 & 60   & $360.36 \pm 18.00$ & This work \\
COLIBRÍ & DDRAGO &  288.0   & i & 17.72 $\pm$ 0.07 & 60   & $270.02 \pm 16.68$ & This work \\
COLIBRÍ & DDRAGO &  432.0   & i & 18.04 $\pm$ 0.11 & 60   & $200.36 \pm 20.93$ & This work \\
COLIBRÍ & DDRAGO &  504.0   & i & 18.30 $\pm$ 0.08 & 60   & $158.09 \pm 11.95$ & This work \\
COLIBRÍ & DDRAGO &  576.0   & i & 18.52 $\pm$ 0.06 & 60   & $129.36 \pm  7.12$ & This work  \\
COLIBRÍ & DDRAGO &  648.0   & i & 18.65 $\pm$ 0.08 & 60   & $114.73 \pm  8.11$ & This work  \\
COLIBRÍ & DDRAGO &  756.0   & i & 18.73 $\pm$ 0.08 & 60   & $106.87 \pm  7.56$ & This work  \\
COLIBRÍ & DDRAGO &  828.0   & i & 18.73 $\pm$ 0.08 & 60   & $106.59 \pm  7.49$ & This work  \\
COLIBRÍ & DDRAGO & 1116.0   & i & 19.02 $\pm$ 0.09 & 300  & $17.89  \pm  6.81$ & This work   \\
COLIBRÍ & DDRAGO & 1512.0   & i & 19.32 $\pm$ 0.19 & 300  & $81.54  \pm  6.78$ & This work   \\
COLIBRÍ & DDRAGO & 1908.0   & i & 19.70 $\pm$ 0.28 & 300  & $62.17  \pm 11.08$ & This work  \\
COLIBRÍ & DDRAGO & 2304.0   & i & 19.92 $\pm$ 0.10 & 300  & $43.60  \pm 11.29$ & This work  \\
COLIBRÍ & DDRAGO & 2988.0   & i & 20.12 $\pm$ 0.11 & 300  & $35.63  \pm  3.40$ & This work   \\
COLIBRÍ & DDRAGO & 3384.0   & i & 20.22 $\pm$ 0.25 & 300  & $29.57  \pm  2.97$ & This work   \\
COLIBRÍ & DDRAGO & 4644.0   & g & 20.69 $\pm$ 0.25 & 300  & $27.15  \pm  6.19$ & This work   \\
COLIBRÍ & DDRAGO & 4752.0   & r & 20.64 $\pm$ 0.41 & 300  & $16.00  \pm  4.00$ & This work   \\
COLIBRÍ & DDRAGO & 4824.0   & i & 20.90 $\pm$ 0.31 & 300  & $14.50  \pm  4.20$ & This work   \\
50cm  &  Harlingten & 28080.0   & r & 21.67 $\pm$ 0.52 & 960  & $6.92 \pm 3.31$& This work   \\
VLT&FORS2      & 76032.0  & i & 21.91 $\pm$ 0.10 & 720  & $4.37 \pm 0.49 $  & This work\\
VLT&FORS2      & 76032.0  & r & 22.17 $\pm$ 0.06 & 720  & $5.70 \pm 0.57 $  & This work\\
COLIBRÍ & DDRAGO &  85860.0  & r & 21.94 $\pm$ 0.27 & 5640 & $5.46   \pm  1.35$   & This work \\
COLIBRÍ & DDRAGO & 172260.0  & r & 21.63 $\pm$ 0.24 & 5160 & $7.15   \pm  1.61$   & This work \\
COLIBRÍ & DDRAGO & 259955.1  & r & 22.28 $\pm$ 0.25 & 4020 & $3.96   \pm  0.93$   & This work \\
VLT&FORS2      & 421632.0 & r & 23.36 $\pm$ 0.07 & 1440 & $1.46 \pm 0.41$ & This work \\
VLT&FORS2      & 423360.0 & i & 23.00 $\pm$ 0.15 & 720  & $2.08 \pm 0.31$ & This work \\
\toprule
\toprule
\multicolumn{8}{c}{\textbf{Radio}}  \\
\toprule
\toprule
Telescope&Instrument & Mid Time [s] & Band & & Exposure Time [s] & Flux [$\mu$Jy] & References \\
\midrule
\midrule
ATCA &     \ldots     &    360.0 & 9~GHz      &      \ldots      & 36000& $<24$ & \cite{39501}\\
VLA &      \ldots     & 168480.0 & X (10~GHz) &      \ldots      & 900  & $63.00 \pm 9.00$   & This work\\
ATCA &     \ldots     & 190080.0 & 9~GHz      &      \ldots      & 14400& $233.00 \pm 15.00$ & \cite{39501}\\
VLA &      \ldots     & 859680.0 & X (10~GHz) &      \ldots      & 900  & $200.00 \pm 12.00$ & This work\\
\bottomrule
\end{tabular}
\end{table*}

\subsection{Radio}
%%%%%%%%%%%%%%%%%%%%%%%%%%%%%%%%%%% 
We observed the field of GRB~250221A with the Karl G. Jansky Very Large Array (VLA) at a central frequency of 10~GHz (X-band), using a total bandwidth of 4~GHz. The first epoch was obtained at a mid-time of $T_0+1.95$~days post-trigger. At the time of this observation, the array was transitioning from configuration A to D. To account for this, the longer baselines were excluded from the visibilities, resulting in an angular resolution of 7.3~\arcsec. The total integration time on source was 15~minutes. Data were flagged and calibrated using the VLA online {\sc CASA} pipeline v6.6.1 \citep{casa}, and imaged and cleaned using the {\sc tclean} task \citep{casa} with a Briggs weighting factor of $0.5$. We detected a weak source at the target position with a flux density of 63$\pm$9~$\mu$Jy \citep{39433}. Given the early epoch of these observations, the measurement may be affected by interstellar scintillation.

As part of our radio campaign, we performed a second observation with the VLA on 2025 March 3 (at a mid-time of $T_0+9.95$~days) with the same frequency set-up. The array configuration was D resulting in an angular resolution of 7.2~\arcsec. The total integration time on the source was 15~minutes. The source flux density increased to (200$\pm$12)~$\mu$Jy, similar to the ATCA value reported eight days before \citep{39501}. For our VLA campaign, the root mean square {\itshape rms} noise in the cleaned radio map and flux density statistical uncertainty were determined using the {\sc imstat} task \citep{casa} in a region of the map far from bright radio emission. 

Figure~\ref{fig:LC} includes the radio photometry described above. In this plot, we notice a significant difference between our measurements at $T_0 + 2.0$~d and the fluxes reported by \cite{39501} at $T_0 + 2.4$~days.
Such large fluctuations are not uncommon in radio afterglows of highly relativistic sources such as GRBs and FXTs \citep{Yadav2025}. 

%%%%%%%%%%%%%%%%%%%%%%%%%%%%%%%%%%%%%%%%%%%%%
\subsection{Data Analysis and Photometry}
\label{sec:dataanalysis}
%%%%%%%%%%%%%%%%%%%%%%%%%%%%%%%%%%%%%%%%%%%%
The afterglow of GRB~250221A is projected close to two galaxies (see Section~\ref{sec:environment}). Therefore, we performed image subtraction using the Saccadic Fast Fourier Transform \texttt{(SFFT)} \citep{sfft} software using as templates the third epoch of observations with FORS2 for $r$ and archival images from Pan-STARRS DR1 \citep{Magnier2020} for $g$ and $i$ (See Figure~\ref{fig:field}, panel c).

The PSF photometry was performed using a custom pipeline and the calibration including \texttt{SExtractor} \citep{Bertin1996}. To minimize systematic errors in the calibration, we calibrated all our photometry using the Pan-STARRS PS1 DR2 Catalog \citep{Magnier2020}.

Our magnitudes are in the natural AB system without the application of any colour terms and are corrected for the Galactic extinction in the direction of the burst $E_{(B-V)}=0.0473 \pm  0.0015$ \citep{Schlegel1998}, which implies extinctions of $A_g=0.18$, $A_r=0.13$, and $A_i=0.09$.

The detailed photometry from COLIBRÍ, Harlingten 50-cm, and VLT is listed in Table~\ref{tab:observations} and shown in Figure~\ref{fig:LC}.

%%%%%%%%%%%%%%%%%%%%%%%%%%%%%%%%%%%%%%%%%%%%%
\section{Host Galaxy}
\label{sec:hg}
%%%%%%%%%%%%%%%%%%%%%%%%%%%%%%%%%%%%%%%%%%%%%
\subsection{Environment}
\label{sec:environment}
%%%%%%%%%%%%%%%%%%%%%%%%%%%%%%%%%%%%%%%%%%%%%
Figure \ref{fig:field} shows the field of GRB~250221A from the VLT observations in $R$ filter at $T_0+0.9$ and $T_0+11.9$ days, along with the corresponding difference image highlighting the afterglow emission.

A bright galaxy is located about 1.15{\arcsec} to the NW of the optical afterglow.  This object is shown with a green circle and will be referred to from now on as {\itshape G1}. According to the Legacy Survey DR10 \citep{Dey2019},  its AB magnitudes are $g=22.17\pm0.02$, $r=20.86\pm0.01$, $i=20.34\pm0.01$, and $z=20.05\pm0.01$.
At the afterglow position, another fainter galaxy is visible  with AB magnitudes of $g=24.02\pm0.06$, $r=24.09\pm0.08$, $i=23.86\pm0.09$, and $z=24.21\pm0.26$ \citep{Dey2019}. This object is shown with a magenta circle and will be referred to from now on as {\itshape G2}. The properties of both objects are summarized in Table~\ref{tab:host}.

Following \cite{Bloom2002} and the galaxy density estimation from deep optical imaging \citep{Metcalfe2001,Kashikawa2004,McCracken2003}, the probability of finding an unrelated galaxy of magnitude $m_r$ or brighter within the vicinity of a GRB can be approximated as:
\begin{equation}
\label{pcc}
    P_{\rm ch} =1-\exp {\left(-\pi r^2_{\rm i} \times 10^{a (m_r - m_0) + b } \right )},
\end{equation}
with $a=0.36$, $b = -2.42$, and $m_0=24$ for galaxies fainter than $m_r\gtrsim19$~mag
and $a=0.56$, $b = -4.80$ and $m_0=18$ for brighter galaxies. 

The effective radius $r_{\rm i}$ depends on the projected angular separation $R_0$ between the GRB and the galaxy and on the half-light radius $R_\textrm{half}$ of the galaxy. We take $r_{\rm i}=2R_{\rm half}$ for G2 since GRB 250221A is localized inside the detected light, whereas for G1 we use $r_{\rm i}=(R^2_{\rm 0}+4R^2_{\rm half})^{1/2}$ because the GRB position is outside its detected light.

Using the photometric magnitudes from the Legacy Survey DR10 \citep{Dey2019}, $r=20.86$ for {\itshape G1} and $r=24.09$ for {\itshape G2}, we estimate chance coincidence probabilities of $P_{\rm ch} = 0.009$ and $P_{\rm ch} = 0.003$, respectively (see Table~\ref{tab:host}), which makes both candidates equivalent possible host galaxies of GRB~250221A. Therefore, using only standard positional arguments, the host galaxy and distance of GRB~250221A would remain uncertain. 

\begin{table*}
	\centering
	\caption{Properties of the galaxies in the vicinity of GRB~250221A. We identify G2 as the true host galaxy of GRB~250221A; see Section~\ref{sec:redshift}.}
	\label{tab:host}
 \begin{tabular}{lcccccccccc}
		\hline
Label&RA&DEC&$z$&$m_r$& $R_\textrm{half}$ [\arcsec] & $R_{\rm 0}$ [\arcsec]&  Offset $\delta R$ [kpc]& $d_{L}$ [Gpc] & $d_{A}$ [Gpc]\\%&$P_{\rm ch}(<R_{\rm 0})$\\
\hline
G1 & 59.4625 & -15.1331 & 0.343 & 20.86 $\pm$ 0.01 &0.77& 1.15 & 11$\pm$5 & 1.95 & 1.06 \\%& 0.009\\
G2 & 59.4628 & -15.1333 & 0.768 & 24.09 $\pm$ 0.08 &0.42\footnote{There is no information of the $R_\textrm{half}$ due to in the LS catalog, this source is classified as a point source. We assume a radial profile and report the FWHM.}& 0.29 & $<7_{-7}^{+12}$ & 4.91 & 1.57 \\%& 0.003\\
\hline
\end{tabular}
\end{table*}

\begin{figure*}
\centering
\includegraphics[width=1.0\textwidth]{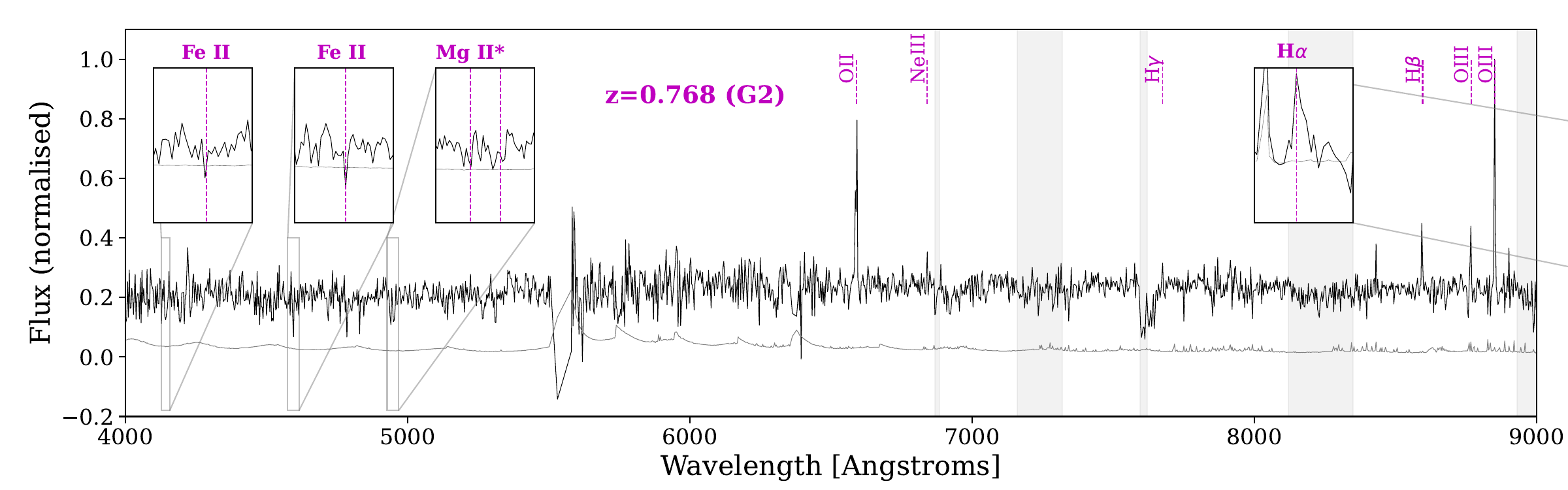}
 \caption{VLT/X-shooter optical spectrum of GRB~250221A (black line) and its corresponding 1$\sigma$ error (gray line), obtained at $T_0 + 21.4$hours. We identify emission lines of [\ion{O}{II}], [\ion{Ne}{III}], H$\gamma$, H$\beta$, and [\ion{O}{III}] along with ISM absorption lines from \ion{Mg}{II}, \ion{Fe}{II}, all consistent with a redshift of $z = 0.768$ (magenta). We also zoom in on the H$\alpha$ emission line (which, for readability, is not shown in the main figure) using the same y-axis interval adopted for the estimation of the star formation rate of {\itshape G2} (see Section~\ref{sec:redshift}).
 We also illustrate the telluric lines (gray shaded regions).
}
 \label{fig:spectrum}
\end{figure*}

%%%%%%%%%%%%%%%%%%%%%%%%%%%%%%%%%%%%%%%%%%%%%
\subsection{Determination of the redshift and identification of the host galaxy}
\label{sec:redshift}
%%%%%%%%%%%%%%%%%%%%%%%%%%%%%%%%%%%%%%%%%%%%%
We determined the redshift by continuum-subtracting the spectra, detecting significant emission and absorption features, and cross-matching them with a set of expected rest-frame lines (\ion{Fe}{II}, \ion{Mg}{II}, \ion{O}{II}, \ion{O}{III}, \ion{Ne}{II}, \ion{Ne}{III}, and Balmer transitions). Candidate redshifts were refined by minimizing residuals between observed and predicted wavelengths, and confirmed through consistency across multiple lines. 

In the combined spectrum, shown in Figure~\ref{fig:spectrum}, we detect strong evidence of the presence of emission lines of \ion{O}{III} (5007~{\AA}), \ion{O}{III} (4959~{\AA}), \ion{H}{$\beta$} (4861~{\AA}), \ion{O}{II} (3726/3729~{\AA}), \ion{Ne}{III} (3859~{\AA}), and the absorption ISM signatures of the lines \ion{Mg}{II} (2796/2803~{\AA}) and \ion{Fe}{II} (2344/2600~{\AA}) at a common redshift of $z = 0.768$. Due to the presence of these features on the GRB afterglow, one can be certain that the burst lies behind the absorber — otherwise, the continuum would not be attenuated. Therefore, we can unambiguously identify {\itshape G2} as the host galaxy of GRB~250221A. 
This result is consistent with the spectroscopic redshift reported by \cite{39418}. 
This result underscores the value of early-time afterglow spectroscopy, which allows an unequivocal identification of the host galaxy, something not always achievable statistically (see Section~\ref{sec:environment}).

Moreover, considering this value of the redshift, this corresponds to a poorly constrained projected offset of $<7$~kpc (see Table~\ref{tab:host}) which is consistent with typical offsets observed for short GRBs \citep[e.g.,][]{Fong2022,Nugent2022,OConnor2022}.

Additionally, the star-formation rate (SFR) of {\itshape G2} was derived from the H$\alpha$ emission line in its rest-frame optical spectrum. Using the measured H$\alpha$ flux of $F(\mathrm{H}\alpha) = (1.31 \pm 0.59) \times 10^{-16}$~erg~s$^{-1}$~cm$^{-2}$, we obtain an H$\alpha$ luminosity of $L(\mathrm{H}\alpha) = 3.8 \times 10^{41}$~erg~s$^{-1}$. The H$\beta$ flux, $F(\mathrm{H}\beta) = (4.67 \pm 1.00) \times 10^{-17}$~erg~s$^{-1}$~cm$^{-2}$, yields a Balmer decrement $\mathrm{H}\alpha/\mathrm{H}\beta = 2.81 \pm 1.40$, consistent with the case~B recombination value of 2.86 and indicating negligible internal extinction \citep{Osterbrock2006}. Following the calibration of \citet[][their Eq.~2]{Kennicutt1998}, we derive an SFR of $\sim3~M_\odot\,\mathrm{yr}^{-1}$ assuming a Salpeter initial mass function \citep{Salpeter1955}. This value is consistent with those typically measured for GRB host galaxies at redshift $z<1.6$ 
\citep[see Fig.~11 in][]{Savaglio2009} and for short duration GRBs \citep{Berger2009,Cucchiara2013,Berger2014,Li2016}.

%%%%%%%%%%%%%%%%%%%%%%%%%%%%%%%%%%%%%%%%%%%%
\section{Data Analysis and Interpretation}
\label{sec:interpretation}
%%%%%%%%%%%%%%%%%%%%%%%%%%%%%%%%%%%%%%%%%%%%

%%%%%%%%%%%%%%%%%%%%%%%%%%%%%%%%%%%%%%%%%%%%
\subsection{GRB Properties: Classification, Energy and Luminosity Estimations}
%%%%%%%%%%%%%%%%%%%%%%%%%%%%%%%%%%%%%%%%%%%%
\subsubsection{Duration}
%%%%%%%%%%%%%%%%%%%%%%%%%%%%%%%%%%%%%%%%%%%%
We retrieved the data products from the {\itshape Swift Burst Analyser} (ObsID: 01290305000) and constructed the {\itshape Swift}/BAT light curve of the prompt emission using the \texttt{HEASoft/FTOOLs} software package\footnote{\url{http://ftools.gsfc.nasa.gov/}} \citep{HEAsoft}. The light curve was generated with the \texttt{batbinevt} routine, which creates a mask-weighted light curve from the {\itshape Swift}/BAT event data. The resulting light curve spans the 15--350~keV energy range with uniform 1~s time bins.

\begin{figure}
	\includegraphics[clip, width=0.98\linewidth]{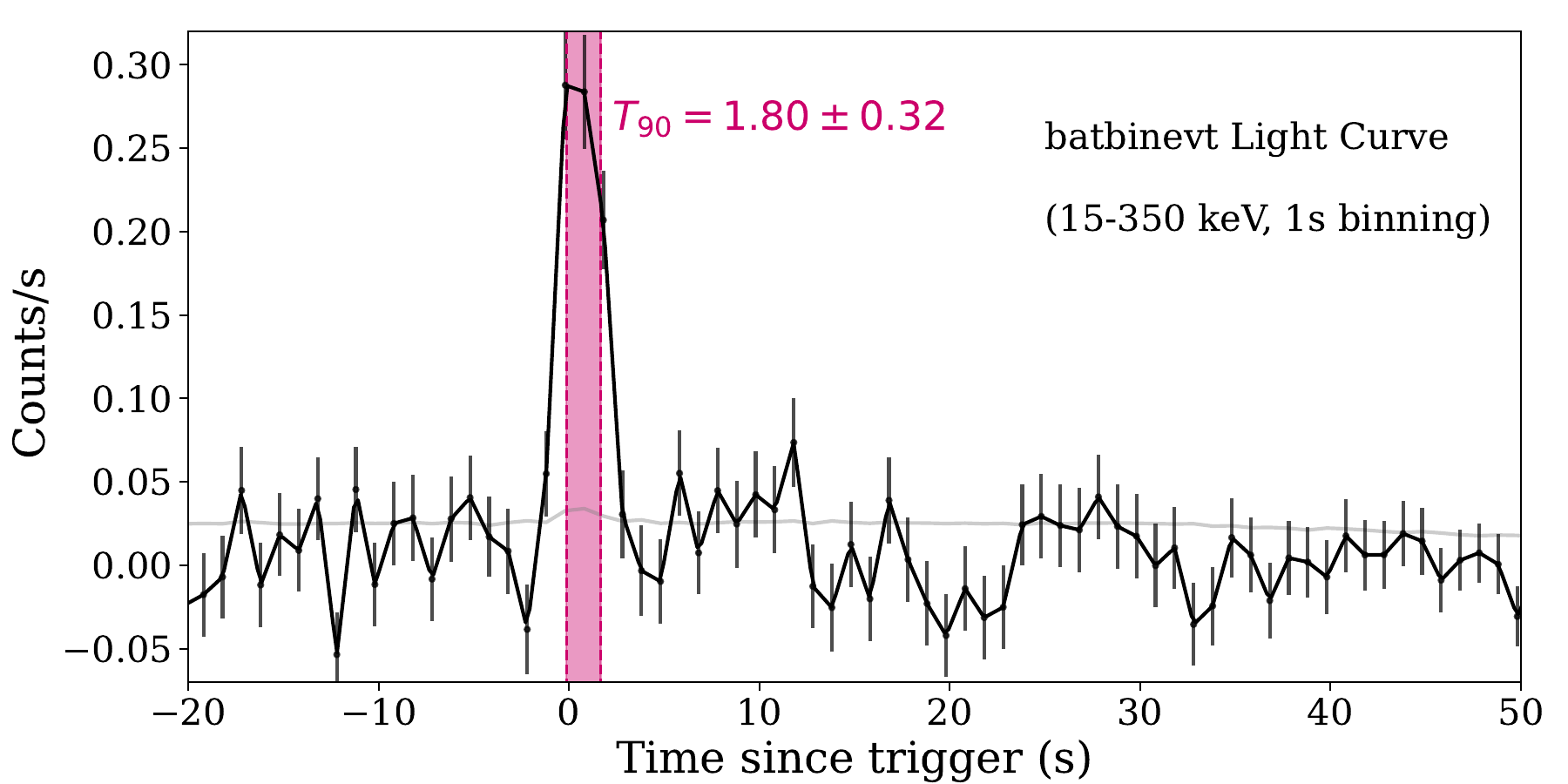}
    \caption{Masked-weighted {\itshape Swift}/BAT (15--350~keV) light curve of GRB~250221A with 1~s binning produced with the \texttt{HEASoft/FTOOLs} software (black solid line). The background count rate is shown in gray. The $T_{90}$ duration of $1.80 \pm 0.32$~s is indicated by the shaded magenta region.
    \label{fig:lcbat}}
\end{figure}

The {\itshape Swift}/BAT light curve (15--350~keV) displayed in Figure~\ref{fig:lcbat} shows a single peak structure, with a duration of $T_{90}=1.80\pm0.32$~seconds.
The fluence was estimated with a value of (3.9$\pm$0.6)$\times10^{-7}$~erg~cm$^{-2}$ in the 15--150~keV band \citep{39471}.

%%%%%%%%%%%%%%%%%%%%%%%%%%%%%%%%%%%%%%%%%%%%
%To be written by Karelle
\subsubsection{Hardness ratio}
\begin{figure}
\centering
\includegraphics[width=0.95\linewidth]{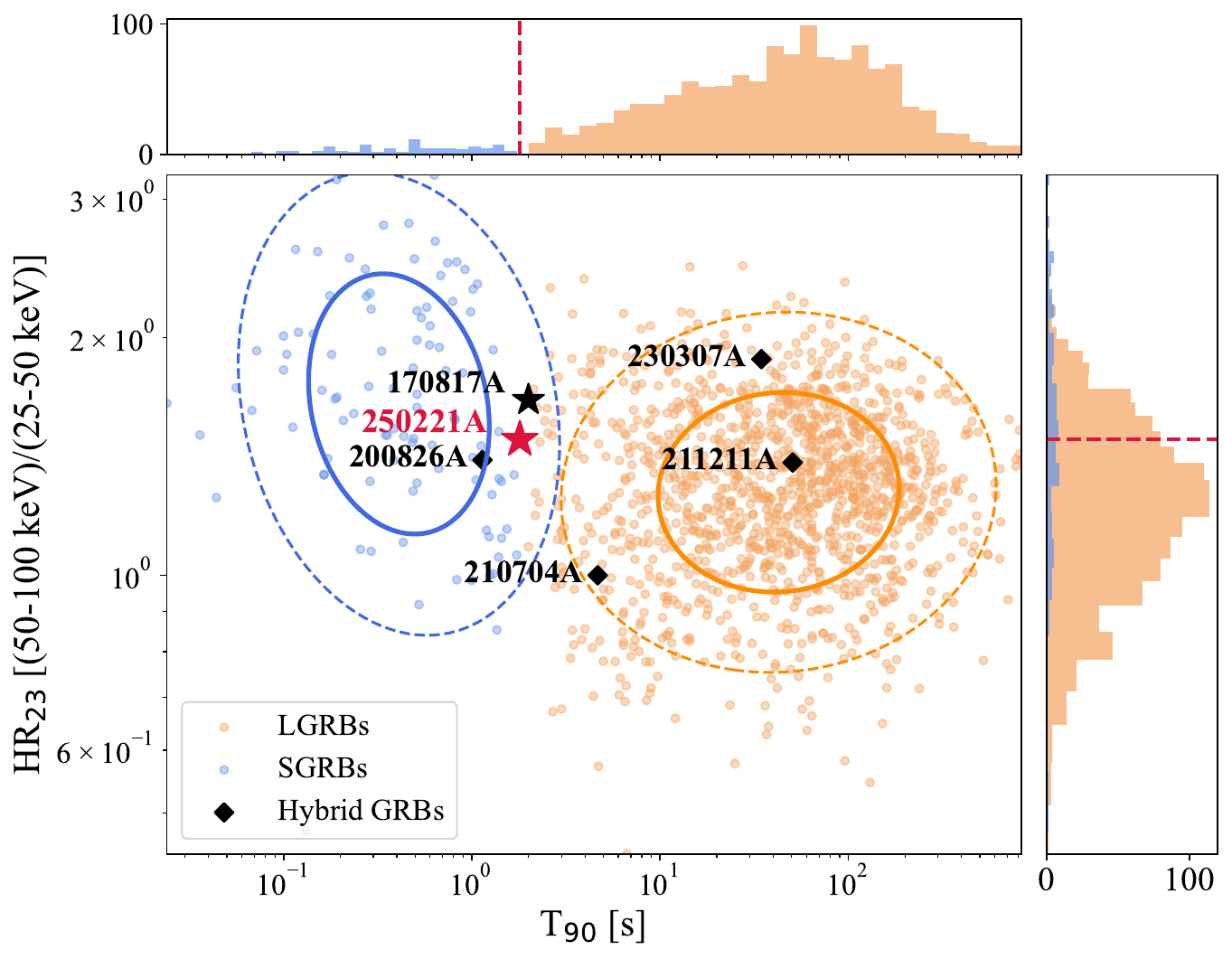}
\caption{HR versus $T_{90}$ diagram for GRB~250221A (crimson star), plotted alongside a sample of {\itshape Swift}/BAT GRBs \citep{Lien2016}. Division between LGRBs (orange) and SGRBs (blue) is done automatically by {\itshape Swift}/BAT based on $T_{90}$ duration of 2~seconds. Hybrid GRB events (black diamonds) are poorly classified using a purely duration based classification scheme, resulting in ambiguity in progenitor assessment. Both GRB~250221A and GRB~170817A (black star) lie in the ambiguous region between the GRB classes in this space, and within the wide spread of the hybrid GRBs.}
\label{fig:hr}
\end{figure}

%To be written by Cassidy
%%%%%%%%%%%%%%%%%%%%%%%%%%%%%%%%%%%%%%%%%%%%
Figure~\ref{fig:hr} shows the HR versus $T_{90}$ diagram for {\itshape Swift}/BAT GRBs with GRB~250221A. The HR value is calculated using the fluence data from the {\itshape Swift}/BAT Gamma-Ray Burst Catalogue \citep{Lien2016} power-law fit, taking the ratio of Band 3 ($50-100$~keV) over Band 2 ($25-50 $~keV).

In the figure, GRB~250221A is indicated by a crimson star. We see that it lies between the peaks of the two distributions, outside the $2\sigma$ (95\%-confidence) region for LGRBs but well within the corresponding region for SGRBs.

Additionally, other notable GRBs are shown in the figure for comparison. GRB~250221A lies close in the diagram to GRB~170817A, the GRB with the first and best observed KN counterpart. Both of these fall within the large distribution of the hybrid GRBs, which scatter widely in both $T_{90}$ and HR, showing again the issue with duration-only classification.

%%%%%%%%%%%%%%%%%%%%%%%%%%%%%%%%%%%%%%%%%%%%
\subsubsection{Amati Correlation}
%%%%%%%%%%%%%%%%%%%%%%%%%%%%%%%%%%%%%%%%%%%%
Using the fluence estimated by \textit{Swift}/BAT of (3.9$\pm$0.6)$\times10^{-7}$~erg~cm$^{-2}$, and following the results from Section~\ref{sec:redshift}, where we determined the redshift, we estimate the isotropic energy and luminosity to be $E_{\rm iso} = (1.1\pm0.2)\times10^{51}$~erg and $L_{\rm iso} = (6.3\pm1.5)\times10^{50}$~erg~s$^{-1}$.

Based on the time-averaged spectral analysis performed with the {\sc batgrbproduct} tool over the interval [-0.220, 1.844]~s, the cutoff power-law peak energy in the time-averaged spectral analysis from \textit{Swift}\footnote{\url{https://swift.gsfc.nasa.gov/results/batgrbcat/BAT_refined_circular/1290305/}}, $E_{\mathrm{p}} = 165.2$~keV. Given the limited sensitivity of BAT above $\sim$100–150 keV, the reported $E_{\rm p}$ should be interpreted as a lower limit, indicating that the burst is spectrally hard.

To better characterise the prompt emission, we therefore adopt the spectral parameters measured by Konus-Wind. They observed the increase in the count rate in its 20--400~keV band from $T_0-1.5$~s to $T_0+4.4$~s. Using a power-law with exponential cutoff (CPL) model, they estimate a fluence of $(1.67\pm0.09)\times10^{-6}$~erg cm$^{-2}$ and a rest-frame peak spectral energy $E_\mathrm{p,z}$ of (428$\pm$44)~keV.

\begin{figure}
\centering
\includegraphics[width=0.95\linewidth]{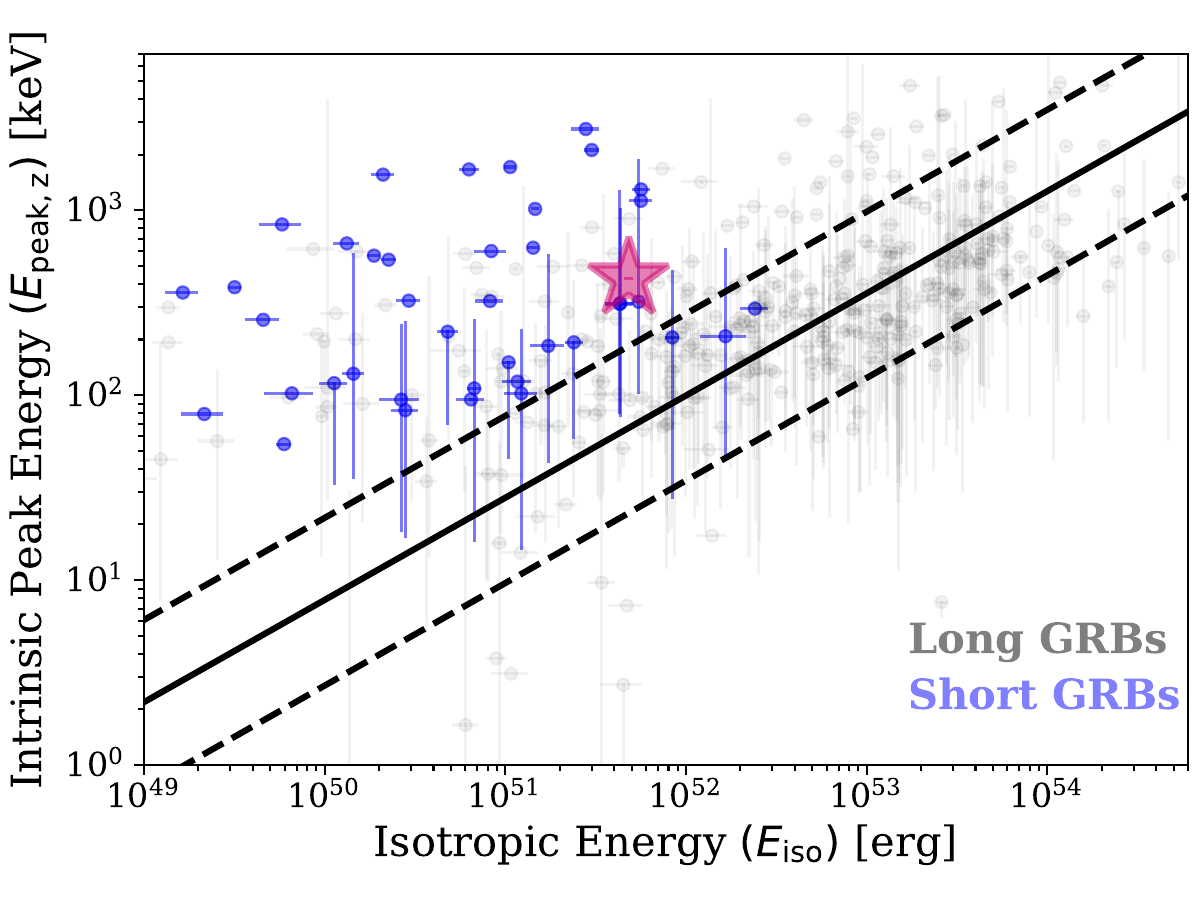}
 \caption{Isotropic energy and intrinsic peak energy for GRB~250221A using the values reported by Konus-Wind \citep{39423} (pink star). Long GRBs ($T_{90}>2$~seconds) are shown in grey and short GRBs ($T_{90}<2$~seconds) are shown in blue. The Amati relation, based on {\itshape Swift}/BAT GRBs with spectroscopic redshifts, is also shown \citep[solid black line,][]{Amati2008,Amati2013}. The $2\sigma$ confidence region of the relation is indicated by dashed lines.} 
 \label{fig:amati}
\end{figure}

Figure~\ref{fig:amati} shows the resulting relations between the isotropic energy and intrinsic peak energy for GRB~250221A. These values are compared to those in the {\itshape Swift}/BAT GRB database. We distinguish between LGRBs (gray points) and SGRBs (blue points). The Amati relation for LGRBs (solid black line) and the $2\sigma$ (95\%) confidence region of the Amati relation for GRBs \citep{Amati2008,Amati2013} are also shown. Redshift values are drawn from the {\itshape Swift}/BAT Gamma-Ray Burst Catalogue\footnote{\url{https://swift.gsfc.nasa.gov/results/batgrbcat/}}, supplemented with spectroscopic measurements from \citet{Fong2022}.
Comparing the properties of our burst, we use the value of $z=0.768$ (see Section~\ref{sec:redshift}) and the values reported by Konus-Wind \citep{39423}. Figure~\ref{fig:amati} shows that GRB~250221A falls outside the 95\%-confidence region for LGRBs and aligns more closely with the short GRB population.

We estimate the isotropic energies, the luminosities and the rest-frame peak energies for {\itshape G2}, the host galaxy, to be $E_{\rm iso}$ = (4.6$\pm$2.6)$\times 10^{51}$~erg, $L_{\rm iso}$ = (1.6$\pm$0.9)$\times 10^{51}$~erg~s$^{-1}$ and $E_{\mathrm{p,z}}$ of 427.9~keV. These values are consistent with the compact object merger progenitor \citep[see Figure~13 from][]{Tsvetkova2017}. \citet{Tsvetkova2017} introduced a Type-I/II \citep[see e.g.][]{Zhang2009} classification that uses several observational diagnostics, including host-galaxy properties, the GRB’s projected offset, the presence or absence of an associated supernova, and its location in the $T_{90}$–hardness ratio plane. Within this framework, Type I bursts are expected to share a common physical origin—namely, compact-object mergers.

%%%%%%%%%%%%%%%%%%%%%%%%%%%%%%%%%%%%%%%%%%%%
\subsection{Afterglow Model}
\label{sec:lcmodel}
%%%%%%%%%%%%%%%%%%%%%%%%%%%%%%%%%%%%%%%%%%%%
Within the fireball model framework \citep{Rees-Meszaros-92,Meszaros-Rees-93,Meszaros1997}, the afterglow emission originates from external shocks produced by the interaction of the relativistic jet with the surrounding circumstellar medium \citep{Sari-Piran-95,Sari-97}. Here we consider an ultra-relativistic, thin spherical shell, propagating inside an external medium with constant number density $n$. Initially, the shell is coasting with a bulk Lorentz factor (LF) $\Gamma_0\gg1$ and is slowed down by sweeping up the external medium in its path. This interaction gives rise to a double collisionless shock structure. A \textit{forward} shock propagates ahead of the shell with $\Gamma_{\rm sh}=\sqrt{2}\Gamma$ and shock-heats the swept up material that moves with bulk LF $\Gamma$ behind it. 
A \textit{reverse} shock propagates through the shell and extracts its kinetic energy while shock-heating the ejecta. When the mass of the swept up material exceeds $M_0/\Gamma_0$, (where $M_0 = E_{\rm k,iso}/(\Gamma_0 c^2)$ is the baryon load of the shell, $E_{\rm k,iso}$ is its kinetic energy, and $c$ the speed of light), the shell starts to decelerate. 
The subsequent dynamical evolution of the forward shock is given by the \citet{Blandford-Mckee-76} self-similar solution, with $\Gamma_{\rm sh}\propto R^{-3/2}$. At this moment most of the kinetic energy of the shell is transferred to the shocked swept up material, resulting in a peak in the afterglow light curve and a power-law decline thereafter.

The forward shock accelerates a fraction of the electrons ($0<\xi_e<1$; \citealt{Eichler-Waxman-05}) entering the shock into a power-law energy distribution, such that their comoving number density behind the shock is $n_e'(\gamma)\propto\gamma^{-p}$ for electrons with $\gamma \geq \gamma_m$ (where $\gamma$ is the LF of the electrons and $\gamma_m$ is the LF of minimal-energy electrons). The collisionless shock 
also compresses any pre-existing magnetic fields in the upstream medium and generates new magnetic fields in-situ. The relativistic electrons behind the shock cool by radiating broadband synchrotron radiation, which is described by a power-law that is broken smoothly at multiple characteristic frequencies \citep{Sari1998}. At late times, the afterglow emission is in the slow-cooling regime where the flux density, $F_{\nu} \propto T^{\alpha} \nu^{\beta}$, peaks at $\nu_m<\nu_c$, where $T$ is the time in the observer frame and $\nu_m$ is the peak synchrotron frequency of minimal-energy electrons \citep{Granot2002}. At higher frequencies ($\nu > \nu_m$) the spectrum breaks at the characteristic cooling-break frequency, $\nu_c$, that corresponds to the peak synchrotron frequency of electrons cooling at the dynamical time. At frequencies $\nu_m < \nu < \nu_c$, the spectral index is $\beta = (1 - p)/2$, while for $\nu > \nu_c$, it becomes $\beta = -p/2$. We interpret the observed afterglow radiation in this framework and use the known scaling relations \citep{Sari1998,Granot2002} to constrain physical parameters.

The early ($T < 3\times10^4$~s) optical and X-ray data can be modelled with simple power-laws having slopes $\alpha_\mathrm{o} = -0.93 \pm 0.01$ and $\alpha_\mathrm{x} = -0.97 \pm 0.04$, respectively (see Figure~\ref{fig:LC}). 
We interpret this emission to come from the external forward shock and the similarity in the temporal decay suggests that both emissions are produced by the same synchrotron power-law segment (which is at $\nu_m<\nu<\nu_c$). Comparing the early optical decay slope $\alpha_\mathrm{o} = -0.93 \pm 0.01$ 
with the expected one in a constant ambient density ($F \propto T^{3(1-p)/4}$), we obtain an electron power-law index ($p = 2.24 \pm 0.01$) that is consistent with values reported for other GRBs \citep[e.g.,][]{Kumar2015}. This result is incompatible with a wind environment, where the expected decay slope ($F \propto T^{(1-3p)/4}$) would imply $p = 1.57 \pm 0.01$, an unusually low value for GRBs \citep[see, e.g.,][]{Zhang2004,Chevalier2000}.

\begin{figure}
    \centering
    \includegraphics[width=1.1\linewidth]{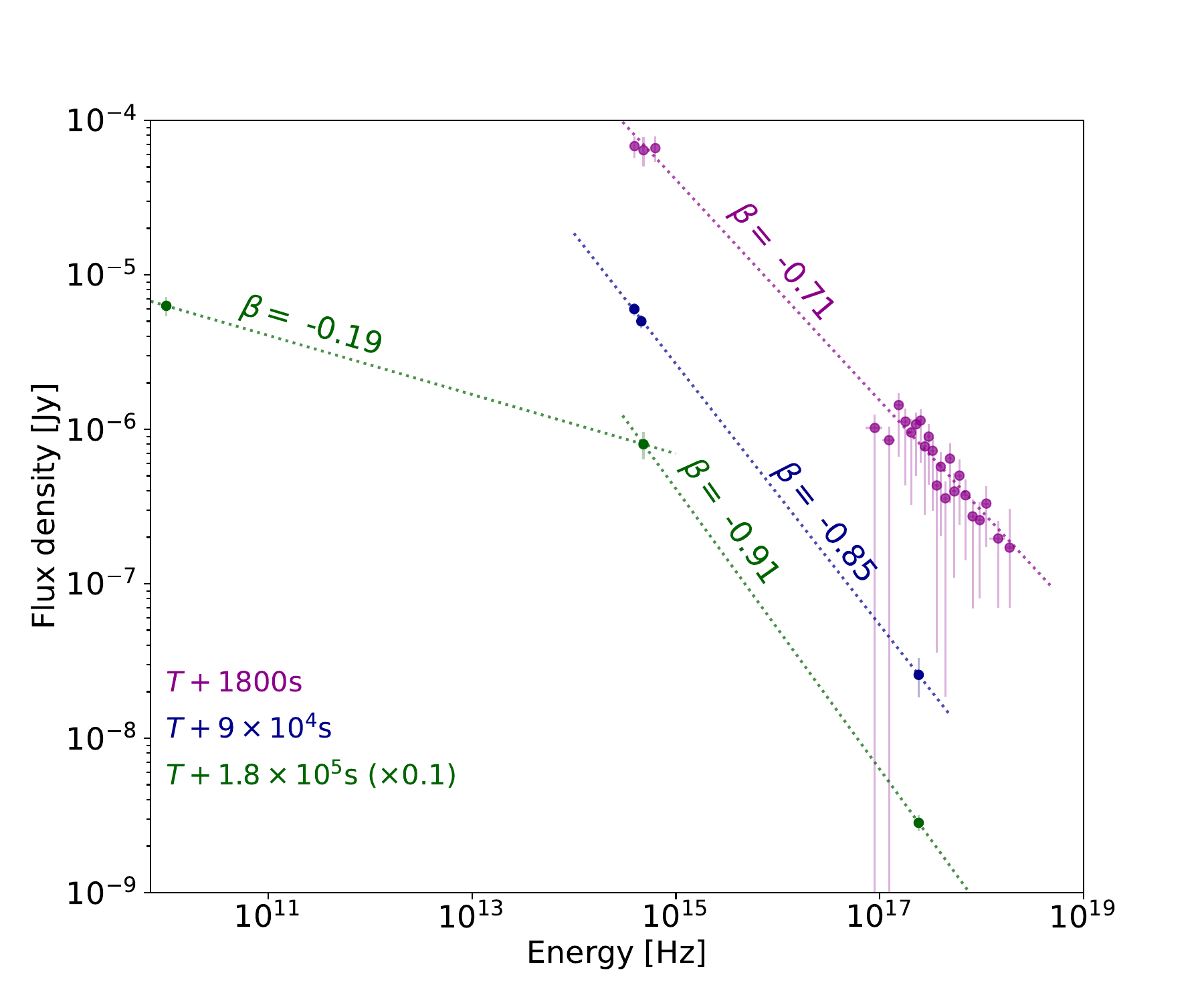}
    \caption{Evolution of the SED of GRB~250221A at three different epochs. The first epoch at an average time of $T_0 + 1800~s$ (purple), the second epoch at $T_0+9\times 10^{4}$~s (blue), and the third epoch at $T_0+1.8 \times 10^{5}$~s (green). } 
    \label{fig:sed}
\end{figure}

Figure~\ref{fig:sed} presents the SED for the early emission phase of GRB~250221A, built using the data from \textit{Swift}/XRT and COLIBRÍ, corresponding to an average photon arrival time of $T = 1800$s for the XRT data. The optical flux densities for the early SED were interpolated from the light curve model described in the previous paragraph, evaluated at the same epoch ($T = 1800$~s). The X-ray data were binned in groups of at most 10 bins with at least a 10~$\sigma$ significance. Two other epochs at $T=9\times 10^{4}$~s and $T=1.8 \times 10^{5}$~s are also shown, and were taken directly from the light curve, as X-ray data were scarce at later times, and a time-sliced spectrum was not possible. 

%%%%%%%% FIGURE %%%%%%%%%%%%%%%%%%%%%%%%%%%%%%%%
%%%%%%%%%%%%%%%%%%%%%%%%%%%%%%%%%%%%%%%%%%%%%%%%

The first two epochs are well described by simple power-laws with spectral index $\beta = -0.71 \pm 0.03$ and $\beta = -0.85 \pm 0.01$, respectively. The third epoch, covering the peak of the rebrightening in optical and the apparent plateau in X-rays, is best described by two power-law segments: the first one going from the radio to optical energies, with a spectral index $\beta = -0.19 \pm 0.02$, the second one going from the optical to X-ray energies, with a spectral index $\beta = -0.91 \pm 0.04$. Given the limited number of data used to infer the radio-to-optical spectral slope, the exact position of the spectral break is uncertain and the low-energy spectral slope (-0.19) is not well constrained and should be interpreted as a lower limit.

Therefore, in a slow-cooling synchrotron regime, with an electron power-law index of $p = 2.24 \pm 0.01$, we obtain an optical spectral index of $\beta = (1-p)/2 =-0.62 \pm 0.01$. This value differs from the spectral fit $\beta = -0.71 \pm 0.03$ at $T+1800$~s, that gives an electron power-law index of $p=2.42 \pm 0.06$. The flux density for $\nu_m < \nu < \nu_c$ at time $T = 1800~s (T_{3.26}$~s\footnote{In this work we use the shorthand notation of $Q_n=Q/10^n$ for $Q$ expressed in cgs units.}) is thus given by \citep{Granot2002}
\begin{eqnarray}
    &F_\nu \simeq 0.087 \left(\frac{1+z}{1.768}\right)^{1.31}\,
    E_{\rm k,iso,52}^{1.31}\,
    n_{-1}^{0.5}\,
    \epsilon_{e,-1}^{1.24}\,
    \epsilon_{B,-3}^{0.81}\,
    T_{3.26}^{-0.93}\,& \\
    & \times\,\nu_{14}^{-0.62}\,
    d_{L,28.18}^{-2}\,{\rm mJy}\,,& \nonumber
\end{eqnarray}
when assuming fiducial values for the model parameters and for $\xi_e=1$ and $z=0.768$ with the corresponding luminosity distance of $d_{L}=1.52\times10^{28}$\,cm. Comparing this scaling relation with optical observations at $T=1800$\,s and $\nu=4.81\times10^{14}$\,Hz, with $F_\nu=6.4\times10^{-2}$\,mJy, allows us to constrain the kinetic energy,
\begin{equation}\label{eq:Ekiso-constraint}
    E_{\rm k,iso} = 1.67\times10^{52}\,
    n_{-1}^{-0.38}\,
    \epsilon_{e,-1}^{-0.95}\,
    \epsilon_{B,-3}^{-0.62}\,
{\rm erg}\,,
\end{equation}
in terms of fiducial parameters whose values are not yet determined from other constraints.

An additional constraint on $E_{\rm k,iso}$ comes from the efficiency of the prompt $\gamma$-ray emission, $\eta_\gamma = [1 + (E_{\rm k,iso}/E_{\gamma,\rm iso})]^{-1}\gtrsim15\%$ \citep[e.g.][]{Beniamini+16}, which would limit the blast wave energy to 
$E_{\rm k,iso}<[(1-\eta_\gamma)/\eta_\gamma]E_{\gamma,\rm iso} = 6.24\times10^{51}$\,erg if this level of efficiency was realized in this GRB.\footnote{However, we do not apply this efficiency constraint (which is not robust enough and can vary among different GRBs, especially those showing a rebrightening).}

%%%%%%%% FIGURE %%%%%%%%%%%%%%%%%%%%%%%%%%%%%%%%
\begin{figure*}
    \centering
    \includegraphics[width=0.48\linewidth]{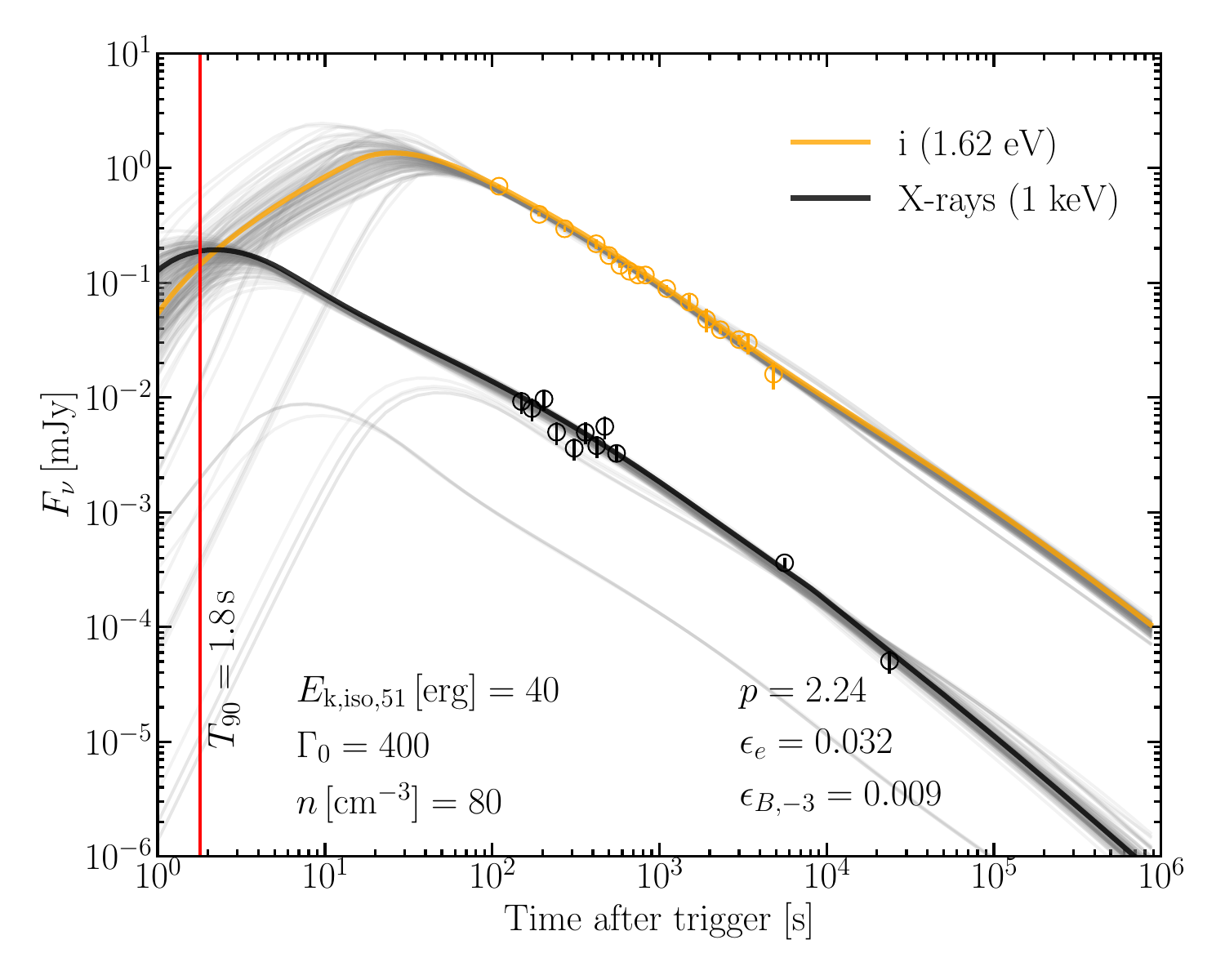}
    \includegraphics[width=0.48\linewidth]{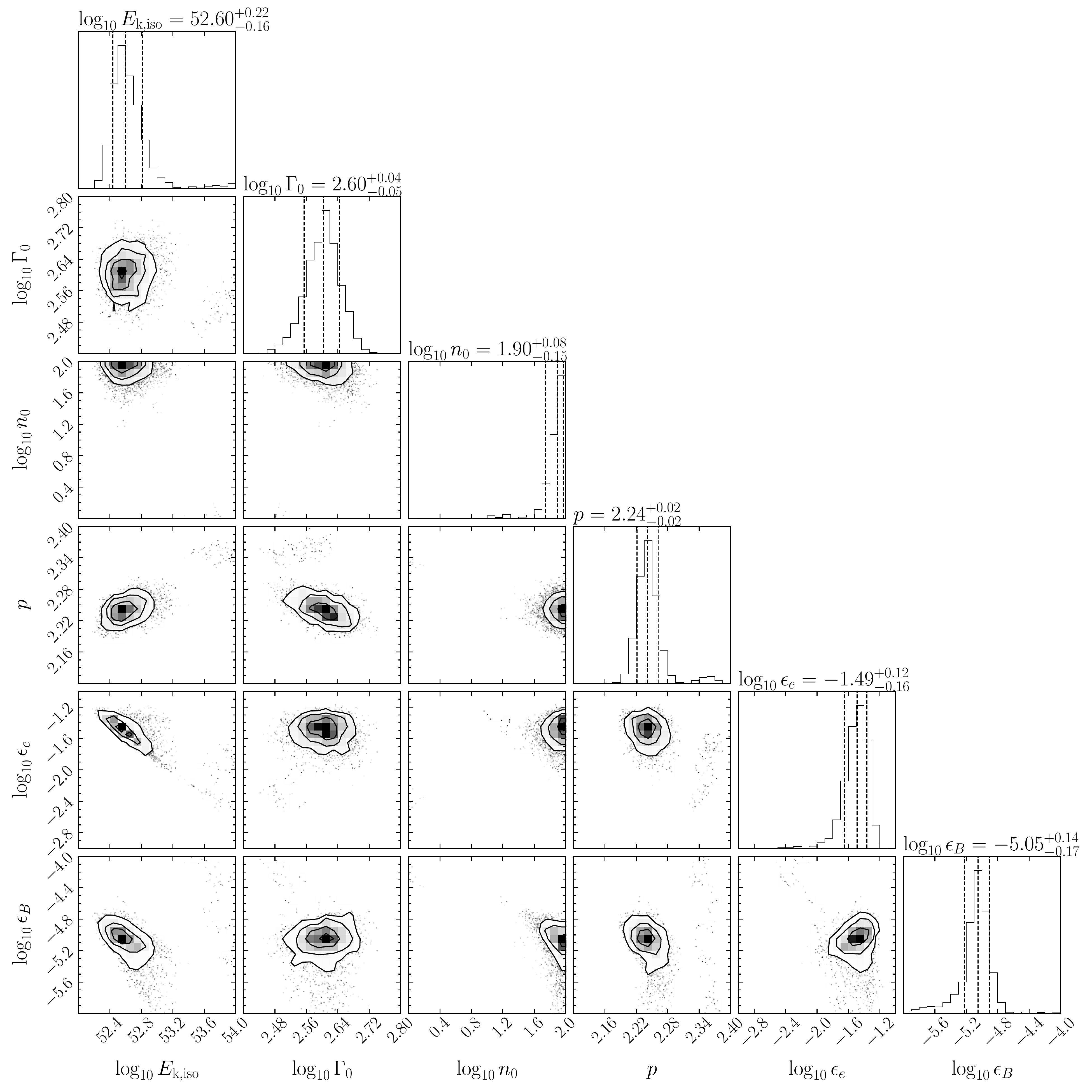}
    \caption{(Left) Afterglow model light curve fit to observations using the model from \citet{Gill-Granot-18} that features a blast wave from a spherical shell moving in a ($k=0$) constant density ISM. The 100 thin gray light curves are obtained from sampling the posterior distribution of 
    model parameters, and the thick black and orange light curves are shown for model parameters that best fit the observations. The red vertical line shows the $T_{90}$ duration of 
    the prompt GRB. 
    (Right) Model parameter posterior distributions obtained from an MCMC light curve fit.}
    \label{fig:lc-fit}
\end{figure*}
%%%%%%%%%%%%%%%%%%%%%%%%%%%%%%%%%%%%%%%%%%%%%%%%

A constraint on the interstellar medium (ISM) density can be obtained from the deceleration time of the blast wave, which also corresponds to the peak time of the afterglow light curve. The blast wave decelerates when the swept-up mass, $M_{\rm sw}(R)=(4\pi/3)R^3m_pn = M_0/\Gamma_0 = E_{\rm k,iso}/\Gamma_0^2c^2$, that yields, $R_{\rm dec} = (3E_{\rm k,iso}/4\pi m_pc^2n\Gamma_0^2)^{1/3}$, where $\Gamma\simeq\Gamma_0$ 
is the coasting bulk LF of the ejecta at $R<R_{\rm dec}$. The photon arrival time in the observer frame corresponding to emission produced at $R=R_{\rm dec}$ is:
\begin{eqnarray}
    T_{\rm dec} &&\simeq (1+z)\frac{R_{\rm dec}}{2\Gamma_0^2c} 
    = (1+z)\left(\frac{3 E_{\rm k,iso}}{32\pi m_pc^5n\Gamma_0^8}\right)^{1/3} \\
    &&\simeq 429 \left(\frac{1+z}{1.768}\right) E_{\rm k,iso,52}^{1/3}n_{-1}^{-1/3}\Gamma_{0,2}^{-8/3}\,{\rm s}\,. \nonumber
\end{eqnarray}
Since the afterglow light curve shown in Figure\,\ref{fig:LC} does not show a peak, the blast wave deceleration time must be $T_{\rm dec}<10^2$\,s. We use this constraint along with Equation~\,(\ref{eq:Ekiso-constraint}), to constrain the ISM density in terms of the shock microphysical parameters, with $n>n_{\min}$, 
\begin{equation}\label{eq:n-constraint}
    n_{\rm min} = 2.1\,
    \epsilon_{e,-1}^{-0.69}\,
    \epsilon_{B,-3}^{-0.45}\,
    \Gamma_{0,2}^{-5.79}\,
    {\rm cm}^{-3}
\end{equation}
assuming that $\Gamma_0=100\,\Gamma_{0,2}$, which is the minimum LF needed for an on-axis burst to overcome the compactness problem and emit a GRB. SGRBs are typically found with significant offsets from the centers of their respective host galaxies, with 50 percent showing an offset in excess of 5~kpc \citep[e.g.][]{Fong-Berger-13,OConnor2022}. The ISM density is then expected to be low with $10^{-3}\lesssim n ({\rm cm}^{-3}) \lesssim 1$ \citep{Berger2014,OConnor2020}. In fact, \citet{OConnor2020} find that the central 60 per cent of SGRBs in their sample have ISM densities in the range $10^{-4}\lesssim n_{\rm ISM}/{\rm cm^{-3}}\lesssim 10^{-1.5}$ and only a minute fraction have densities above $1\,{\rm cm^{-3}}$. Therefore, the minimum ISM density found for this source (see MCMC fit below) represents the extreme and may even disfavour a compact binary merger scenario.

The optical band is above the peak synchrotron frequency of the minimal energy electrons, which is given by \citet{Granot2002} as
\begin{equation}
    \nu_m = 6.43\times10^{10}\,(1+z)^{1/2}\,
    E_{\rm k,iso,52}^{1/2}\,
    \epsilon_{e,-1}^2\,
    \epsilon_{B,-3}^{1/2}\,
    T_{\rm 1\,day}^{-3/2}\,{\rm Hz}\,,
\end{equation}
and this constraint gives a weak lower limit on $\epsilon_B > \epsilon_{B,\min}$, 
\begin{equation}
    \epsilon_{B,\min} = 1.8\times10^{-10}\,
    \epsilon_{e,-1}^{-1.53}\,
    \Gamma_{0,2}^{4.94}\,.
\end{equation}
An additional constraint on $\epsilon_B$ can be obtained from noting that $\nu_c>\nu_X=10$\,keV, where
\begin{equation}
    \nu_c = 2.68\times10^{17}\,(1+z)^{-1/2}\,
    E_{\rm k,iso,52}^{-1/2}\,
    n_0^{-1}\,
    \epsilon_{B,-3}^{-3/2}\,
    T_{\rm 1\,day}^{-1/2}\,{\rm Hz}\,,
\end{equation}
which yields $\epsilon_B < \epsilon_{B,\max}$, 
\begin{equation}
    \epsilon_{B,\max} = 3.1\times10^{-4}\,
    \epsilon_{e,-1}^{1.24}\,
    \Gamma_{0,2}^{5.65}\,.
\end{equation}
When assuming the minimum value of $\epsilon_B$ we constrain the minimum ISM density using Eq.(\ref{eq:n-constraint}) to $n_{\rm min}\simeq2.3\times10^3\Gamma_{0,2}^{-8.01}\,{\rm cm^{-3}}$. This density is extremely high and never realized for GRBs. It can be reduced to a more acceptable value by raising $\Gamma_0$ to a larger value, e.g. $\Gamma_0=400$ (as obtained below from our MCMC fit), which yields $n_{\min}\simeq0.034\Gamma_{0,2.6}^{-8.01}\,{\rm cm^{-3}}$. We then use the minimum ISM density to obtain a constraint on the isotropic-equivalent kinetic energy of the blast wave from above using Eq.(\ref{eq:Ekiso-constraint}), $E_{\rm k,iso,max}\simeq5.5\times10^{54}\Gamma_{0,2.6}^{-0.02}$\,erg (Here and above we have ignored the dependence on $\epsilon_e$ as it has a very small power). Alternatively, we consider the maximum value of $\epsilon_B$ and find $n_{\min}\simeq3.5\epsilon_{e,-1}^{-1.25}\Gamma_{0,2}^{-8.3}\,{\rm cm^{-3}}$, which further yields $E_{\rm k,iso,max}\simeq8.9\times10^{51}\epsilon_{e,-1}^{-1.24}\Gamma_{0,2}^{-0.34}$\,erg.
The available data is not able to put strong constraints on the model parameters. To understand the spread in the values of the different parameters, we perform a light curve fit to the optical ($\nu_O=3.93\times10^{14}$\,Hz) and X-ray ($\nu_X=2$\,keV) data using  Markov-chain Monte-Carlo (MCMC) methods and a model of a spherical blast wave propagating in a constant density ISM ($k=0$) from \citet{Gill-Granot-18}. The fit is obtained using the data before the re-brightening episode at $T\lesssim3\times10^4$\,s and the results are shown 
in Figure\,\ref{fig:lc-fit}.

The MCMC fit finds an ultrarelativistic blast wave with a coasting bulk LF of $\Gamma_0\sim400$ and an isotropic-equivalent kinetic energy of $E_{\rm k,iso}\gtrsim10^{52}$\,erg. Given that the radiated isotropic-equivalent energy in $\gamma$-rays is around $E_{\gamma,\rm iso}\sim10^{51}$\,erg, it yields a $\gamma$-ray efficiency of $\eta_\gamma\lesssim10\%$. The fit also demands the jet moves in a dense environment with ISM density $n\gtrsim80\,{\rm cm}^{-3}$, which is large for a binary merger scenario 
\citep[e.g.][]{OConnor2020,Fong+15} and more in line with Collapsar produced long GRBs. The fit also finds the peak time of the afterglow, and hence the deceleration time to be $T_{\rm dec}\sim T_{90}$ as can be inferred from the peak of the X-ray afterglow; the optical light curve shows a delayed peak due to the passage of $\nu_m\propto T^{-3/2}$ across the optical band.

%%%%%%%%%%%%%%%%%%%%%%%%%%%%%%%%%%%%%%%%%%%%
\subsection{Nature and Progenitor of GRB~250221A}
\label{sec:nature}
%%%%%%%%%%%%%%%%%%%%%%%%%%%%%%%%%%%%%%%%%%%%
The properties of the GRB environment, such as its redshift, offset from the host galaxy and evidence for star-formation (see Figure~\ref{fig:spectrum}), are consistent with both long and short GRBs. 
However, based on its duration, temporal evolution, the presence of a single sharp spike in the prompt emission, its peak energy, and its position in the Amati correlation, the properties of GRB~250221A fall within the typical range expected for SGRBs \citep{Berger2014}.
These considerations suggest that GRB~250221A may have originated from the coalescence of two compact objects.

Nevertheless, the atypically large ISM density required by the afterglow modelling (Section~\ref{sec:lcmodel}) appears more in line with a Collapsar exploding within its dense progenitor cloud. 

If this was the case, the event would lie at the extreme end of the LGRB population in terms of $T_{90}$, hardness ratio, and peak energy. We will further explore this possibility in the following sections.

%%%%%%%%%%%%%%%%%%%%%%%%%%%%%%%%%%%%%%%%%%%%
\section{Origin of Optical Rebrightening}
\label{sec:discussion}
%%%%%%%%%%%%%%%%%%%%%%%%%%%%%%%%%%%%%%%%%%%%
Figure~\ref{fig:LC} shows evidence for a change in the fading evolution and a rebrightening at $T > 0.6$~d.

Late-time rebrightenings have been reported in several long GRBs: GRB~100814A \citep{Nardini2014} and the ultra-long GRB~111209A \citep{Kann2019,Gendre2013,Greiner2015}. In addition, the {\itshape Einstein Probe} mission \citep{Yuan2025} has recently identified rebrightening events without associated GRBs, for example EP241021a \citep{Gianfagna2025,Yadav2025} and EP240414a \citep{vanDalen2025}, likely representing off-axis explosions. While these may constitute a different phenomenological class, they demonstrate that such late-time rises are not unprecedented among long-duration or SN-associated transients.

In this section, we explore and discuss possible scenarios that could account for the excess in GRB~250221A. 
Given the ambiguity surrounding the origin of the event, we first consider whether the observed excess can be explained by a kilonova, a fast blue optical transient, or a supernova associated with the death of a massive star (Section~\ref{sec:sn}), then we turn to scenarios related to the host galaxies and the surrounding environment, including the possibility of gravitational microlensing (Section~\ref{sec:gm}) or density variations in the CBM (Section~\ref{sec:dv}).
Finally, we investigate whether the excess could be produced by a refreshed shock within the main jet of GRB~250221A (Section~\ref{sec:refreshed-shock}).

%%%%%%%%%%%%%%%%%%%%%%%%%%%%%%%%%%%%%%%%%%%%
\subsection{Kilonova}
\label{sec:kn}
%%%%%%%%%%%%%%%%%%%%%%%%%%%%%%%%%%%%%%%%%%%%

Scenarios involving kilonova origin for the optical excess can be ruled out. An AT2017gfo-like \citep{Abbott2017} event placed at $z=0.768$ would appear at $r\sim28.6$ mag, far below our detection limits. Even the brightest magnetar-powered kilonova models are undetectable at this distance \citep[see e.g.][]{Metzger2010,Metzger2012}.

%%%%%%%%%%%%%%%%%%%%%%%%%%%%%%%%%%%%%%%%%%%%
\subsection{FBOT}
\label{sec:fbot}
%%%%%%%%%%%%%%%%%%%%%%%%%%%%%%%%%%%%%%%%%%%%

Similarly, a fast blue optical transient (FBOT) \citep[see e.g.][]{Perley2019,Perley2021} origin is very unlikely. While these transients reach high luminosities and rise rapidly, the observed negative spectral slopes ($\beta<0$) of the afterglow of GRB 250221A are inconsistent with the expected blue continua of FBOTs ($\beta>0$). Thus, neither a kilonova nor an FBOT provides a plausible explanation for the excess emission.

%%%%%%%%%%%%%%%%%%%%%%%%%%%%%%%%%%%%%%%%%%%%
\subsection{Supernova}
\label{sec:sn}
%%%%%%%%%%%%%%%%%%%%%%%%%%%%%%%%%%%%%%%%%%%%
Firstly, we consider the case where the GRB~250221A event resulted from the collapse of a rapidly rotating, stripped-envelope massive star \citep{Woosley1993}, in which case one might think that the late optical excess observed after $T_0+0.6$~d could be attributed to the emergence of an underlying supernova. 

Broad-lined Type Ic SNe, have been associated with LGRBs through both temporal and spatial coincidence, as well as spectroscopic confirmation \citep[see e.g.][]{Galama1998, Hjorth2003, Pian2006, Mirabal2006, Becerra2017}. These SNe usually present peak absolute magnitudes of $M_V \approx -19$ to $-20$ around 10 to 15 days after the burst in the rest frame \citep[e.g.][]{Cano2017}. This timescale is determined by the radioactive decay of $^{56}$Ni and $^{56}$Co.

At the luminosity distance of $d = 4.9$~Gpc, standard type Ibc supernovae \citep[SN 1998bw-like;][]{Galama1999} would exhibit apparent peak magnitudes in the range of $\sim~24.5.0-25.5$, making them several orders of magnitude fainter than the late-time component observed in GRB~250221A. On the other hand, a very luminous supernova associated with a magnetar such as SN2011kl could be brighter by about one magnitude \cite{Greiner2015,Kann2019} and therefore comparable with the values exhibited for GRB~250221A in the rebrightening.

However, the excess exhibited by GRB~250221A emerges on timescales of hours rather than the typical rise time of several days expected for supernovae \citep{Cano2017}. This discrepancy in both luminosity and temporal behaviour renders a supernova origin an unlikely explanation for the late-time photometric evolution of the transient (see Figure~\ref{fig:LC}).

%%%%%%%%%%%%%%%%%%%%%%%%%%%%%%%%%%%%%%%%%%%%
\subsection{Gravitational Microlensing}
\label{sec:gm}
%%%%%%%%%%%%%%%%%%%%%%%%%%%%%%%%%%%%%%%%%%%%
It is tempting to associate the rebrightening with a microlensing event, assuming that {\itshape G2} is the host and {\itshape G1} is a galaxy on the line-of-sight.  In the following, we briefly describe the formalism for GRB microlensing first introduced by \cite{LoebPerna98}. As discussed in section\,\ref{sec:lcmodel}, the temporal evolution of the spherical  blast wave is described by the \citet{Blandford-Mckee-76} self-similar solution, with $\Gamma\propto R^{-3/2}$, after it has been decelerated by its interaction with 
the ISM.

The afterglow image on the plane of the sky is predicted to appear as a ring with a narrow fractional width, $W$, that depends on the observed frequency and whose radius grows with apparent time $T$ as a power law, with
$R_s(T)\propto R/\Gamma \propto R^{5/2} \propto T^{5/8}$ where $R\propto \Gamma^2T \propto T^{1/4}$.  
In the optical band, $W$ is predicted to be of the order of a few percent \citep{LoebPerna98,Granot99}.
After about a day after the trigger, the radius of the ring, 
$R_s(T)\approx1.1\times10^{17}~E_{\rm k,iso,52}^{1/8}~n_1^{-1/8} T_{\rm day}^{5/8}~{\rm cm}$ 
\citep{Piran1999}, becomes comparable to the Einstein radius of a solar mass lens, $r_E=\sqrt{(4GM/c^2)(D_LD_{LS}/D_S)}$ % 
where $D_L$, $D_{LS}$, and $D_S$ are the lens-observer, lens-source and source-observer angular distances, respectively, $M$ is the mass of the lens. 

The magnification of a uniform ring of fractional width $W$ and impact factor $b$ is given by \citealt{LoebPerna98},
\begin{equation}\label{ring}
\mu(R_s,\ W,\ b)={\Psi(R_s,\ b)-(1-W)^2\Psi[(1-W)R_s,\ b]\over
  1-(1-W)^2},
\end{equation}
where $\Psi(R_s,\ b)$ is the magnification of a uniform disk of radius $R_s$,

\begin{align}
\Psi(R_s,\ b)={2 \over \pi R_s^2} \Big[ \int_{|b-R_s|}^{b+R_s} dR
{{R^2+2}\over{\sqrt{R^2+4}}} {\rm arccos}{{b^2+R^2-R_s^2}\over{2Rb}}\nonumber\\+
H(R_s-b){\pi\over2}(R_s-b)\sqrt{(R_s-b)^2+4}\Big],
\end{align}
and $H(x)$ is the Heaviside step function. The lensed flux is given by $F_\nu^{\rm lensed}(T)=\mu[R_s(T),\ W,\ b] F_\nu (T)$.

\begin{figure}
    \centering
    \includegraphics[width=\linewidth]{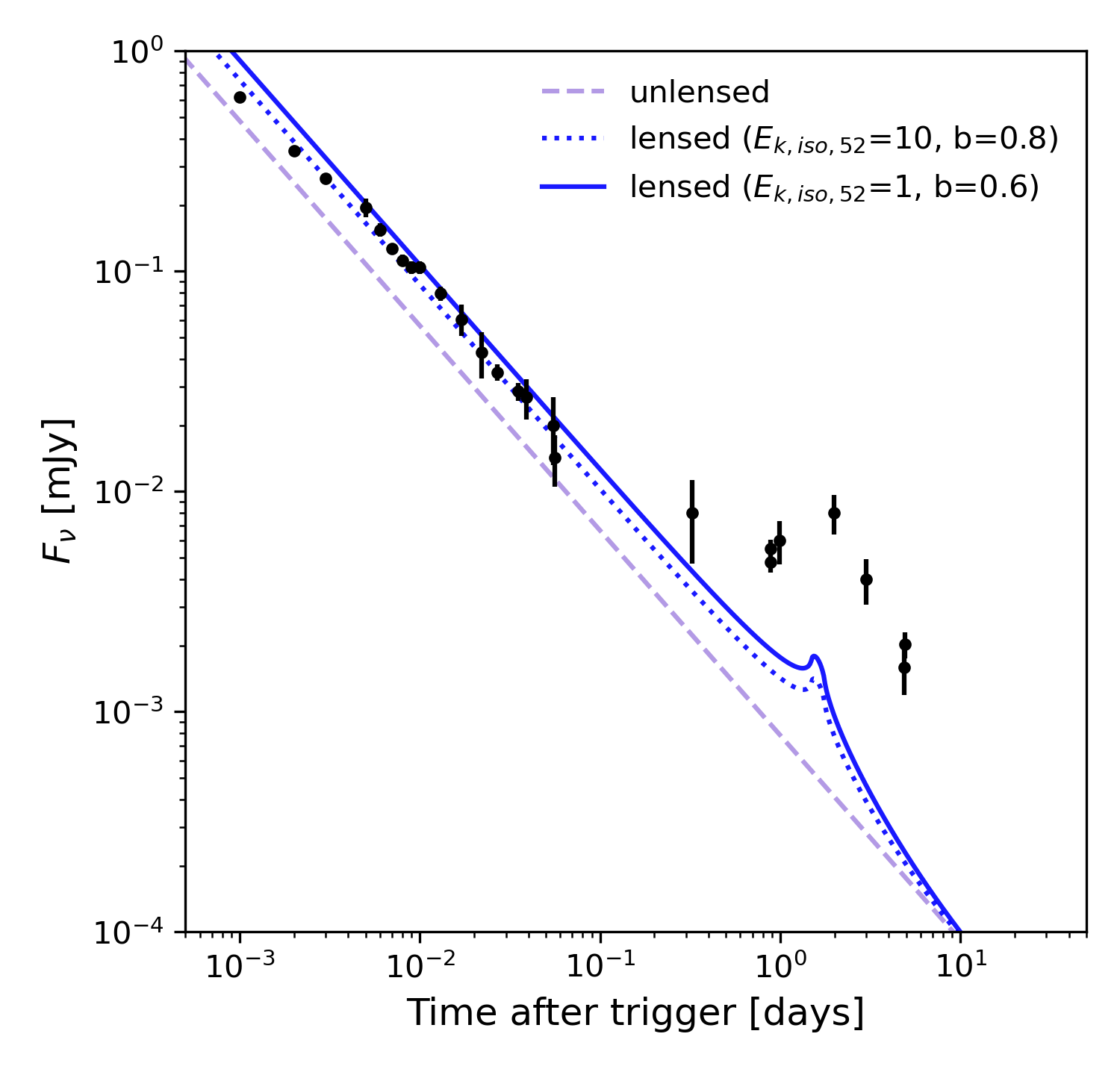}
    \caption{Unlensed (dashed line) and lensed flux from a GRB afterglow with $b=0.6$, $E_{\rm k,iso,52}=1$ (solid line, corresponding to a magnification of $\sim3.5$) and $b=0.8$, $E_{\rm k,iso,52}=10$ (dotted line, corresponding to a magnification of $\sim2.7$) at a frequency of $\sim4.8\times10^{14}$~Hz (optical data; black points). The density is $n_1=5$. The fractional width of the emission ring is W=10\%. The lens mass is M = 50 $M_\odot$, and its redshift is 0.343. The source redshift is 0.768.
    }
    \label{fig:microlensing}
\end{figure}

Therefore, assuming that the lens is in {\itshape G1} and the GRB is in {\itshape G2}, we get $D_L\approx1037$~Mpc, $D_{LS}\approx783$~Mpc, and $D_S\approx1571$~Mpc, respectively \citep{Planck2020} and hence, $r_E\approx 3.1 \times 10^{16}(M/M_\odot)^{1/2}~{\rm cm}$. 
Since the radius of the ring depends on the kinetic energy and the density of the medium, we adopt here a range of values around the values shown in Figure\,\ref{fig:lc-fit}.
We find that for the following ranges $1 \lesssim n_1 \lesssim 5$ and $0.1 \lesssim E_{\rm k,iso,52} \lesssim 10$, that are broadly consistent with those inferred in Sec.\,\ref{sec:lcmodel}, the timescale for the excess demands  $b<1$ and $M\sim50~M_\odot$. The width of the excess demands a fractional width of the ring $W\sim0.1$. However, this modelling does not allow accounting for the amplitude of the excess, as can be seen in Figure~\ref{fig:microlensing}. We note that these microlensing calculations are  done considering a uniform ring for the afterglow image \citep{LoebPerna98}. 

%%%%%%%%%%%%%%%%%%%%%%%%%%%%%%%%%%%%%%%%%%%%
\subsection{ISM Local Density Enhancement}
\label{sec:dv}
%%%%%%%%%%%%%%%%%%%%%%%%%%%%%%%%%%%%%%%%%%%%
As we describe in Section~~\ref{sec:lcmodel}, the fireball model describes the afterglow emission as a smooth light curve that can be modelled with power-law segments. However, in reality, light curve variability is observed in several GRBs \citep{Nakar2003,deUgarte2005}, and can be attributed to an increase in external density that the fireball interacts with, experiencing additional deceleration \citep{Nakar2007}.

In the case of LGRBs, which are predominantly found in star-forming galaxies \citep{Perley2016}, variations in the external density profile may arise due to several factors. These include the presence of massive stellar winds, the transition between the low-density bubble created by the stellar wind and the outer molecular cloud, and density structures within the molecular cloud itself. The interaction of the relativistic jet with such inhomogeneities could produce variability in the afterglow light curve, potentially as early as $T \sim 1$~day post-burst \citep{Tam2005}.

In contrast, for SGRBs, and likely GRB~250221A, this explanation does not apply. Their compact-object progenitors merge long after both stars have exploded as supernovae, leaving behind a CBM free of stellar-wind disturbances. Thus, any external density variations must stem from inhomogeneities in the ISM \citep{Nakar2007}.

Nevertheless, \cite{Nakar2007} demonstrated that density fluctuations in the external medium have a minimal impact on the forward shock emission as the dynamics of the relativistic blast wave tend to suppress any resulting re-brightening. 
Furthermore, \cite{vanErten2009} performed hydrodynamical simulations of a relativistic blast wave interacting with a wind termination shock, finding no significant variability in the optical light curve, even in the presence of large density contrasts. Therefore, these results favour scenarios involving refreshed shocks or multiple ejection episodes as more plausible explanations for the observed variability.

%%%%%%%%%%%%%%%%%%%%%%%%%%%%%%%%%%%%%%%%%%%%
\subsection{Refreshed Shock}\label{sec:refreshed-shock}
%%%%%%%%%%%%%%%%%%%%%%%%%%%%%%%%%%%%%%%%%%%%

%%%%%%%%%%%%%%%%%%%%%%%%%%%%%%%%%%%%%%%%%%%%
\subsubsection{Reverse Shock Emission from Two-Shell Collision}
\label{sec:RS_shell_coll}
%%%%%%%%%%%%%%%%%%%%%%%%%%%%%%%%%%%%%%%%%%%%
To explain the rebrightening of the afterglow emission around starting from $T_0\sim1$\,day (hereafter $T_{\rm bright}$) , we first consider a scenario where a second (inner) shell of mass emitted by the central engine, with some delay of $t_{\rm em}$ in the engine frame, catches up from behind with the decelerating first (outer) shell and refreshes the shock by injecting more energy 
\citep[e.g.][]{Kumar-Piran-00,Sari-Meszaros-00,Zhang-Meszaros-02,Vlasis+11} (see Figure~\ref{fig:shells}). We develop the model of two spherical shell collision in Appendix\,(\ref{sec:ref-shock}), where we calculate characteristic radii, timescales, and dynamics of the system \citep[also see, e.g.,][]{Anderson+25}. As the two shells collide (see Figure\,\ref{fig:shells}) a forward shock (FS) propagates into the outer shell and a reverse shock (RS) propagates into the inner shell. The shock jump conditions at the two shocks are different, as the outer shell is already relativistically hot while the inner shell is cold. The main parameters in the problem include the ratio of the kinetic energies of the two shells ($E_{(6),\rm k,iso}/E_{(3),\rm k,iso}$), the ratio of their initial bulk Lorentz factors ($\Gamma_{(3),0}/\Gamma_{(6),0}$), and the shock microphysical parameters at the two shocks. The parenthesized numbers in the subscript refer to the regions shown in Fig.\,\ref{fig:shells}.

\begin{figure}
\centering\includegraphics[width=\linewidth]{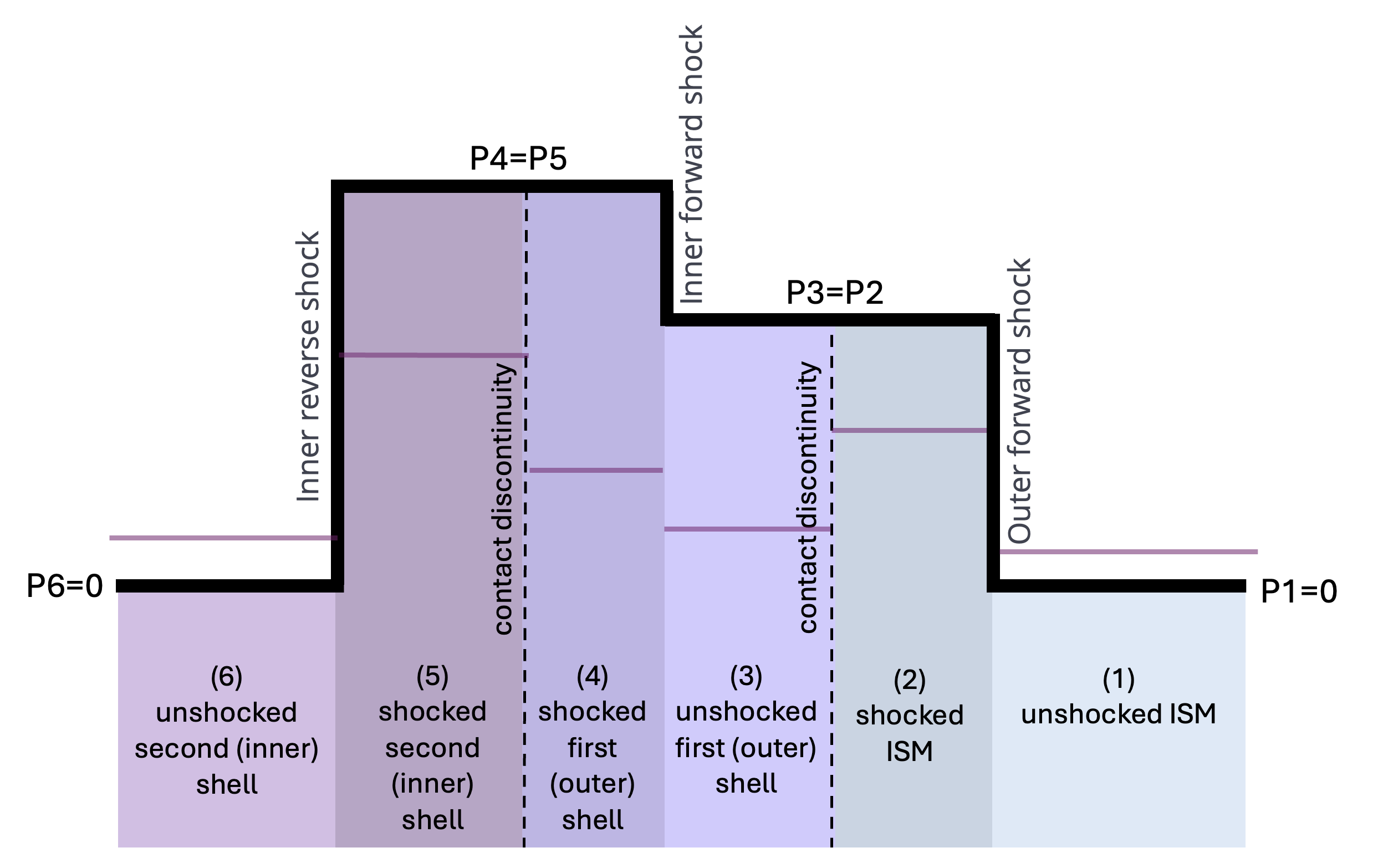}
    \caption{
    Sketch (not to scale) of the different shocked and unshocked regions that arise in the collision 
    of two  
    kinetic-energy-dominated shells. 
    The thick solid line indicates the pressure, or the internal energy density, in each region. 
    The thin solid line indicates the comoving mass density. Figure adapted from \citet{Zhang-Meszaros-02}.
    }
    \label{fig:shells}
\end{figure}

In the case of a binary merger, the two shells must be emitted with negligible delay, such that $t_{\rm em}\ll T_{\rm 90}$, and therefore the inner shell must have $\Gamma_{(6),0}<\Gamma_{(3),0}$. This condition also guarantees that the inner shell must be spreading radially before the collision, i.e. it is in the thin-shell regime. Also, in general, if $t_{\rm em}\ll T_{\rm bright}$, then the ratio of the initial bulk Lorentz factors is fixed (see Equation\,\ref{eq:Gamma_0-ratio}), such that
\begin{eqnarray}
    \frac{\Gamma_{(3),0}}{\Gamma_{(6),0}}&&\approx\frac{[4(T_{\rm bright}/T_{\rm dec})]^{3/8}}{2} \\
    &&\simeq 56 \left(\frac{1+z}{1.768}\right)^{-3/8}E_{\rm k,iso,52.6}^{-1/8}n_{1.9}^{1/8}T_{\rm bright,4.8}^{3/8}
    \quad({\rm G2})\,. \nonumber
\end{eqnarray}
Since $\Gamma_{(3),0}\sim400$ is obtained from the MCMC fits, this yields $\Gamma_{(6),0}\sim8$. Figure\,\ref{fig:shell-dynamics} shows the dynamical evolution of the different shocked and unshocked regions, where we find that the RS remains non-relativistic as it crosses the inner shell, as expected for a thin shell.

As shown in \citet{Zhang-Meszaros-02}, the RS emission typically dominates over that coming from the other two shocked regions before it declines rapidly. The shock jump conditions in Equation\,\ref{eq:shock-jump-BM76} yield the energy and number density of the shocked medium, which we then use to calculate the characteristic synchrotron frequencies and the peak flux \citep[e.g.][]{Sari1998, Gill-Granot-18}. In calculating the observed emission, we simplify to obtain radiation emitted only along the line-of-sight (LOS) without performing equal-arrival-time-surface (EATS) integration, with photon arrival time given by $T = (1+z)R/4\Gamma^2c$.

\begin{figure}
    \centering
    \includegraphics[width=0.48\textwidth]{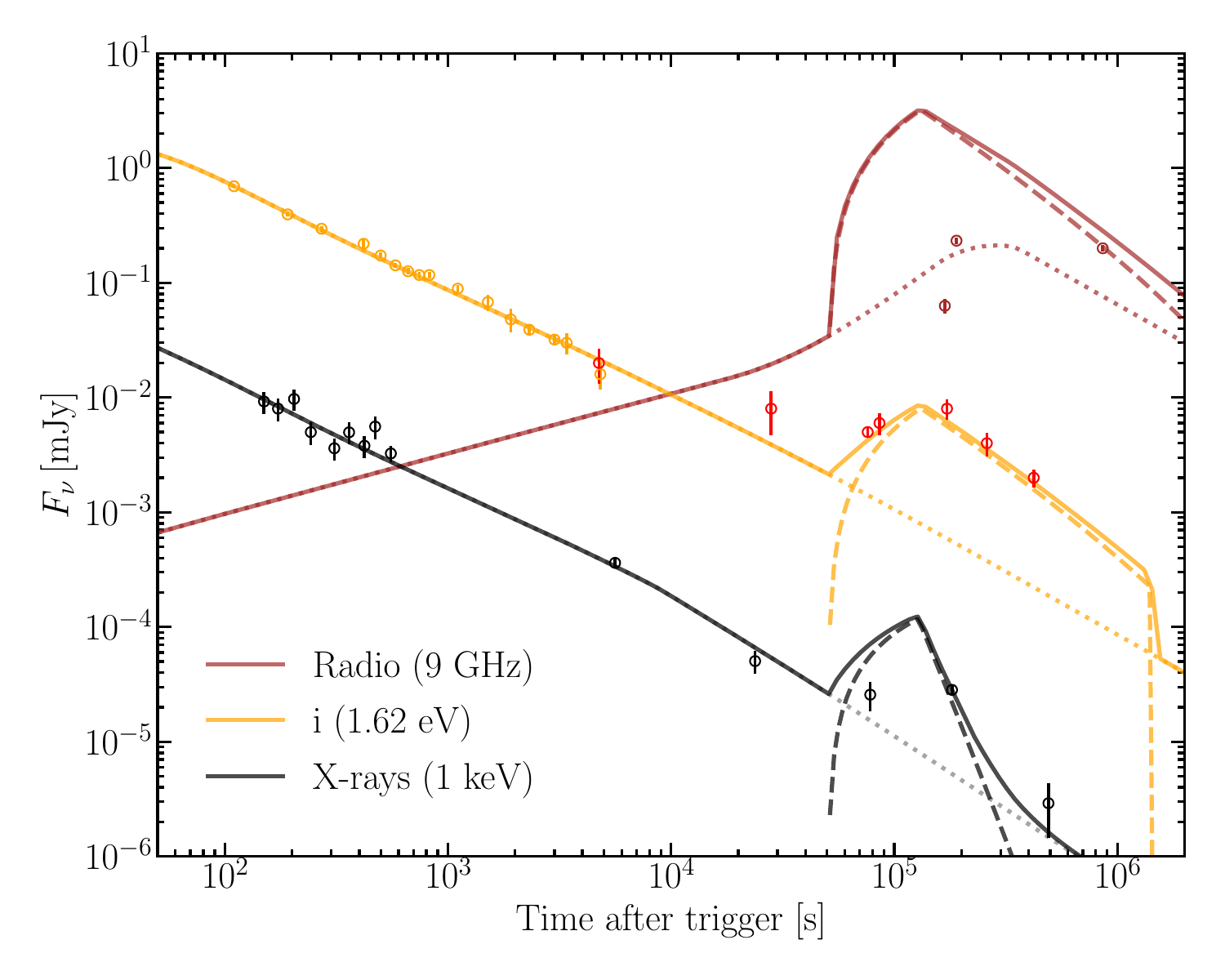}
    \caption{
    Best-fit light curve as obtained in Figure\,\ref{fig:lc-fit} (dotted), now with the addition of reverse shock emission arising from the two spherical shell collision (dashed), with the sum of the two emissions shown with a solid line. The shock-microphysical parameters for the emission from the refreshed shock are $\epsilon_e=0.04$ and $\epsilon_B = 3\times10^{-5}$ with $p=2.1$.
    }
    \label{fig:shell-emission}
\end{figure}

Figure\,\ref{fig:shell-emission} shows the RS emission (dashed) from the inner shocked shell in addition to that produced at the FS (dotted) from the outer shell. The RS emission rises to a peak that marks the instant when the shock finishes crossing the inner shell. After RS passage, the shock is no longer active and no newly accelerated particles are being injected behind the shock. Therefore, the emission declines and also cuts off for $\nu>\nu_{\rm cut}$ \citep{Kobayashi-00}, which affects the X-rays first and then the optical at a later time, where both show a sudden drop in the light curve. In reality, the light curve does not drop so suddenly as emission from angles away from the LOS still arrives with an angular delay of $T_\theta\simeq (1+z)R/4\Gamma^2c\simeq1.3\times10^5(1+z)R_{17.6}\Gamma_{0.7}^{-2}$\,s, causing the light curve to be smeared over this timescale. Nevertheless, this model has a drawback: it overproduces the radio emission by approximately an order of magnitude compared to the observations (see Figure~\ref{fig:shell-emission}). We will investigate this apparent disagreement in the next section.

The two shell collision scenario, as described here, demands a significantly large amount of energy to be residing in the inner shell. In this case, we find that $E_{(6),\rm k,iso}/E_{(3),\rm k,iso}\simeq20$ is required in order to obtain the correct radius at which the two colliding shells are shocked (see Fig.\,\ref{fig:shell-dynamics}) and produce the rebrightening, which for $\Gamma_{(6),0}\sim8$ (see above) implies a large density in the inner shell compared to the outer one. Since the RS is non-relativistic, large densities are needed to obtain a bright emission. When such a large amount of energy is injected into the blast wave, it should also lead to very bright FS emission (not calculated in this model) after the RS emission fades, which would significantly overproduce the emission during the decaying phase of the rebrightening episode. Therefore, this model is disfavoured by the observations.

%%%%%%%%%%%%%%%%%%%%%%%%%%%%%%%%%%%%%%%%%%%%
\subsubsection{Energy Injection into the Forward Shock of a Jet}
\label{sec:FS_energy_inj}
%%%%%%%%%%%%%%%%%%%%%%%%%%%%%%%%%%%%%%%%%%%%
Next we consider an alternative model with continuous energy injection into the blast wave produced by a jet. Two different ways of doing so have been discussed in the literature for a spherical flow, where (i) the ejecta comprises a radial velocity stratification, with progressively slower moving inner shells trailing behind the faster outer shell \citep{Rees-Meszaros-98,Sari-Meszaros-00}, or (ii) a rapidly spinning central engine, e.g. a millisecond magnetar, continuously injects energy into the blast wave via a magneto-hydrodynamical (MHD) wind as it spins down \citep{Dai-Lu-98a, Dai-Lu-98b, Zhang-Meszaros-01, Zhang-Meszaros-02}. In both scenarios, the afterglow light curve shows an achromatic bump or rebrightening as the energy is injected, with the post-injection light curve following the same temporal decay as that of the pre-injection one in the absence of spectral break passage. Both scenarios have been used in many earlier works \citep[e.g.,][]{Laskar+18, Schroeder+24, deWet+24, Schroeder+25} to explain rebrightening features in afterglow light curves, with the assumption that the trailing ejecta in model (i) only catches up with the faster moving blast wave at the time of the rebrightening.

Here we consider energy injection over a narrow radial width $\Delta R=R_2-R_1$, where injection commences at radius $R_1$ and ceases at $R_2$. Total energy $E_{\rm inj}$ is injected as a power-law in radius at the rate 
\begin{equation}
    \frac{dE}{dR} = \frac{(1+\ell)}{\Delta R}E_{\rm inj}\left(\frac{R-R_1}{\Delta R}\right)^\ell \propto R^\ell\,,
\end{equation}
so that the energy of the blast wave grows as $E(R)\propto R^{1+\ell}$. From energy conservation in the blast wave \citep{Blandford-Mckee-76,Chiang-Dermer-99,Huang+99,Zhang-Meszaros-02}, we get
\begin{equation}
    E_{\rm inj} + (\Gamma_0-\Gamma)M_0c^2 = (\Gamma^2-1)M_{\rm sw}c^2\,,
\end{equation}
where $M_0$ and $M_{\rm sw}$ are the masses of the ejecta and swept-up external medium, which simplifies to the known result of $E_{\rm k,iso}=\Gamma_0M_0c^2=\Gamma^2M_{\rm sw}c^2$ for 
$\Gamma_0\gg\Gamma$ and $\Gamma\gg1$ in the absence of energy injection. When $E_{\rm inj}$ dominates over $E_{\rm k,iso}$, the above equation simplifies further to $E_{\rm inj}\sim\Gamma^2M_{\rm sw}c^2\propto\Gamma^2R^3$ for an ISM external medium. This yields the scaling $\Gamma(R)\propto R^{(\ell-2)/2}$, and since $R\propto\Gamma^2T$, the apparent time scales with radius as $T\propto R^{3-\ell}$, which yields $E_{\rm inj}\propto T^{(1+\ell)/(3-\ell)}$. 

\begin{figure}
    \centering
    \includegraphics[width=0.48\textwidth]{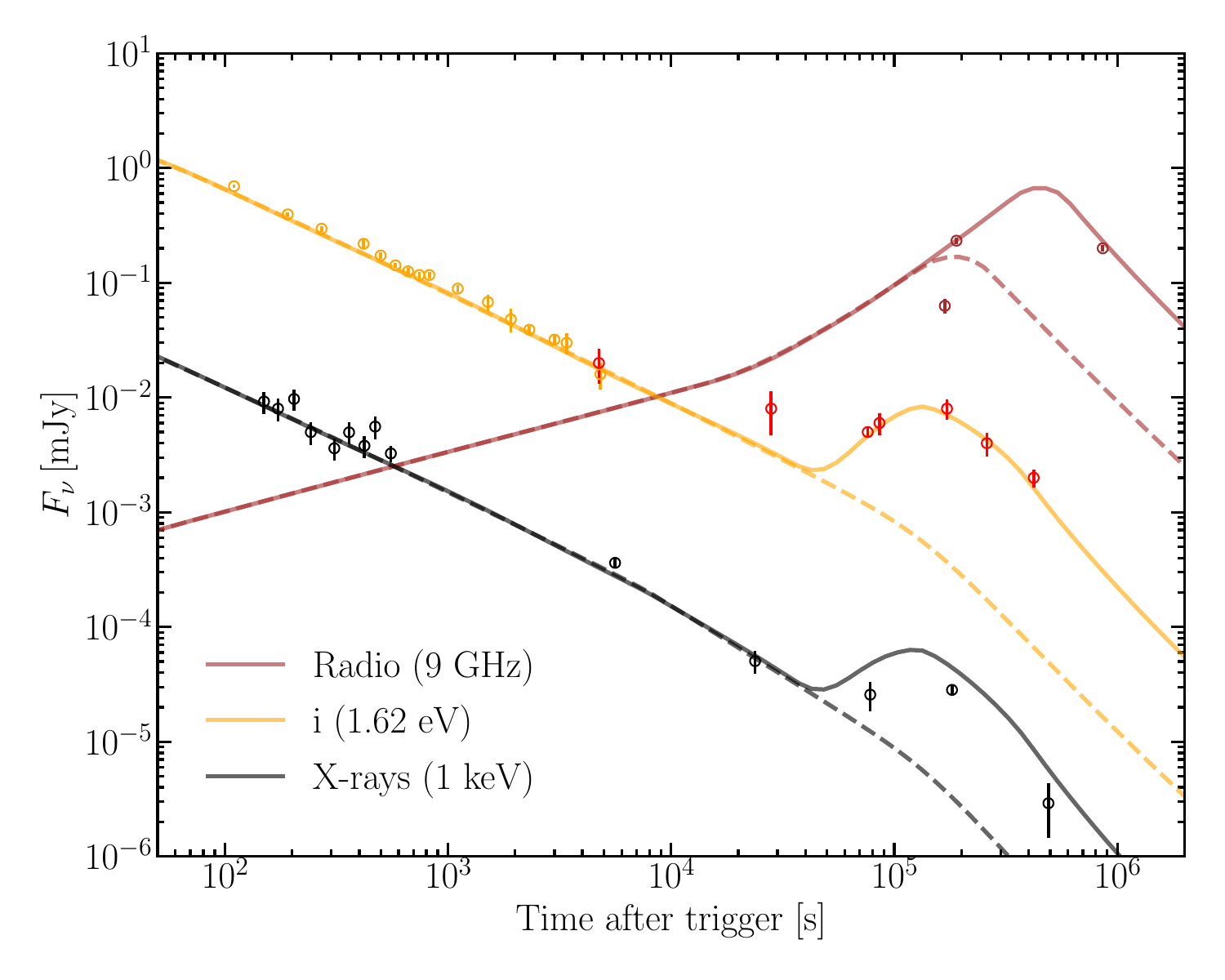}
    \caption{
    Best-fit light curve as obtained in Figure\,\ref{fig:lc-fit} (dashed), now with energy injection (solid) into a jet with half-opening angle $\theta_j=11.5$\,deg. The injected energy is $E_{\rm inj}=4.5E_{\rm k,iso}$, with $R_1=1.3\times10^{17}$\,cm, $R_2 = 1.82R_1$, $\ell=0.1$, and $\epsilon_e=0.027$, $\epsilon_B=10^{-5}$. 
    }
    \label{fig:energy-injection}
\end{figure}

To solve for the dynamical evolution of the spherical blast wave with energy injection, we use the numerical code of \citet{Gill-Granot-23} that produces EATS-integrated afterglow emission. Figure\,\ref{fig:energy-injection} shows the afterglow light curve where we take the best-fit solution from Figure\,\ref{fig:lc-fit} and inject $E_{\rm inj}\simeq4.5E_{\rm k,iso}$ into the blast wave with $\ell=0.1$. During the energy injection phase, the light curves for the radio ($\nu_m<\nu_R<\nu_a<\nu_c$), optical ($\nu_a<\nu_O<\nu_c$), and X-rays ($\nu_c<\nu_X$) show distinct temporal trends since all three are located on different power-law segments of the synchrotron spectrum. Here $\nu_m$ and $\nu_c$ are the characteristic injection and cooling-break frequencies, as defined earlier, and $\nu_a$ is the self-absorption frequency. Since the radio emission is self-absorbed, the rebrightening shows a temporal lag of $\Delta T\sim4$\,days with respect to that seen in optical and X-rays. 

To explain the late-time radio and X-ray emission, a jet break in the light curve at $T\sim10^5$\,s is required, with a jet half-opening angle of $\theta_j=11.5^{\circ}$. We introduce the jet break in our spherical model by manually integrating the emission over the surface of the flow up to the sharp edge of the jet at $\theta_j$, for the same parameters obtained from the MCMC fit of the spherical model. An on-axis ($\theta_{\rm obs}=0$) observer would see a jet break in the afterglow light curve when the angular size of the beaming cone around the line-of-sight exceeds the opening angle of the jet, i.e. when $1/\Gamma=\theta_j$, which would occur at the time $T_j=(\Gamma_0\theta_j)^{8/3}T_{\rm dec}\simeq2\times10^5$\,s for $\Gamma_0=400$ and $T_{\rm dec}\simeq2$\,s. The true jet kinetic energy in this scenario can be estimated from the total isotropic-equivalent energy, with $E_k\simeq(\theta_j^2/2)E_{\rm k,iso,tot}\simeq4.4\times10^{51}$\,erg where $E_{\rm k,iso,tot} = \mathbf{E_{\rm inj}+E_{\rm k,iso}}\simeq5.5E_{\rm k,iso}$, and $E_{\rm k,iso}$ is the energy of the fastest moving material. 

This model is able to explain both the optical data during the rebrightening phase quite well, but mildly overproduces the X-ray emission while maintaining the correct temporal decay trend. The reason for this discrepancy may lie in the assumption of fixed shock microphysical parameters over the entire duration of the afterglow. Energy injection may break this assumption and require different shock microphysical parameters during the rebrightening episode. The model also overproduces the radio emission at the time of the first detection. The sharp rise in the observed radio emission over $\Delta T\sim2.2\times10^4$\,s is difficult to reproduce in this model. As discussed in the previous section, the angular delay time over which any sharp pulse gets smeared is $T_\theta\sim10^5$\,s when $T\sim10^5$\,s, so that $\Delta T/T\sim1$. Therefore, the radio emission is not expected to rise so sharply over $\Delta T/T\ll1$ during the afterglow.

In the first scenario, where the ejecta has a radial gradient in velocity, so that the mass above a given bulk-$\Gamma$ scales as $M(>\Gamma)\propto\Gamma^{-s}$ \citep{Rees-Meszaros-98,Sari-Meszaros-00}, the energy in the blast wave grows as $E(>\Gamma)\propto\Gamma^{1-s}\propto T^{-3(1-s)\over(7+s)}$ after the fastest moving material decelerates. The scaling of the blast wave energy with $T$ in this model corresponds to $s=(4+\ell)/(2-\ell)\simeq2.2$. Since $s>1$, both the mass and energy in the ejecta are dominated by the slower moving material. In the second scenario, energy is injected by the spinning down magnetar with power $L(T)\propto T^q$, where the blast wave energy grows as $E\propto T^{1+q}$ \citep{Zhang-Meszaros-01}. Comparison of this scaling with our model gives $q=2(\ell-1)/(3-\ell)=-0.62$. {When $q>-1$ the injected energy is expected to dominate the blast wave dynamics over the impulsive, i.e. without energy injection, blast wave as the injected kinetic energy will exceed that of the impulsive shell (see above). In this scenario, the injected power is produced by electromagnetic losses due to dipolar spindown of the magnetar, and this power is expected to follow the scaling with observer-frame time, with $L(T)\propto T^0$ for $T\ll T_{\rm sd}$ and $L(T)\propto T^{-2}$ for $T \gg T_{\rm sd}$. The characteristic spin-down time is given by $T_{\rm sd} = 3c^3I/B_p^2R_{\rm NS}^6\Omega_0^2=2.05\times10^3 I_{45}B_{p,15}^{-2}P_{0,-3}^2R_{\rm NS,6}^{-6}$\,s, where $B_p$ is the surface polar magnetic field, $P_0$ is the initial spin period, and $I$ and $R_{\rm NS}$ are the moment of inertia and radius of the neutron star. The obtained value of $q = -0.62$ does not match the expected asymptotic power-law slopes of the injected power both before and after $T_{\rm sd}$. Therefore, this simple description of the magnetar scenario is not strongly favoured, however, a more detailed MCMC fit is required to rule it out.

%%%%%%%%%%%%%%%%%%%%%%%%%%%%%%%%%%%%%%%%%%%%
\section{Summary \& Discussion}
\label{sec:summary}
%%%%%%%%%%%%%%%%%%%%%%%%%%%%%%%%%%%%%%%%%%%%
We have presented multi-wavelength observations of the short duration GRB~250221A obtained with COLIBRÍ, the Harlingten 50~cm Telescope, VLT, and VLA from 1~hour after the burst up to 11~days later. We have supplemented our data with public data from {\itshape Swift}/BAT, {\itshape Swift}/XRT, {\itshape Einstein Probe}/FXT and photometry from the Legacy Survey catalogue DR10 \citep{Dey2019}.

Deep imaging shows the presence of two candidate host galaxies in the field of GRB~250221A. Our analysis of the X-Shooter afterglow spectrum firmly places the event at a redshift $z=0.768$  in a galaxy with evidence of star formation. 

The light curve shows a typical decay in the optical and X-ray frequencies before $T<3 \times 10^4$~s for synchrotron emission, with the characteristic frequencies being $\nu_m \lesssim \nu_{\rm opt} \lesssim \nu_X\lesssim \nu_c $ indicating a slow-cooling synchrotron regime in a circumstellar profile of constant density. At later times, after $T\sim0.6$~d, we observe an excess in the X-ray, optical, and radio light curves.

We have ruled out several scenarios to explain this excess, including a supernova, gravitational microlensing, and a local density enhancement.

The direct observational evidence, particularly the short duration, peak energy and hardness ratio as well as the Amati correlation more strongly supports a merger-driven coalescence scenario as the progenitor channel for GRB~250221A. Nevertheless, the  best fit parameters obtained for the afterglow model before the refreshed shock component suggest a high-density environment, which is more consistent with a collapsar scenario.

Regardless, the central engine can be argued to be either a millisecond magnetar \citep{Metzger2011} or a hyperaccreting black hole (BH) \citep{1999MacFadyenWoosley} that should be able to provide a total energy $E_{\rm tot}=E_k + E_\gamma \approx E_k \simeq 4.5\times10^{51}$\,erg to power the GRB and its afterglow. In the Collapsar scenario, the total duration over which the outflow is powered is obtained from $t_{\rm jet}=t_{\rm bo} + T_{90}/(1+z)\sim10$\,s where we take $t_{\rm bo}\sim9$\,s for the jet to break out of the stellar progenitor \citep{Bromberg-Tchekhovskoy-16}. The mean jet power that the engine must be able to provide is $L_{\rm jet}=E_{\rm tot}/\eta_jt_{\rm jet}\sim4.5\times10^{51}\eta_{j,-1}^{-1}t_{\rm jet,1}^{-1}\,{\rm erg\,s}^{-1}$ for an assumed efficiency of $\eta_j=0.1\eta_{\rm j,-1}$. In the case of a binary merger of two NSs, the jet breakout time out of the dynamical ejecta is $t_{\rm bo}\sim0.5$\,s \citep[e.g.,][]{Moharana-Piran-17,Gill+19}, which would demand a mean jet power $L_{\rm jet}\sim3\times10^{52}\eta_{j,-1}^{-1}(t_{\rm jet}/1.5\,{\rm s})^{-1}\,{\rm erg\,s}^{-1}$. 
The total energy and mean jet power in this event can be supplied by any of the various known jet launching mechanisms, namely $\nu\bar{\nu}$ annihilation from accretion onto a NS or BH \citep[see, e.g.][]{1999MacFadyenWoosley,Zalamea2011, globus2014}, an MHD wind from a millisecond magnetar \citep{Metzger2011}, a Kerr BH with a Blandford-Znajek engine \citep{1977BZ,2000Brown}, or a Schwarzschild/Kerr BH with a Blandford–Payne engine \citep{1982BlandfordPayne}. 

We attribute the rebrightening in the afterglow light curve at $T>0.6$~days to energy injection by the central engine into the forward shock produced by a relativistic jet. Three popular scenarios of energy injection in which a single and isolated rebrightening episode is realized are refreshing of the forward shock due to a mild collision (as opposed to a violent one) between two matter shells \citep{Kumar-Piran-00}, the ejecta having a radial velocity stratification \citep{Rees-Meszaros-98,Sari-Meszaros-00}, or energy injected by a spinning down millisecond magnetar via an MHD wind \citep{Zhang-Meszaros-01}. In all of these three scenarios, the energy of the blast wave is gently increased without the formation of strong shocks propagating through both the initial blast wave and energizing material. The main result is an increase in the emission from the forward shock produced by the initial blast wave.
Alternatively, the merger of two NSs produces a hypermassive (beyond the TOV limit) millisecond-magnetar central engine which can power the initial jet.  Energy can be later injected into the GRB jet by the gravitational collapse into a Schwarzschild BH, as the magnetar spins down and loses centrifugal support, thus releasing the remaining binding energy $E_B \simeq 5 \times 10^{53} {\rm erg}$ \citep[see][for further discussion and details on this model; see also Appendix~\ref{App:CentralEngine}]{2015MM} into a second shock, channelled into the pathway left by the initial jet, that eventually collides with the slowed-down head of the jet.

Rebrightenings of this kind are extremely rare among short GRBs. Although some short bursts show early-time X-ray or optical plateaus—often interpreted as signatures of long-lived central engines—clear late-time rises in the afterglow flux are not typically observed. A few events, such as GRB~060614 \citep{Yang2015} and GRB~231117A \citep{Schroeder2025}, exhibit shallow decay phases lasting up to $\sim$1 day, but these do not display the distinct, pronounced rebrightening seen in GRB~250221A; instead, their afterglows remain nearly constant or transition smoothly into the standard decay regime. In addition to this unusual behaviour, GRB~250221A ranks among the most luminous and energetic short-GRB afterglows at $T=1$ day, further distinguishing it from the known population. This behaviour contrasts sharply with that of several long GRBs, where rises on similar timescales have been linked to supernova emission, refreshed shocks, or structured jets. If GRB~250221A indeed belongs to the Type I class according to its prompt properties, the late-time rise we observe would represent the first clear instance of such behaviour within the short-GRB population \citep[see Fig. 10 from][]{Kann2011}, underscoring the unusual nature of this event and suggesting that an atypical energy-injection or density-structure scenario may be required.

The complete multi-wavelength photometric follow-up and modelling of GRB~250221A provide key insights into its progenitor, central engine, and energy injection mechanisms, while afterglow spectroscopy proved crucial for the identification of the host galaxy (and therefore properties such as the SFR and distance). 

Together, these observations add to the growing number of long and short GRBs that exhibit rebrightening episodes in their afterglows, commonly attributed to refreshed shocks and late-time central engine activity \citep[e.g.][]{Granot+03,Fan-Xu-06,Soderberg+06,de-Ugarte+07,Troja2007,Rowlinson+13,Hascoet+12,Laskar2015,Laskar+18,Lamb2019,deWet+23,Moss+23,deWet+24,Schroeder+24,Schroeder+25}. The case of GRB~250221A demonstrates how coordinated, multi-wavelength follow-up and prompt afterglow spectroscopy can reveal new facets of GRB physics and challenge our current understanding of their progenitors.

%%%%%%%%%%%%%%%%%%%%%%%%%%%%%%%%%%%%%%%%%%%%
\section*{Acknowledgements}
%%%%%%%%%%%%%%%%%%%%%%%%%%%%%%%%%%%%%%%%%%%%

We are grateful to the referee for their thorough reading of the manuscript and for their insightful and constructive comments, which helped improve the presentation and robustness of our results.

We thank Antonio Castellanos-Ramírez for his useful comments.

We thank the staff of the Observatorio Astronómico Nacional on Sierra San Pedro Mártir.

Some of the data used in this paper were acquired with the DDRAGO instrument on the COLIBRÍ telescope at the Observatorio Astronómico Nacional on the Sierra de San Pedro Mártir. COLIBRÍ and DDRAGO are funded by the Universidad Nacional Autónoma de México (CIC and DGAPA/PAPIIT IN109418 and IN109224), and SECIHTI/CONACyT (277901, Ciencias de Frontera 1046632 and Laboratorios Nacionales). COLIBRÍ received financial support from the French government under the France 2030 investment plan, as part of the Initiative d’Excellence d’Aix-Marseille Université-A*MIDEX (ANR-11-LABX-0060 -- OCEVU and AMX-19-IET-008 -- IPhU), from LabEx FOCUS (ANR-11-LABX-0013), from the CSAA-INSU-CNRS support program, and from the International Research Program ERIDANUS from CNRS. COLIBRÍ and DDRAGO are operated and maintained by the Observatorio Astronómico Nacional and the Instituto de Astronomía of the Universidad Nacional Autónoma de México.

This work made use of data supplied by the UK {\itshape Swift} Science Data Centre at the University of Leicester. 

Based on observations collected at the European Organisation for Astronomical Research in the Southern Hemisphere under ESO program 1114.D-0276(M).

The National Radio Astronomy Observatory and Green Bank Observatory are facilities of the U.S. National Science Foundation operated under cooperative agreement by Associated Universities, Inc. 

This work is based on the data obtained with Einstein Probe, a space mission supported by the Strategic Priority Program on Space Science of Chinese Academy of Sciences, in collaboration with the European Space Agency, the Max-Planck-Institute for extraterrestrial Physics (Germany), and the Centre National d'Études Spatiales (France).

CAV acknowledges support from a SECIHTI fellowship. 

RLB, RR, ET, MY and YY acknowledge support from the European Research Council through the Consolidator grant BHianca (grant agreement ID~101002761).

AMW is grateful for support from UNAM/DGAPA project IN109224.

NG and LGG gratefully acknowledge the support of the Simons Foundation (MP-SCMPS-00001470, N.G., L.G.G.).

%%%%%%%%%%%%%%%%%%%%%%%%%%%%%%%%%%%%%%%%%%%%%%%%%%
\section*{Data Availability}
%%%%%%%%%%%%%%%%%%%%%%%%%%%%%%%%%%%%%%%%%%%%
The data underlying this article will be shared on reasonable request to the corresponding author.


\bibliographystyle{mnras}
\bibliography{references} 

@ARTICLE{Zhang2009,
       author = {{Zhang}, Bing and {Zhang}, Bin-Bin and {Virgili}, Francisco J. and {Liang}, En-Wei and {Kann}, D. Alexander and {Wu}, Xue-Feng and {Proga}, Daniel and {Lv}, Hou-Jun and {Toma}, Kenji and {M{\'e}sz{\'a}ros}, Peter and {Burrows}, David N. and {Roming}, Peter W.~A. and {Gehrels}, Neil},
        title = "{Discerning the Physical Origins of Cosmological Gamma-ray Bursts Based on Multiple Observational Criteria: The Cases of z = 6.7 GRB 080913, z = 8.2 GRB 090423, and Some Short/Hard GRBs}",
      journal = {\apj},
         year = 2009,
        month = oct,
       volume = {703},
       number = {2},
        pages = {1696-1724},
          doi = {10.1088/0004-637X/703/2/1696},
archivePrefix = {arXiv},
       eprint = {0902.2419},
 primaryClass = {astro-ph.HE},
       adsurl = {https://ui.adsabs.harvard.edu/abs/2009ApJ...703.1696Z}
}

@ARTICLE{Kann2010,
       author = {{Kann}, D.~A. and {Klose}, S. and {Zhang}, B. and {Malesani}, D. and {Nakar}, E. and {Pozanenko}, A. and {Wilson}, A.~C. and {Butler}, N.~R. and {Jakobsson}, P. and {Schulze}, S. and {Andreev}, M. and {Antonelli}, L.~A. and {Bikmaev}, I.~F. and {Biryukov}, V. and {B{\"o}ttcher}, M. and {Burenin}, R.~A. and {Castro Cer{\'o}n}, J.~M. and {Castro-Tirado}, A.~J. and {Chincarini}, G. and {Cobb}, B.~E. and {Covino}, S. and {D'Avanzo}, P. and {D'Elia}, V. and {Della Valle}, M. and {de Ugarte Postigo}, A. and {Efimov}, Yu. and {Ferrero}, P. and {Fugazza}, D. and {Fynbo}, J.~P.~U. and {G{\r{a}}lfalk}, M. and {Grundahl}, F. and {Gorosabel}, J. and {Gupta}, S. and {Guziy}, S. and {Hafizov}, B. and {Hjorth}, J. and {Holhjem}, K. and {Ibrahimov}, M. and {Im}, M. and {Israel}, G.~L. and {Je{\'l}inek}, M. and {Jensen}, B.~L. and {Karimov}, R. and {Khamitov}, I.~M. and {Kizilo{\v{g}}lu}, {\"U}. and {Klunko}, E. and {Kub{\'a}nek}, P. and {Kutyrev}, A.~S. and {Laursen}, P. and {Levan}, A.~J. and {Mannucci}, F. and {Martin}, C.~M. and {Mescheryakov}, A. and {Mirabal}, N. and {Norris}, J.~P. and {Ovaldsen}, J.-E. and {Paraficz}, D. and {Pavlenko}, E. and {Piranomonte}, S. and {Rossi}, A. and {Rumyantsev}, V. and {Salinas}, R. and {Sergeev}, A. and {Sharapov}, D. and {Sollerman}, J. and {Stecklum}, B. and {Stella}, L. and {Tagliaferri}, G. and {Tanvir}, N.~R. and {Telting}, J. and {Testa}, V. and {Updike}, A.~C. and {Volnova}, A. and {Watson}, D. and {Wiersema}, K. and {Xu}, D.},
        title = "{The Afterglows of Swift-era Gamma-ray Bursts. I. Comparing pre-Swift and Swift-era Long/Soft (Type II) GRB Optical Afterglows}",
      journal = {\apj},
         year = 2010,
        month = sep,
       volume = {720},
       number = {2},
        pages = {1513-1558},
          doi = {10.1088/0004-637X/720/2/1513},
archivePrefix = {arXiv},
       eprint = {0712.2186},
 primaryClass = {astro-ph},
       adsurl = {https://ui.adsabs.harvard.edu/abs/2010ApJ...720.1513K}
}

@ARTICLE{Savaglio2009,
       author = {{Savaglio}, S. and {Glazebrook}, K. and {Le Borgne}, D.},
        title = "{The Galaxy Population Hosting Gamma-Ray Bursts}",
      journal = {\apj},
         year = 2009,
        month = jan,
       volume = {691},
       number = {1},
        pages = {182-211},
          doi = {10.1088/0004-637X/691/1/182},
archivePrefix = {arXiv},
       eprint = {0803.2718},
 primaryClass = {astro-ph},
       adsurl = {https://ui.adsabs.harvard.edu/abs/2009ApJ...691..182S}
}

@ARTICLE{Lien2016,
       author = {{Lien}, Amy and {Sakamoto}, Takanori and {Barthelmy}, Scott D. and {Baumgartner}, Wayne H. and {Cannizzo}, John K. and {Chen}, Kevin and {Collins}, Nicholas R. and {Cummings}, Jay R. and {Gehrels}, Neil and {Krimm}, Hans A. and {Markwardt}, Craig. B. and {Palmer}, David M. and {Stamatikos}, Michael and {Troja}, Eleonora and {Ukwatta}, T.~N.},
        title = "{The Third Swift Burst Alert Telescope Gamma-Ray Burst Catalog}",
      journal = {\apj},
         year = 2016,
        month = sep,
       volume = {829},
       number = {1},
          eid = {7},
        pages = {7},
          doi = {10.3847/0004-637X/829/1/7},
archivePrefix = {arXiv},
       eprint = {1606.01956},
 primaryClass = {astro-ph.HE},
       adsurl = {https://ui.adsabs.harvard.edu/abs/2016ApJ...829....7L}
}

@BOOK{Osterbrock2006,
       author = {{Osterbrock}, Donald E. and {Ferland}, Gary J.},
        title = "{Astrophysics of gaseous nebulae and active galactic nuclei}",
         year = 2006,
       adsurl = {https://ui.adsabs.harvard.edu/abs/2006agna.book.....O}
}

@ARTICLE{Salpeter1955,
       author = {{Salpeter}, Edwin E.},
        title = "{The Luminosity Function and Stellar Evolution.}",
      journal = {\apj},
         year = 1955,
        month = jan,
       volume = {121},
        pages = {161},
          doi = {10.1086/145971},
       adsurl = {https://ui.adsabs.harvard.edu/abs/1955ApJ...121..161S}
}

@ARTICLE{Kennicutt1998,
       author = {{Kennicutt}, Jr., Robert C.},
        title = "{Star Formation in Galaxies Along the Hubble Sequence}",
      journal = {\araa},
         year = 1998,
        month = jan,
       volume = {36},
        pages = {189-232},
          doi = {10.1146/annurev.astro.36.1.189},
archivePrefix = {arXiv},
       eprint = {astro-ph/9807187},
 primaryClass = {astro-ph},
       adsurl = {https://ui.adsabs.harvard.edu/abs/1998ARA&A..36..189K}
}

@ARTICLE{OConnor2020,
       author = {{O'Connor}, Brendan and {Beniamini}, Paz and {Kouveliotou}, Chryssa},
        title = "{Constraints on the circumburst environments of short gamma-ray bursts}",
      journal = {\mnras},
         year = 2020,
        month = jul,
       volume = {495},
       number = {4},
        pages = {4782-4799},
          doi = {10.1093/mnras/staa1433},
archivePrefix = {arXiv},
       eprint = {2004.00031},
 primaryClass = {astro-ph.HE},
       adsurl = {https://ui.adsabs.harvard.edu/abs/2020MNRAS.495.4782O}
}

@INPROCEEDINGS{Modigliani2010,
       author = {{Modigliani}, Andrea and {Goldoni}, Paolo and {Royer}, Fr{\'e}d{\'e}ric and {Haigron}, Regis and {Guglielmi}, Laurent and {Fran{\c{c}}ois}, Patrick and {Horrobin}, Matthew and {Bristow}, Paul and {Vernet}, Joel and {Moehler}, Sabine and {Kerber}, Florian and {Ballester}, Pascal and {Mason}, Elena and {Christensen}, Lise},
        title = "{The X-shooter pipeline}",
    booktitle = {Observatory Operations: Strategies, Processes, and Systems III},
         year = 2010,
       editor = {{Silva}, David R. and {Peck}, Alison B. and {Soifer}, B. Thomas},
       series = {Society of Photo-Optical Instrumentation Engineers (SPIE) Conference Series},
       volume = {7737},
        month = jul,
          eid = {773728},
        pages = {773728},
          doi = {10.1117/12.857211},
       adsurl = {https://ui.adsabs.harvard.edu/abs/2010SPIE.7737E..28M}
}

@ARTICLE{Clocchiatti2011,
       author = {{Clocchiatti}, Alejandro and {Suntzeff}, Nicholas B. and {Covarrubias}, Ricardo and {Candia}, Pablo},
        title = "{The Ultimate Light Curve of SN 1998bw/GRB 980425}",
      journal = {\aj},
         year = 2011,
        month = may,
       volume = {141},
       number = {5},
          eid = {163},
        pages = {163},
          doi = {10.1088/0004-6256/141/5/163},
archivePrefix = {arXiv},
       eprint = {1106.1695},
 primaryClass = {astro-ph.HE},
       adsurl = {https://ui.adsabs.harvard.edu/abs/2011AJ....141..163C}
}

@software{Bertin2010,
       author = {{Bertin}, Emmanuel},
        title = "{SWarp: Resampling and Co-adding FITS Images Together}",
 howpublished = {Astrophysics Source Code Library, record ascl:1010.068},
         year = 2010,
        month = oct,
          eid = {ascl:1010.068},
archivePrefix = {ascl},
       eprint = {1010.068},
       adsurl = {https://ui.adsabs.harvard.edu/abs/2010ascl.soft10068B}
}

@INPROCEEDINGS{Bertin2006,
       author = {{Bertin}, E.},
        title = "{Automatic Astrometric and Photometric Calibration with SCAMP}",
    booktitle = {Astronomical Data Analysis Software and Systems XV},
         year = 2006,
       editor = {{Gabriel}, C. and {Arviset}, C. and {Ponz}, D. and {Enrique}, S.},
       series = {Astronomical Society of the Pacific Conference Series},
       volume = {351},
        month = jul,
        pages = {112},
       adsurl = {https://ui.adsabs.harvard.edu/abs/2006ASPC..351..112B}
}

@ARTICLE{Yadav2025,
       author = {{Yadav}, Muskan and {Troja}, Eleonora and {Ricci}, Roberto and {Yang}, Yu-Han and {Wieringa}, Mark H. and {O'Connor}, Brendan and {Kang}, Yacheng and {Becerra}, Rosa L. and {Ryan}, Geoffrey and {Busmann}, Malte},
        title = "{Radio Observations Point to a Moderately Relativistic Outflow in the Fast X-Ray Transient EP241021a}",
      journal = {\apj},
         year = 2025,
        month = dec,
       volume = {995},
       number = {2},
          eid = {216},
        pages = {216},
          doi = {10.3847/1538-4357/ae1746},
       adsurl = {https://ui.adsabs.harvard.edu/abs/2025ApJ...995..216Y}
}

@ARTICLE{Berger2009,
       author = {{Berger}, E.},
        title = "{The Host Galaxies of Short-Duration Gamma-Ray Bursts: Luminosities, Metallicities, and Star-Formation Rates}",
      journal = {\apj},
         year = 2009,
        month = jan,
       volume = {690},
       number = {1},
        pages = {231-237},
          doi = {10.1088/0004-637X/690/1/231},
archivePrefix = {arXiv},
       eprint = {0805.0306},
 primaryClass = {astro-ph},
       adsurl = {https://ui.adsabs.harvard.edu/abs/2009ApJ...690..231B}
}

@ARTICLE{Li2016,
       author = {{Li}, Ye and {Zhang}, Bing and {L{\"u}}, Hou-Jun},
        title = "{A Comparative Study of Long and Short GRBs. I. Overlapping Properties}",
      journal = {\apjs},
         year = 2016,
        month = nov,
       volume = {227},
       number = {1},
          eid = {7},
        pages = {7},
          doi = {10.3847/0067-0049/227/1/7},
archivePrefix = {arXiv},
       eprint = {1608.03383},
 primaryClass = {astro-ph.HE},
       adsurl = {https://ui.adsabs.harvard.edu/abs/2016ApJS..227....7L}
}

@ARTICLE{AguiFernandez2023,
       author = {{Ag{\"u}{\'\i} Fern{\'a}ndez}, J.~F. and {Th{\"o}ne}, C.~C. and {Kann}, D.~A. and {de Ugarte Postigo}, A. and {Selsing}, J. and {Schady}, P. and {Yates}, R.~M. and {Greiner}, J. and {Oates}, S.~R. and {Malesani}, D.~B. and {Xu}, D. and {Klotz}, A. and {Campana}, S. and {Rossi}, A. and {Perley}, D.~A. and {Bla{\v{z}}ek}, M. and {D'Avanzo}, P. and {Giunta}, A. and {Hartmann}, D. and {Heintz}, K.~E. and {Jakobsson}, P. and {Kirkpatrick}, IV, C.~C. and {Kouveliotou}, C. and {Melandri}, A. and {Pugliese}, G. and {Salvaterra}, R. and {Starling}, R.~L.~C. and {Tanvir}, N.~R. and {Vergani}, S.~D. and {Wiersema}, K.},
        title = "{GRB 160410A: The first chemical study of the interstellar medium of a short GRB}",
      journal = {\mnras},
         year = 2023,
        month = mar,
       volume = {520},
       number = {1},
        pages = {613-636},
          doi = {10.1093/mnras/stad099},
archivePrefix = {arXiv},
       eprint = {2109.13838},
 primaryClass = {astro-ph.HE},
       adsurl = {https://ui.adsabs.harvard.edu/abs/2023MNRAS.520..613A}
}

@ARTICLE{Schlegel1998,
       author = {{Schlegel}, David J. and {Finkbeiner}, Douglas P. and {Davis}, Marc},
        title = "{Maps of Dust Infrared Emission for Use in Estimation of Reddening and Cosmic Microwave Background Radiation Foregrounds}",
      journal = {\apj},
         year = 1998,
        month = jun,
       volume = {500},
       number = {2},
        pages = {525-553},
          doi = {10.1086/305772},
archivePrefix = {arXiv},
       eprint = {astro-ph/9710327},
 primaryClass = {astro-ph},
       adsurl = {https://ui.adsabs.harvard.edu/abs/1998ApJ...500..525S}
}

@ARTICLE{Kann2011,
       author = {{Kann}, D.~A. and {Klose}, S. and {Zhang}, B. and {Covino}, S. and {Butler}, N.~R. and {Malesani}, D. and {Nakar}, E. and {Wilson}, A.~C. and {Antonelli}, L.~A. and {Chincarini}, G. and {Cobb}, B.~E. and {D'Avanzo}, P. and {D'Elia}, V. and {Della Valle}, M. and {Ferrero}, P. and {Fugazza}, D. and {Gorosabel}, J. and {Israel}, G.~L. and {Mannucci}, F. and {Piranomonte}, S. and {Schulze}, S. and {Stella}, L. and {Tagliaferri}, G. and {Wiersema}, K.},
        title = "{The Afterglows of Swift-era Gamma-Ray Bursts. II. Type I GRB versus Type II GRB Optical Afterglows}",
      journal = {\apj},
         year = 2011,
        month = jun,
       volume = {734},
       number = {2},
          eid = {96},
        pages = {96},
          doi = {10.1088/0004-637X/734/2/96},
archivePrefix = {arXiv},
       eprint = {0804.1959},
 primaryClass = {astro-ph},
       adsurl = {https://ui.adsabs.harvard.edu/abs/2011ApJ...734...96K}
}

@article{Prochaska2007,
  author = {Prochaska, J. X. and Chen, H.-W. and Bloom, J. S.},
  title = {Dissecting the Circumburst and Intervening Absorption Line Systems of GRB Afterglows},
  journal = {ApJ},
  volume = {648},
  pages = {95-115},
  year = {2007},
  doi = {10.1086/505740}
}

@article{Vreeswijk2013,
  author = {Vreeswijk, P. M. and Ledoux, C. and Raassen, A. J. J. and et al.},
  title = {Rapid-response mode VLT/UVES spectroscopy of GRB afterglows},
  journal = {A\&A},
  volume = {549},
  pages = {A22},
  year = {2013},
  doi = {10.1051/0004-6361/201219477}
}

@article{deUgartePostigo2014,
  author = {de Ugarte Postigo, A. and Thöne, C. C. and Rowlinson, A. and et al.},
  title = {Spectroscopy of GRB 130603B: A short GRB in a star-forming galaxy at z=0.356},
  journal = {A\&A},
  volume = {563},
  pages = {A62},
  year = {2014},
  doi = {10.1051/0004-6361/201322782}
}

@article{Selsing2018,
  author = {Selsing, J. and Malesani, D. B. and Goldoni, P. and et al.},
  title = {X-shooter GRB afterglow legacy sample (XS-GRB)},
  journal = {A\&A},
  volume = {616},
  pages = {A48},
  year = {2018},
  doi = {10.1051/0004-6361/201732259}
}

@article{Cucchiara2013,
  author = {Cucchiara, A. and Prochaska, J. X. and Perley, D. and et al.},
  title = {The Afterglow and Early-type Host Galaxy of the Short GRB 130603B},
  journal = {ApJ},
  volume = {777},
  pages = {94},
  year = {2013},
  doi = {10.1088/0004-637X/777/2/94}
}

@ARTICLE{Zhang2004,
       author = {{Zhang}, Bing and {M{\'e}sz{\'a}ros}, Peter},
        title = "{Gamma-Ray Bursts: progress, problems \& prospects}",
      journal = {International Journal of Modern Physics A},
         year = 2004,
        month = jan,
       volume = {19},
       number = {15},
        pages = {2385-2472},
          doi = {10.1142/S0217751X0401746X},
archivePrefix = {arXiv},
       eprint = {astro-ph/0311321},
 primaryClass = {astro-ph},
       adsurl = {https://ui.adsabs.harvard.edu/abs/2004IJMPA..19.2385Z}
}

@ARTICLE{Chevalier2000,
       author = {{Chevalier}, Roger A. and {Li}, Zhi-Yun},
        title = "{Wind Interaction Models for Gamma-Ray Burst Afterglows: The Case for Two Types of Progenitors}",
      journal = {\apj},
         year = 2000,
        month = jun,
       volume = {536},
       number = {1},
        pages = {195-212},
          doi = {10.1086/308914},
archivePrefix = {arXiv},
       eprint = {astro-ph/9908272},
 primaryClass = {astro-ph},
       adsurl = {https://ui.adsabs.harvard.edu/abs/2000ApJ...536..195C}
}

@ARTICLE{Kobayashi2003,
       author = {{Kobayashi}, Shiho and {Zhang}, Bing},
        title = "{GRB 021004: Reverse Shock Emission}",
      journal = {\apjl},
         year = 2003,
        month = jan,
       volume = {582},
       number = {2},
        pages = {L75-L78},
          doi = {10.1086/367691},
archivePrefix = {arXiv},
       eprint = {astro-ph/0210584},
 primaryClass = {astro-ph},
       adsurl = {https://ui.adsabs.harvard.edu/abs/2003ApJ...582L..75K}
}

@ARTICLE{deUgarte2005,
       author = {{de Ugarte Postigo}, A. and {Castro-Tirado}, A.~J. and {Gorosabel}, J. and {J{\'o}hannesson}, G. and {Bj{\"o}rnsson}, G. and {Gudmundsson}, E.~H. and {Bremer}, M. and {Pak}, S. and {Tanvir}, N. and {Castro Cer{\'o}n}, J.~M. and {Guzyi}, S. and {Jel{\'\i}nek}, M. and {Klose}, S. and {P{\'e}rez-Ram{\'\i}rez}, D. and {Aceituno}, J. and {Campo Bagat{\'\i}n}, A. and {Covino}, S. and {Cardiel}, N. and {Fathkullin}, T. and {Henden}, A.~A. and {Huferath}, S. and {Kurata}, Y. and {Malesani}, D. and {Mannucci}, F. and {Ruiz-Lapuente}, P. and {Sokolov}, V. and {Thiele}, U. and {Wisotzki}, L. and {Antonelli}, L.~A. and {Bartolini}, C. and {Boattini}, A. and {Guarnieri}, A. and {Piccioni}, A. and {Pizzichini}, G. and {del Principe}, M. and {di Paola}, A. and {Fugazza}, D. and {Ghisellini}, G. and {Hunt}, L. and {Konstantinova}, T. and {Masetti}, N. and {Palazzi}, E. and {Pian}, E. and {Stefanon}, M. and {Testa}, V. and {Tristram}, P.~J.},
        title = "{GRB 021004 modelled by multiple energy injections}",
      journal = {\aap},
         year = 2005,
        month = dec,
       volume = {443},
       number = {3},
        pages = {841-849},
          doi = {10.1051/0004-6361:20052898},
archivePrefix = {arXiv},
       eprint = {astro-ph/0506544},
 primaryClass = {astro-ph},
       adsurl = {https://ui.adsabs.harvard.edu/abs/2005A&A...443..841D}
}

@ARTICLE{Shemi1990,
       author = {{Shemi}, Amotz and {Piran}, Tsvi},
        title = "{The Appearance of Cosmic Fireballs}",
      journal = {\apjl},
         year = 1990,
        month = dec,
       volume = {365},
        pages = {L55},
          doi = {10.1086/185887},
       adsurl = {https://ui.adsabs.harvard.edu/abs/1990ApJ...365L..55S}
}

@ARTICLE{Nardini2011,
       author = {{Nardini}, M. and {Greiner}, J. and {Kr{\"u}hler}, T. and {Filgas}, R. and {Klose}, S. and {Afonso}, P. and {Clemens}, C. and {Guelbenzu}, A.~N. and {Olivares E.}, F. and {Rau}, A. and {Rossi}, A. and {Updike}, A. and {K{\"u}pc{\"u} Yolda{\c{s}}}, A. and {Yolda{\c{s}}}, A. and {Burlon}, D. and {Elliott}, J. and {Kann}, D.~A.},
        title = "{On the nature of the extremely fast optical rebrightening of the afterglow of GRB 081029}",
      journal = {\aap},
         year = 2011,
        month = jul,
       volume = {531},
          eid = {A39},
        pages = {A39},
          doi = {10.1051/0004-6361/201116814},
archivePrefix = {arXiv},
       eprint = {1105.0917},
 primaryClass = {astro-ph.HE},
       adsurl = {https://ui.adsabs.harvard.edu/abs/2011A&A...531A..39N}
}

@ARTICLE{Laskar2015,
       author = {{Laskar}, Tanmoy and {Berger}, Edo and {Margutti}, Raffaella and {Perley}, Daniel and {Zauderer}, B. Ashley and {Sari}, Re'em and {Fong}, Wen-fai},
        title = "{Energy Injection in Gamma-Ray Burst Afterglows}",
      journal = {\apj},
         year = 2015,
        month = nov,
       volume = {814},
       number = {1},
          eid = {1},
        pages = {1},
          doi = {10.1088/0004-637X/814/1/1},
archivePrefix = {arXiv},
       eprint = {1504.03702},
 primaryClass = {astro-ph.HE},
       adsurl = {https://ui.adsabs.harvard.edu/abs/2015ApJ...814....1L}
}

@ARTICLE{Troja2007,
       author = {{Troja}, E. and {Cusumano}, G. and {O'Brien}, P.~T. and {Zhang}, B. and {Sbarufatti}, B. and {Mangano}, V. and {Willingale}, R. and {Chincarini}, G. and {Osborne}, J.~P. and {Marshall}, F.~E. and {Burrows}, D.~N. and {Campana}, S. and {Gehrels}, N. and {Guidorzi}, C. and {Krimm}, H.~A. and {La Parola}, V. and {Liang}, E.~W. and {Mineo}, T. and {Moretti}, A. and {Page}, K.~L. and {Romano}, P. and {Tagliaferri}, G. and {Zhang}, B.~B. and {Page}, M.~J. and {Schady}, P.},
        title = "{Swift Observations of GRB 070110: An Extraordinary X-Ray Afterglow Powered by the Central Engine}",
      journal = {\apj},
         year = 2007,
        month = aug,
       volume = {665},
       number = {1},
        pages = {599-607},
          doi = {10.1086/519450},
archivePrefix = {arXiv},
       eprint = {astro-ph/0702220},
 primaryClass = {astro-ph},
       adsurl = {https://ui.adsabs.harvard.edu/abs/2007ApJ...665..599T}
}

@ARTICLE{Lamb2019,
       author = {{Lamb}, G.~P. and {Tanvir}, N.~R. and {Levan}, A.~J. and {de Ugarte Postigo}, A. and {Kawaguchi}, K. and {Corsi}, A. and {Evans}, P.~A. and {Gompertz}, B. and {Malesani}, D.~B. and {Page}, K.~L. and {Wiersema}, K. and {Rosswog}, S. and {Shibata}, M. and {Tanaka}, M. and {van der Horst}, A.~J. and {Cano}, Z. and {Fynbo}, J.~P.~U. and {Fruchter}, A.~S. and {Greiner}, J. and {Heintz}, K.~E. and {Higgins}, A. and {Hjorth}, J. and {Izzo}, L. and {Jakobsson}, P. and {Kann}, D.~A. and {O'Brien}, P.~T. and {Perley}, D.~A. and {Pian}, E. and {Pugliese}, G. and {Starling}, R.~L.~C. and {Th{\"o}ne}, C.~C. and {Watson}, D. and {Wijers}, R.~A.~M.~J. and {Xu}, D.},
        title = "{Short GRB 160821B: A Reverse Shock, a Refreshed Shock, and a Well-sampled Kilonova}",
      journal = {\apj},
         year = 2019,
        month = sep,
       volume = {883},
       number = {1},
          eid = {48},
        pages = {48},
          doi = {10.3847/1538-4357/ab38bb},
archivePrefix = {arXiv},
       eprint = {1905.02159},
 primaryClass = {astro-ph.HE},
       adsurl = {https://ui.adsabs.harvard.edu/abs/2019ApJ...883...48L}
}

@ARTICLE{Goodman1986,
       author = {{Goodman}, J.},
        title = "{Are gamma-ray bursts optically thick?}",
      journal = {\apjl},
         year = 1986,
        month = sep,
       volume = {308},
        pages = {L47},
          doi = {10.1086/184741},
       adsurl = {https://ui.adsabs.harvard.edu/abs/1986ApJ...308L..47G}
}

@ARTICLE{Yang2022,
       author = {{Yang}, Jun and {Ai}, Shunke and {Zhang}, Bin-Bin and {Zhang}, Bing and {Liu}, Zi-Ke and {Wang}, Xiangyu Ivy and {Yang}, Yu-Han and {Yin}, Yi-Han and {Li}, Ye and {L{\"u}}, Hou-Jun},
        title = "{A long-duration gamma-ray burst with a peculiar origin}",
      journal = {\nat},
         year = 2022,
        month = dec,
       volume = {612},
       number = {7939},
        pages = {232-235},
          doi = {10.1038/s41586-022-05403-8},
archivePrefix = {arXiv},
       eprint = {2204.12771},
 primaryClass = {astro-ph.HE},
       adsurl = {https://ui.adsabs.harvard.edu/abs/2022Natur.612..232Y}
}

@ARTICLE{Petropoulou2020,
       author = {{Petropoulou}, M. and {Beniamini}, P. and {Vasilopoulos}, G. and {Giannios}, D. and {Barniol Duran}, R.},
        title = "{Deciphering the properties of the central engine in GRB collapsars}",
      journal = {\mnras},
         year = 2020,
        month = aug,
       volume = {496},
       number = {3},
        pages = {2910-2921},
          doi = {10.1093/mnras/staa1695},
archivePrefix = {arXiv},
       eprint = {2006.07482},
 primaryClass = {astro-ph.HE},
       adsurl = {https://ui.adsabs.harvard.edu/abs/2020MNRAS.496.2910P}
}

@ARTICLE{Dainotti2024,
       author = {{Dainotti}, M.~G. and {De Simone}, B. and {Mohideen Malik}, R.~F. and {Pasumarti}, V. and {Levine}, D. and {Saha}, N. and {Gendre}, B. and {Kido}, D. and {Watson}, A.~M. and {Becerra}, R.~L. and {Belkin}, S. and {Desai}, S. and {Pedreira do E.~S.}, A.~C.~C. and {Das}, U. and {Li}, L. and {Oates}, S.~R. and {Cenko}, S.~B. and {Pozanenko}, A. and {Volnova}, A. and {Hu}, Y. -D. and {Castro-Tirado}, A.~J. and {Orange}, N.~B. and {Moriya}, T.~J. and {Fraija}, N. and {Niino}, Y. and {Rinaldi}, E. and {Butler}, N.~R. and {Gonz{\'a}lez}, J. d. J.~G. and {Kutyrev}, A.~S. and {Lee}, W.~H. and {Prochaska}, X. and {Ramirez-Ruiz}, E. and {Richer}, M. and {Siegel}, M.~H. and {Misra}, K. and {Rossi}, A. and {Lopresti}, C. and {Quadri}, U. and {Strabla}, L. and {Ruocco}, N. and {Leonini}, S. and {Conti}, M. and {Rosi}, P. and {Ramirez}, L.~M.~T. and {Zola}, S. and {Jindal}, I. and {Kumar}, R. and {Chan}, L. and {Fuentes}, M. and {Lambiase}, G. and {Kalinowski}, K.~K. and {Jamal}, W.},
        title = "{An optical gamma-ray burst catalogue with measured redshift - I. Data release of 535 gamma-ray bursts and colour evolution}",
      journal = {\mnras},
         year = 2024,
        month = oct,
       volume = {533},
       number = {4},
        pages = {4023-4043},
          doi = {10.1093/mnras/stae1484},
archivePrefix = {arXiv},
       eprint = {2405.02263},
 primaryClass = {astro-ph.HE},
       adsurl = {https://ui.adsabs.harvard.edu/abs/2024MNRAS.533.4023D}
}

@ARTICLE{Gendre2025,
       author = {{Gendre}, Bruce},
        title = "{A Review of Long-Lasting Activities of the Central Engine of Gamma-Ray Bursts}",
      journal = {Galaxies},
         year = 2025,
        month = jan,
       volume = {13},
       number = {1},
          eid = {7},
        pages = {7},
          doi = {10.3390/galaxies13010007},
archivePrefix = {arXiv},
       eprint = {2501.01857},
 primaryClass = {astro-ph.HE},
       adsurl = {https://ui.adsabs.harvard.edu/abs/2025Galax..13....7G}
}

@INCOLLECTION{Hjorth2012,
       author = {{Hjorth}, Jens and {Bloom}, Joshua S.},
        title = "{The Gamma-Ray Burst - Supernova Connection}",
    booktitle = {Chapter 9 in ''Gamma-Ray Bursts},
         year = 2012,
       editor = {{Kouveliotou}, Chryssa and {Wijers}, Ralph A.~M.~J. and {Woosley}, Stan},
        pages = {169-190},
          doi = {10.48550/arXiv.1104.2274},
       adsurl = {https://ui.adsabs.harvard.edu/abs/2012grb..book..169H}
}

@ARTICLE{MacFadyen1999ApJ,
       author = {{MacFadyen}, A.~I. and {Woosley}, S.~E.},
        title = "{Collapsars: Gamma-Ray Bursts and Explosions in ``Failed Supernovae''}",
      journal = {\apj},
         year = 1999,
        month = oct,
       volume = {524},
       number = {1},
        pages = {262-289},
          doi = {10.1086/307790},
archivePrefix = {arXiv},
       eprint = {astro-ph/9810274},
 primaryClass = {astro-ph},
       adsurl = {https://ui.adsabs.harvard.edu/abs/1999ApJ...524..262M}
}

@ARTICLE{Tsvetkova2017,
       author = {{Tsvetkova}, A. and {Frederiks}, D. and {Golenetskii}, S. and {Lysenko}, A. and {Oleynik}, P. and {Pal'shin}, V. and {Svinkin}, D. and {Ulanov}, M. and {Cline}, T. and {Hurley}, K. and {Aptekar}, R.},
        title = "{The Konus-Wind Catalog of Gamma-Ray Bursts with Known Redshifts. I. Bursts Detected in the Triggered Mode}",
      journal = {\apj},
         year = 2017,
        month = dec,
       volume = {850},
       number = {2},
          eid = {161},
        pages = {161},
          doi = {10.3847/1538-4357/aa96af},
archivePrefix = {arXiv},
       eprint = {1710.08746},
 primaryClass = {astro-ph.HE},
       adsurl = {https://ui.adsabs.harvard.edu/abs/2017ApJ...850..161T}
}

@ARTICLE{Planck2020,
       author = {{Planck Collaboration} and {Aghanim}, N. and {Akrami}, Y. and {Arroja}, F. and {Ashdown}, M. and {Aumont}, J. and {Baccigalupi}, C. and {Ballardini}, M. and {Banday}, A.~J. and {Barreiro}, R.~B. and {Bartolo}, N. and {Basak}, S. and {Battye}, R. and {Benabed}, K. and {Bernard}, J. -P. and {Bersanelli}, M. and {Bielewicz}, P. and {Bock}, J.~J. and {Bond}, J.~R. and {Borrill}, J. and {Bouchet}, F.~R. and {Boulanger}, F. and {Bucher}, M. and {Burigana}, C. and {Butler}, R.~C. and {Calabrese}, E. and {Cardoso}, J. -F. and {Carron}, J. and {Casaponsa}, B. and {Challinor}, A. and {Chiang}, H.~C. and {Colombo}, L.~P.~L. and {Combet}, C. and {Contreras}, D. and {Crill}, B.~P. and {Cuttaia}, F. and {de Bernardis}, P. and {de Zotti}, G. and {Delabrouille}, J. and {Delouis}, J. -M. and {D{\'e}sert}, F. -X. and {Di Valentino}, E. and {Dickinson}, C. and {Diego}, J.~M. and {Donzelli}, S. and {Dor{\'e}}, O. and {Douspis}, M. and {Ducout}, A. and {Dupac}, X. and {Efstathiou}, G. and {Elsner}, F. and {En{\ss}lin}, T.~A. and {Eriksen}, H.~K. and {Falgarone}, E. and {Fantaye}, Y. and {Fergusson}, J. and {Fernandez-Cobos}, R. and {Finelli}, F. and {Forastieri}, F. and {Frailis}, M. and {Franceschi}, E. and {Frolov}, A. and {Galeotta}, S. and {Galli}, S. and {Ganga}, K. and {G{\'e}nova-Santos}, R.~T. and {Gerbino}, M. and {Ghosh}, T. and {Gonz{\'a}lez-Nuevo}, J. and {G{\'o}rski}, K.~M. and {Gratton}, S. and {Gruppuso}, A. and {Gudmundsson}, J.~E. and {Hamann}, J. and {Handley}, W. and {Hansen}, F.~K. and {Helou}, G. and {Herranz}, D. and {Hildebrandt}, S.~R. and {Hivon}, E. and {Huang}, Z. and {Jaffe}, A.~H. and {Jones}, W.~C. and {Karakci}, A. and {Keih{\"a}nen}, E. and {Keskitalo}, R. and {Kiiveri}, K. and {Kim}, J. and {Kisner}, T.~S. and {Knox}, L. and {Krachmalnicoff}, N. and {Kunz}, M. and {Kurki-Suonio}, H. and {Lagache}, G. and {Lamarre}, J. -M. and {Langer}, M. and {Lasenby}, A. and {Lattanzi}, M. and {Lawrence}, C.~R. and {Le Jeune}, M. and {Leahy}, J.~P. and {Lesgourgues}, J. and {Levrier}, F. and {Lewis}, A. and {Liguori}, M. and {Lilje}, P.~B. and {Lilley}, M. and {Lindholm}, V. and {L{\'o}pez-Caniego}, M. and {Lubin}, P.~M. and {Ma}, Y. -Z. and {Mac{\'\i}as-P{\'e}rez}, J.~F. and {Maggio}, G. and {Maino}, D. and {Mandolesi}, N. and {Mangilli}, A. and {Marcos-Caballero}, A. and {Maris}, M. and {Martin}, P.~G. and {Martinelli}, M. and {Mart{\'\i}nez-Gonz{\'a}lez}, E. and {Matarrese}, S. and {Mauri}, N. and {McEwen}, J.~D. and {Meerburg}, P.~D. and {Meinhold}, P.~R. and {Melchiorri}, A. and {Mennella}, A. and {Migliaccio}, M. and {Millea}, M. and {Mitra}, S. and {Miville-Desch{\^e}nes}, M. -A. and {Molinari}, D. and {Moneti}, A. and {Montier}, L. and {Morgante}, G. and {Moss}, A. and {Mottet}, S. and {M{\"u}nchmeyer}, M. and {Natoli}, P. and {N{\o}rgaard-Nielsen}, H.~U. and {Oxborrow}, C.~A. and {Pagano}, L. and {Paoletti}, D. and {Partridge}, B. and {Patanchon}, G. and {Pearson}, T.~J. and {Peel}, M. and {Peiris}, H.~V. and {Perrotta}, F. and {Pettorino}, V. and {Piacentini}, F. and {Polastri}, L. and {Polenta}, G. and {Puget}, J. -L. and {Rachen}, J.~P. and {Reinecke}, M. and {Remazeilles}, M. and {Renault}, C. and {Renzi}, A. and {Rocha}, G. and {Rosset}, C. and {Roudier}, G. and {Rubi{\~n}o-Mart{\'\i}n}, J.~A. and {Ruiz-Granados}, B. and {Salvati}, L. and {Sandri}, M. and {Savelainen}, M. and {Scott}, D. and {Shellard}, E.~P.~S. and {Shiraishi}, M. and {Sirignano}, C. and {Sirri}, G. and {Spencer}, L.~D. and {Sunyaev}, R. and {Suur-Uski}, A. -S. and {Tauber}, J.~A. and {Tavagnacco}, D. and {Tenti}, M. and {Terenzi}, L. and {Toffolatti}, L. and {Tomasi}, M. and {Trombetti}, T. and {Valiviita}, J. and {Van Tent}, B. and {Vibert}, L. and {Vielva}, P. and {Villa}, F. and {Vittorio}, N. and {Wandelt}, B.~D. and {Wehus}, I.~K. and {White}, M. and {White}, S.~D.~M. and {Zacchei}, A. and {Zonca}, A.},
        title = "{Planck 2018 results. I. Overview and the cosmological legacy of Planck}",
      journal = {\aap},
         year = 2020,
        month = sep,
       volume = {641},
          eid = {A1},
        pages = {A1},
          doi = {10.1051/0004-6361/201833880},
archivePrefix = {arXiv},
       eprint = {1807.06205},
 primaryClass = {astro-ph.CO},
       adsurl = {https://ui.adsabs.harvard.edu/abs/2020A&A...641A...1P}
}

@ARTICLE{Perley2016,
       author = {{Perley}, Daniel A. and {Niino}, Yuu and {Tanvir}, Nial R. and {Vergani}, Susanna D. and {Fynbo}, Johan P.~U.},
        title = "{Long-Duration Gamma-Ray Burst Host Galaxies in Emission and Absorption}",
      journal = {\ssr},
         year = 2016,
        month = dec,
       volume = {202},
       number = {1-4},
        pages = {111-142},
          doi = {10.1007/s11214-016-0237-4},
archivePrefix = {arXiv},
       eprint = {1602.00770},
 primaryClass = {astro-ph.HE},
       adsurl = {https://ui.adsabs.harvard.edu/abs/2016SSRv..202..111P}
}

@ARTICLE{vanErten2009,
       author = {{van Eerten}, H.~J. and {Meliani}, Z. and {Wijers}, R.~A.~M.~J. and {Keppens}, R.},
        title = "{No visible optical variability from a relativistic blast wave encountering a wind termination shock}",
      journal = {\mnras},
         year = 2009,
        month = sep,
       volume = {398},
       number = {1},
        pages = {L63-L67},
          doi = {10.1111/j.1745-3933.2009.00711.x},
archivePrefix = {arXiv},
       eprint = {0906.3629},
 primaryClass = {astro-ph.HE},
       adsurl = {https://ui.adsabs.harvard.edu/abs/2009MNRAS.398L..63V}
}

@ARTICLE{Nakar2007,
       author = {{Nakar}, Ehud and {Granot}, Jonathan},
        title = "{Smooth light curves from a bumpy ride: relativistic blast wave encounters a density jump}",
      journal = {\mnras},
         year = 2007,
        month = oct,
       volume = {380},
       number = {4},
        pages = {1744-1760},
          doi = {10.1111/j.1365-2966.2007.12245.x},
archivePrefix = {arXiv},
       eprint = {astro-ph/0606011},
 primaryClass = {astro-ph},
       adsurl = {https://ui.adsabs.harvard.edu/abs/2007MNRAS.380.1744N}
}

@article{Tam2005,
title = {Early re-brightenings in GRB afterglows as signatures of low-to-high density boundary},
journal = {New Astronomy},
volume = {10},
number = {7},
pages = {535-544},
year = {2005},
issn = {1384-1076},
doi = {https://doi.org/10.1016/j.newast.2005.03.003},
url = {https://www.sciencedirect.com/science/article/pii/S1384107605000345},
author = {P.H. Tam and C.S.J. Pun and Y.F. Huang and K.S. Cheng}
}

@ARTICLE{Mirabal2006,
       author = {{Mirabal}, N. and {Halpern}, J.~P. and {An}, D. and {Thorstensen}, J.~R. and {Terndrup}, D.~M.},
        title = "{GRB 060218/SN 2006aj: A Gamma-Ray Burst and Prompt Supernova at z = 0.0335}",
      journal = {\apjl},
         year = 2006,
        month = jun,
       volume = {643},
       number = {2},
        pages = {L99-L102},
          doi = {10.1086/505177},
archivePrefix = {arXiv},
       eprint = {astro-ph/0603686},
 primaryClass = {astro-ph},
       adsurl = {https://ui.adsabs.harvard.edu/abs/2006ApJ...643L..99M}
}

@ARTICLE{Pian2006,
       author = {{Pian}, Elena and {Mazzali}, P.~A.},
        title = "{Anisotropies in Core Collapse Supernovae}",
      journal = {Chinese Journal of Astronomy and Astrophysics Supplement},
         year = 2006,
        month = dec,
       volume = {6},
       number = {S1},
        pages = {335-341},
          doi = {10.1088/1009-9271/6/S1/43},
       adsurl = {https://ui.adsabs.harvard.edu/abs/2006ChJAS...6a.335P}
}

@ARTICLE{LoebPerna98,
       author = {{Loeb}, Abraham and {Perna}, Rosalba},
        title = "{Microlensing of Gamma-Ray Burst Afterglows}",
      journal = {\apj},
         year = 1998,
        month = mar,
       volume = {495},
       number = {2},
        pages = {597-603},
          doi = {10.1086/305337},
archivePrefix = {arXiv},
       eprint = {astro-ph/9708159},
 primaryClass = {astro-ph},
       adsurl = {https://ui.adsabs.harvard.edu/abs/1998ApJ...495..597L}
}

@ARTICLE{Cano2017,
       author = {{Cano}, Zach and {Wang}, Shan-Qin and {Dai}, Zi-Gao and {Wu}, Xue-Feng},
        title = "{The Observer's Guide to the Gamma-Ray Burst Supernova Connection}",
      journal = {Advances in Astronomy},
         year = 2017,
        month = jan,
       volume = {2017},
          eid = {8929054},
        pages = {8929054},
          doi = {10.1155/2017/8929054},
archivePrefix = {arXiv},
       eprint = {1604.03549},
 primaryClass = {astro-ph.HE},
       adsurl = {https://ui.adsabs.harvard.edu/abs/2017AdAst2017E...5C}
}

@ARTICLE{Galama1998,
       author = {{Galama}, T.~J. and {Vreeswijk}, P.~M. and {van Paradijs}, J. and {Kouveliotou}, C. and {Augusteijn}, T. and {B{\"o}hnhardt}, H. and {Brewer}, J.~P. and {Doublier}, V. and {Gonzalez}, J. -F. and {Leibundgut}, B. and {Lidman}, C. and {Hainaut}, O.~R. and {Patat}, F. and {Heise}, J. and {in't Zand}, J. and {Hurley}, K. and {Groot}, P.~J. and {Strom}, R.~G. and {Mazzali}, P.~A. and {Iwamoto}, K. and {Nomoto}, K. and {Umeda}, H. and {Nakamura}, T. and {Young}, T.~R. and {Suzuki}, T. and {Shigeyama}, T. and {Koshut}, T. and {Kippen}, M. and {Robinson}, C. and {de Wildt}, P. and {Wijers}, R.~A.~M.~J. and {Tanvir}, N. and {Greiner}, J. and {Pian}, E. and {Palazzi}, E. and {Frontera}, F. and {Masetti}, N. and {Nicastro}, L. and {Feroci}, M. and {Costa}, E. and {Piro}, L. and {Peterson}, B.~A. and {Tinney}, C. and {Boyle}, B. and {Cannon}, R. and {Stathakis}, R. and {Sadler}, E. and {Begam}, M.~C. and {Ianna}, P.},
        title = "{An unusual supernova in the error box of the {\ensuremath{\gamma}}-ray burst of 25 April 1998}",
      journal = {\nat},
         year = 1998,
        month = oct,
       volume = {395},
       number = {6703},
        pages = {670-672},
          doi = {10.1038/27150},
archivePrefix = {arXiv},
       eprint = {astro-ph/9806175},
 primaryClass = {astro-ph},
       adsurl = {https://ui.adsabs.harvard.edu/abs/1998Natur.395..670G}
}

@ARTICLE{Perley2019,
       author = {{Perley}, Daniel A. and {Mazzali}, Paolo A. and {Yan}, Lin and {Cenko}, S. Bradley and {Gezari}, Suvi and {Taggart}, Kirsty and {Blagorodnova}, Nadia and {Fremling}, Christoffer and {Mockler}, Brenna and {Singh}, Avinash and {Tominaga}, Nozomu and {Tanaka}, Masaomi and {Watson}, Alan M. and {Ahumada}, Tom{\'a}s and {Anupama}, G.~C. and {Ashall}, Chris and {Becerra}, Rosa L. and {Bersier}, David and {Bhalerao}, Varun and {Bloom}, Joshua S. and {Butler}, Nathaniel R. and {Copperwheat}, Chris and {Coughlin}, Michael W. and {De}, Kishalay and {Drake}, Andrew J. and {Duev}, Dmitry A. and {Frederick}, Sara and {Gonz{\'a}lez}, J. Jes{\'u}s and {Goobar}, Ariel and {Heida}, Marianne and {Ho}, Anna Y.~Q. and {Horst}, John and {Hung}, Tiara and {Itoh}, Ryosuke and {Jencson}, Jacob E. and {Kasliwal}, Mansi M. and {Kawai}, Nobuyuki and {Khanam}, Tanazza and {Kulkarni}, Shrinivas R. and {Kumar}, Brajesh and {Kumar}, Harsh and {Kutyrev}, Alexander S. and {Lee}, William H. and {Maeda}, Keiichi and {Mahabal}, Ashish and {Murata}, Katsuhiro L. and {Neill}, James D. and {Ngeow}, Chow-Choong and {Penprase}, Bryan and {Pian}, Elena and {Quimby}, Robert and {Ramirez-Ruiz}, Enrico and {Richer}, Michael G. and {Rom{\'a}n-Z{\'u}{\~n}iga}, Carlos G. and {Sahu}, D.~K. and {Srivastav}, Shubham and {Socia}, Quentin and {Sollerman}, Jesper and {Tachibana}, Yutaro and {Taddia}, Francesco and {Tinyanont}, Samaporn and {Troja}, Eleonora and {Ward}, Charlotte and {Wee}, Jerrick and {Yu}, Po-Chieh},
        title = "{The fast, luminous ultraviolet transient AT2018cow: extreme supernova, or disruption of a star by an intermediate-mass black hole?}",
      journal = {\mnras},
         year = 2019,
        month = mar,
       volume = {484},
       number = {1},
        pages = {1031-1049},
          doi = {10.1093/mnras/sty3420},
archivePrefix = {arXiv},
       eprint = {1808.00969},
 primaryClass = {astro-ph.HE},
       adsurl = {https://ui.adsabs.harvard.edu/abs/2019MNRAS.484.1031P}
}

@ARTICLE{Perley2021,
       author = {{Perley}, Daniel A. and {Ho}, Anna Y.~Q. and {Yao}, Yuhan and {Fremling}, Christoffer and {Anderson}, Joseph P. and {Schulze}, Steve and {Kumar}, Harsh and {Anupama}, G.~C. and {Barway}, Sudhanshu and {Bellm}, Eric C. and {Bhalerao}, Varun and {Chen}, Ting-Wan and {Duev}, Dmitry A. and {Galbany}, Llu{\'\i}s and {Graham}, Matthew J. and {Gromadzki}, Mariusz and {Guti{\'e}rrez}, Claudia P. and {Ihanec}, Nada and {Inserra}, Cosimo and {Kasliwal}, Mansi M. and {Kool}, Erik C. and {Kulkarni}, S.~R. and {Laher}, Russ R. and {Masci}, Frank J. and {Neill}, James D. and {Nicholl}, Matt and {Pursiainen}, Miika and {van Roestel}, Joannes and {Sharma}, Yashvi and {Sollerman}, Jesper and {Walters}, Richard and {Wiseman}, Philip},
        title = "{Real-time discovery of AT2020xnd: a fast, luminous ultraviolet transient with minimal radioactive ejecta}",
      journal = {\mnras},
         year = 2021,
        month = dec,
       volume = {508},
       number = {4},
        pages = {5138-5147},
          doi = {10.1093/mnras/stab2785},
archivePrefix = {arXiv},
       eprint = {2103.01968},
 primaryClass = {astro-ph.HE},
       adsurl = {https://ui.adsabs.harvard.edu/abs/2021MNRAS.508.5138P}
}

@ARTICLE{OConnor2022,
       author = {{O'Connor}, B. and {Troja}, E. and {Dichiara}, S. and {Beniamini}, P. and {Cenko}, S.~B. and {Kouveliotou}, C. and {Gonz{\'a}lez}, J.~B. and {Durbak}, J. and {Gatkine}, P. and {Kutyrev}, A. and {Sakamoto}, T. and {S{\'a}nchez-Ram{\'\i}rez}, R. and {Veilleux}, S.},
        title = "{A deep survey of short GRB host galaxies over z 0-2: implications for offsets, redshifts, and environments}",
      journal = {\mnras},
         year = 2022,
        month = oct,
       volume = {515},
       number = {4},
        pages = {4890-4928},
          doi = {10.1093/mnras/stac1982},
archivePrefix = {arXiv},
       eprint = {2204.09059},
 primaryClass = {astro-ph.HE},
       adsurl = {https://ui.adsabs.harvard.edu/abs/2022MNRAS.515.4890O}
}

@ARTICLE{Nugent2022,
       author = {{Nugent}, Anya E. and {Fong}, Wen-Fai and {Dong}, Yuxin and {Leja}, Joel and {Berger}, Edo and {Zevin}, Michael and {Chornock}, Ryan and {Cobb}, Bethany E. and {Kelley}, Luke Zoltan and {Kilpatrick}, Charles D. and {Levan}, Andrew and {Margutti}, Raffaella and {Paterson}, Kerry and {Perley}, Daniel and {Escorial}, Alicia Rouco and {Smith}, Nathan and {Tanvir}, Nial},
        title = "{Short GRB Host Galaxies. II. A Legacy Sample of Redshifts, Stellar Population Properties, and Implications for Their Neutron Star Merger Origins}",
      journal = {\apj},
         year = 2022,
        month = nov,
       volume = {940},
       number = {1},
          eid = {57},
        pages = {57},
          doi = {10.3847/1538-4357/ac91d1},
archivePrefix = {arXiv},
       eprint = {2206.01764},
 primaryClass = {astro-ph.GA},
       adsurl = {https://ui.adsabs.harvard.edu/abs/2022ApJ...940...57N}
}

@ARTICLE{Fong2022,
       author = {{Fong}, Wen-fai and {Nugent}, Anya E. and {Dong}, Yuxin and {Berger}, Edo and {Paterson}, Kerry and {Chornock}, Ryan and {Levan}, Andrew and {Blanchard}, Peter and {Alexander}, Kate D. and {Andrews}, Jennifer and {Cobb}, Bethany E. and {Cucchiara}, Antonino and {Fox}, Derek and {Fryer}, Chris L. and {Gordon}, Alexa C. and {Kilpatrick}, Charles D. and {Lunnan}, Ragnhild and {Margutti}, Raffaella and {Miller}, Adam and {Milne}, Peter and {Nicholl}, Matt and {Perley}, Daniel and {Rastinejad}, Jillian and {Escorial}, Alicia Rouco and {Schroeder}, Genevieve and {Smith}, Nathan and {Tanvir}, Nial and {Terreran}, Giacomo},
        title = "{Short GRB Host Galaxies. I. Photometric and Spectroscopic Catalogs, Host Associations, and Galactocentric Offsets}",
      journal = {\apj},
         year = 2022,
        month = nov,
       volume = {940},
       number = {1},
          eid = {56},
        pages = {56},
          doi = {10.3847/1538-4357/ac91d0},
archivePrefix = {arXiv},
       eprint = {2206.01763},
 primaryClass = {astro-ph.GA},
       adsurl = {https://ui.adsabs.harvard.edu/abs/2022ApJ...940...56F}
}

@ARTICLE{casa,
       author = {{CASA Team} and {Bean}, Ben and {Bhatnagar}, Sanjay and {Castro}, Sandra and {Donovan Meyer}, Jennifer and {Emonts}, Bjorn and {Garcia}, Enrique and {Garwood}, Robert and {Golap}, Kumar and {Gonzalez Villalba}, Justo and {Harris}, Pamela and {Hayashi}, Yohei and {Hoskins}, Josh and {Hsieh}, Mingyu and {Jagannathan}, Preshanth and {Kawasaki}, Wataru and {Keimpema}, Aard and {Kettenis}, Mark and {Lopez}, Jorge and {Marvil}, Joshua and {Masters}, Joseph and {McNichols}, Andrew and {Mehringer}, David and {Miel}, Renaud and {Moellenbrock}, George and {Montesino}, Federico and {Nakazato}, Takeshi and {Ott}, Juergen and {Petry}, Dirk and {Pokorny}, Martin and {Raba}, Ryan and {Rau}, Urvashi and {Schiebel}, Darrell and {Schweighart}, Neal and {Sekhar}, Srikrishna and {Shimada}, Kazuhiko and {Small}, Des and {Steeb}, Jan-Willem and {Sugimoto}, Kanako and {Suoranta}, Ville and {Tsutsumi}, Takahiro and {van Bemmel}, Ilse M. and {Verkouter}, Marjolein and {Wells}, Akeem and {Xiong}, Wei and {Szomoru}, Arpad and {Griffith}, Morgan and {Glendenning}, Brian and {Kern}, Jeff},
        title = "{CASA, the Common Astronomy Software Applications for Radio Astronomy}",
      journal = {\pasp},
         year = 2022,
        month = nov,
       volume = {134},
       number = {1041},
          eid = {114501},
        pages = {114501},
          doi = {10.1088/1538-3873/ac9642},
archivePrefix = {arXiv},
       eprint = {2210.02276},
 primaryClass = {astro-ph.IM},
       adsurl = {https://ui.adsabs.harvard.edu/abs/2022PASP..134k4501C}
}

@software{HEAsoft,
       author = {{Nasa High Energy Astrophysics Science Archive Research Center (Heasarc)}},
        title = "{HEAsoft: Unified Release of FTOOLS and XANADU}",
 howpublished = {Astrophysics Source Code Library, record ascl:1408.004},
         year = 2014,
        month = aug,
          eid = {ascl:1408.004},
       adsurl = {https://ui.adsabs.harvard.edu/abs/2014ascl.soft08004N}
}

@ARTICLE{sfft,
       author = {{Hu}, Lei and {Wang}, Lifan and {Chen}, Xingzhuo and {Yang}, Jiawen},
        title = "{Image Subtraction in Fourier Space}",
      journal = {\apj},
         year = 2022,
        month = sep,
       volume = {936},
       number = {2},
          eid = {157},
        pages = {157},
          doi = {10.3847/1538-4357/ac7394},
archivePrefix = {arXiv},
       eprint = {2109.09334},
 primaryClass = {astro-ph.IM},
       adsurl = {https://ui.adsabs.harvard.edu/abs/2022ApJ...936..157H}
}

@ARTICLE{Metzger2010,
       author = {{Metzger}, B.~D. and {Mart{\'\i}nez-Pinedo}, G. and {Darbha}, S. and {Quataert}, E. and {Arcones}, A. and {Kasen}, D. and {Thomas}, R. and {Nugent}, P. and {Panov}, I.~V. and {Zinner}, N.~T.},
        title = "{Electromagnetic counterparts of compact object mergers powered by the radioactive decay of r-process nuclei}",
      journal = {\mnras},
         year = 2010,
        month = aug,
       volume = {406},
       number = {4},
        pages = {2650-2662},
          doi = {10.1111/j.1365-2966.2010.16864.x},
archivePrefix = {arXiv},
       eprint = {1001.5029},
 primaryClass = {astro-ph.HE},
       adsurl = {https://ui.adsabs.harvard.edu/abs/2010MNRAS.406.2650M}
}

@ARTICLE{Metzger2012,
       author = {{Metzger}, B.~D. and {Berger}, E.},
        title = "{What is the Most Promising Electromagnetic Counterpart of a Neutron Star Binary Merger?}",
      journal = {\apj},
         year = 2012,
        month = feb,
       volume = {746},
       number = {1},
          eid = {48},
        pages = {48},
          doi = {10.1088/0004-637X/746/1/48},
archivePrefix = {arXiv},
       eprint = {1108.6056},
 primaryClass = {astro-ph.HE},
       adsurl = {https://ui.adsabs.harvard.edu/abs/2012ApJ...746...48M}
}

@ARTICLE{globus2014,
       author = {{Globus}, Noemie and {Levinson}, Amir},
        title = "{Jet Formation in GRBs: A Semi-analytic Model of MHD Flow in Kerr Geometry with Realistic Plasma Injection}",
      journal = {\apj},
         year = 2014,
        month = nov,
       volume = {796},
       number = {1},
          eid = {26},
        pages = {26},
          doi = {10.1088/0004-637X/796/1/26},
archivePrefix = {arXiv},
       eprint = {1408.0126},
 primaryClass = {astro-ph.HE},
       adsurl = {https://ui.adsabs.harvard.edu/abs/2014ApJ...796...26G}
}

@ARTICLE{Amati2008,
       author = {{Amati}, Lorenzo and {Guidorzi}, Cristiano and {Frontera}, Filippo and {Della Valle}, Massimo and {Finelli}, Fabio and {Landi}, Raffaella and {Montanari}, Enrico},
        title = "{Measuring the cosmological parameters with the E$_{p,i}$-E$_{iso}$ correlation of gamma-ray bursts}",
      journal = {\mnras},
         year = 2008,
        month = dec,
       volume = {391},
       number = {2},
        pages = {577-584},
          doi = {10.1111/j.1365-2966.2008.13943.x},
archivePrefix = {arXiv},
       eprint = {0805.0377},
 primaryClass = {astro-ph},
       adsurl = {https://ui.adsabs.harvard.edu/abs/2008MNRAS.391..577A}
}

@ARTICLE{Amati2013,
       author = {{Amati}, Lorenzo and {Della Valle}, Massimo},
        title = "{Measuring Cosmological Parameters with Gamma Ray Bursts}",
      journal = {International Journal of Modern Physics D},
         year = 2013,
        month = dec,
       volume = {22},
       number = {14},
          eid = {1330028},
        pages = {1330028},
          doi = {10.1142/S0218271813300280},
archivePrefix = {arXiv},
       eprint = {1310.3141},
 primaryClass = {astro-ph.CO},
       adsurl = {https://ui.adsabs.harvard.edu/abs/2013IJMPD..2230028A}
}

@ARTICLE{39446,
       author = {{Iskandar}, A. and {Zhu}, H. -C. and {Wang}, X. -F. and {Wang}, L. -T. and {Yan}, Shengyu},
        title = "{GRB 250221A: TNOT detection of the optical counterpart}",
      journal = {GRB Coordinates Network},
         year = 2025,
        month = feb,
       volume = {39446},
        pages = {1},
       adsurl = {https://ui.adsabs.harvard.edu/abs/2025GCN.39446....1I}
}

@ARTICLE{39425,
       author = {{Ghosh}, Ankur and {Razzaque}, Soebur and {Moskvitin}, Alexander and {Sotnikova}, Yulia and {Dukiya}, Naveen and {Gupta}, Rahul},
        title = "{GRB 250221A: Optical counterpart detection by LCO.}",
      journal = {GRB Coordinates Network},
         year = 2025,
        month = feb,
       volume = {39425},
        pages = {1},
       adsurl = {https://ui.adsabs.harvard.edu/abs/2025GCN.39425....1G}
}

@ARTICLE{39424,
       author = {{Pankov}, N. and {Pozanenko}, A. and {Klunlo}, E. and {Volnova}, A. and {IKI-GRB-FuN Collaboration}},
        title = "{GRB 250221A: Mondy optical observations}",
      journal = {GRB Coordinates Network},
         year = 2025,
        month = feb,
       volume = {39424},
        pages = {1},
       adsurl = {https://ui.adsabs.harvard.edu/abs/2025GCN.39424....1P}
}

@ARTICLE{39423,
       author = {{Frederiks}, D. and {Lysenko}, A. and {Ridnaya}, A. and {Svinkin}, D. and {Tsvetkova}, A. and {Ulanov}, M. and {Cline}, T. and {Konus-Wind Team}},
        title = "{Konus-Wind detection of GRB 250221A}",
      journal = {GRB Coordinates Network},
         year = 2025,
        month = feb,
       volume = {39423},
        pages = {1},
       adsurl = {https://ui.adsabs.harvard.edu/abs/2025GCN.39423....1F}
}

@ARTICLE{39422,
       author = {{Pankov}, N. and {Kim}, V. and {Krugov}, M. and {Aimuratov}, Y. and {Volnova}, A. and {Pozanenko}, A. and {IKI-GRB-FuN Collaboration}},
        title = "{GRB 250221A: Assy optical observations}",
      journal = {GRB Coordinates Network},
         year = 2025,
        month = feb,
       volume = {39422},
        pages = {1},
       adsurl = {https://ui.adsabs.harvard.edu/abs/2025GCN.39422....1P}
}

@ARTICLE{39417,
       author = {{Muenter}, H. and {Magnani}, F. and {Antier}, S. and {Rajabov}, Y. and {Andrade}, C. and {Klotz}, A. and {Limonta}, C. and {Andre}, Q. and {Durroux}, A. and {Coughlin}, M. and {Karpov}, S. and {Hello}, P. and {Duverne}, P. -A. and {Pradier}, T. and {Guessoum}, N. and {Grandma Collaboration}},
        title = "{GRB 250221A: GRANDMA/TAROT Detection}",
      journal = {GRB Coordinates Network},
         year = 2025,
        month = feb,
       volume = {39417},
        pages = {1},
       adsurl = {https://ui.adsabs.harvard.edu/abs/2025GCN.39417....1M}
}

@ARTICLE{39413,
       author = {{Cotter}, L. and {Malesani}, D.~B. and {Palmerio}, J. and {de Ugarte Postigo}, A. and {Martin-Carrillo}, A.},
        title = "{GRB 250221A: NOT optical observations}",
      journal = {GRB Coordinates Network},
         year = 2025,
        month = feb,
       volume = {39413},
        pages = {1},
       adsurl = {https://ui.adsabs.harvard.edu/abs/2025GCN.39413....1C}
}

@ARTICLE{39412,
       author = {{Guo}, Helong and {Du}, Guowang and {Chen}, Xinlei and {Chen}, Xiaotong and {Xie}, Yiheng and {Kumar}, Brajesh and {Fang}, Yuan and {Zou}, Xingzhu and {Pan}, Yu and {Han}, Xuhui and {Zhang}, Pinpin and {Xin}, Liping and {Wu}, Chao and {Yang}, Yuanpei and {Zhang}, Jinghua and {Liu}, Xiangkun and {Liu}, Xiaowei and {Mephisto Team}},
        title = "{GRB 250221A: 1.6m Mephisto optical detection}",
      journal = {GRB Coordinates Network},
         year = 2025,
        month = feb,
       volume = {39412},
        pages = {1},
       adsurl = {https://ui.adsabs.harvard.edu/abs/2025GCN.39412....1G}
}

@ARTICLE{Zalamea2011,
       author = {{Zalamea}, Ivan and {Beloborodov}, Andrei M.},
        title = "{Neutrino heating near hyper-accreting black holes}",
      journal = {\mnras},
         year = 2011,
        month = feb,
       volume = {410},
       number = {4},
        pages = {2302-2308},
          doi = {10.1111/j.1365-2966.2010.17600.x},
archivePrefix = {arXiv},
       eprint = {1003.0710},
 primaryClass = {astro-ph.HE},
       adsurl = {https://ui.adsabs.harvard.edu/abs/2011MNRAS.410.2302Z}
}

@ARTICLE{39410,
       author = {{Odeh}, Mohammad and {Alshamsi}, Shaikha and {Manal Pattani}, Nuha and {Guessoum}, Nidhal},
        title = "{GRB 250221A: AKO Optical Upper Limit}",
      journal = {GRB Coordinates Network},
         year = 2025,
        month = feb,
       volume = {39410},
        pages = {1},
       adsurl = {https://ui.adsabs.harvard.edu/abs/2025GCN.39410....1O}
}

@ARTICLE{39406,
       author = {{Melandri}, A. and {Brivio}, R. and {Ferro}, M. and {D'Avanzo}, P. and {Covino}, S. and {Fugazza}, D. and {REM Team}},
        title = "{GRB 250221A: REM optical/NIR afterglow detection}",
      journal = {GRB Coordinates Network},
         year = 2025,
        month = feb,
       volume = {39406},
        pages = {1},
       adsurl = {https://ui.adsabs.harvard.edu/abs/2025GCN.39406....1M}
}

@ARTICLE{39418,
       author = {{Palmerio}, J.~T. and {Saccardi}, A. and {Rayson}, B. and {Malesani}, D.~B. and {Levan}, A.~J. and {Tanvir}, N.~R. and {Stargate Collaboration}},
        title = "{GRB 250221A: VLT/X-shooter redshift z = 0.768}",
      journal = {GRB Coordinates Network},
         year = 2025,
        month = feb,
       volume = {39418},
        pages = {1},
       adsurl = {https://ui.adsabs.harvard.edu/abs/2025GCN.39418....1P}
}

@ARTICLE{39501,
       author = {{Gulati}, A. and {Anderson}, G.~E. and {Morley}, Claire and {Chastain}, S. and {Leung}, J.~K. and {van der Horst}, A.~J. and {Rhodes}, L. and {ATCA PanRadio GRB Collaboration}},
        title = "{GRB 250221A: ATCA Detections and Upper Limits}",
      journal = {GRB Coordinates Network},
         year = 2025,
        month = feb,
       volume = {39501},
        pages = {1},
       adsurl = {https://ui.adsabs.harvard.edu/abs/2025GCN.39501....1G}
}

@ARTICLE{39433,
       author = {{Ricci}, R. and {Troja}, E.},
        title = "{GRB 250221A: VLA radio detection}",
      journal = {GRB Coordinates Network},
         year = 2025,
        month = feb,
       volume = {39433},
        pages = {1},
       adsurl = {https://ui.adsabs.harvard.edu/abs/2025GCN.39433....1R}
}

@ARTICLE{McCracken2003,
       author = {{McCracken}, H.~J. and {Radovich}, M. and {Bertin}, E. and {Mellier}, Y. and {Dantel-Fort}, M. and {Le F{\`e}vre}, O. and {Cuillandre}, J.~C. and {Gwyn}, S. and {Foucaud}, S. and {Zamorani}, G.},
        title = "{The VIRMOS deep imaging survey.  II: CFH12K BVRI optical data for the 0226-04 deep field}",
      journal = {\aap},
         year = 2003,
        month = oct,
       volume = {410},
        pages = {17-32},
          doi = {10.1051/0004-6361:20031081},
archivePrefix = {arXiv},
       eprint = {astro-ph/0306254},
 primaryClass = {astro-ph},
       adsurl = {https://ui.adsabs.harvard.edu/abs/2003A&A...410...17M}
}

@ARTICLE{Kashikawa2004,
       author = {{Kashikawa}, Nobunari and {Shimasaku}, Kazuhiro and {Yasuda}, Naoki and {Ajiki}, Masaru and {Akiyama}, Masayuki and {Ando}, Hiroyasu and {Aoki}, Kentaro and {Doi}, Mamoru and {Fujita}, Shinobu S. and {Furusawa}, Hisanori and {Hayashino}, Tomoki and {Iwamuro}, Fumihide and {Iye}, Masanori and {Karoji}, Hiroshi and {Kobayashi}, Naoto and {Kodaira}, Keiichi and {Kodama}, Tadayuki and {Komiyama}, Yutaka and {Matsuda}, Yuichi and {Miyazaki}, Satoshi and {Mizumoto}, Yoshihiko and {Morokuma}, Tomoki and {Motohara}, Kentaro and {Murayama}, Takashi and {Nagao}, Tohru and {Nariai}, Kyoji and {Ohta}, Kouji and {Okamura}, Sadanori and {Ouchi}, Masami and {Sasaki}, Toshiyuki and {Sato}, Yasunori and {Sekiguchi}, Kazuhiro and {Shioya}, Yasunori and {Tamura}, Hajime and {Taniguchi}, Yoshiaki and {Umemura}, Masayuki and {Yamada}, Toru and {Yoshida}, Makiko},
        title = "{The Subaru Deep Field: The Optical Imaging Data}",
      journal = {\pasj},
         year = 2004,
        month = dec,
       volume = {56},
        pages = {1011-1023},
          doi = {10.1093/pasj/56.6.1011},
archivePrefix = {arXiv},
       eprint = {astro-ph/0410005},
 primaryClass = {astro-ph},
       adsurl = {https://ui.adsabs.harvard.edu/abs/2004PASJ...56.1011K}
}

@ARTICLE{Piran1999,
       author = {{Piran}, T.},
        title = "{Gamma-ray bursts and the fireball model}",
      journal = {\physrep},
         year = 1999,
        month = jun,
       volume = {314},
       number = {6},
        pages = {575-667},
          doi = {10.1016/S0370-1573(98)00127-6},
archivePrefix = {arXiv},
       eprint = {astro-ph/9810256},
 primaryClass = {astro-ph},
       adsurl = {https://ui.adsabs.harvard.edu/abs/1999PhR...314..575P}
}

@ARTICLE{Metcalfe2001,
       author = {{Metcalfe}, N. and {Shanks}, T. and {Campos}, A. and {McCracken}, H.~J. and {Fong}, R.},
        title = "{Galaxy number counts - V. Ultradeep counts: the Herschel and Hubble Deep Fields}",
      journal = {\mnras},
         year = 2001,
        month = may,
       volume = {323},
       number = {4},
        pages = {795-830},
          doi = {10.1046/j.1365-8711.2001.04168.x},
archivePrefix = {arXiv},
       eprint = {astro-ph/0010153},
 primaryClass = {astro-ph},
       adsurl = {https://ui.adsabs.harvard.edu/abs/2001MNRAS.323..795M}
}

@ARTICLE{Granot99,
       author = {{Granot}, Jonathan and {Piran}, Tsvi and {Sari}, Re'em},
        title = "{Images and Spectra from the Interior of a Relativistic Fireball}",
      journal = {\apj},
         year = 1999,
        month = mar,
       volume = {513},
       number = {2},
        pages = {679-689},
          doi = {10.1086/306884},
archivePrefix = {arXiv},
       eprint = {astro-ph/9806192},
 primaryClass = {astro-ph},
       adsurl = {https://ui.adsabs.harvard.edu/abs/1999ApJ...513..679G}
}

@ARTICLE{Bloom2002,
       author = {{Bloom}, J.~S. and {Kulkarni}, S.~R. and {Djorgovski}, S.~G.},
        title = "{The Observed Offset Distribution of Gamma-Ray Bursts from Their Host Galaxies: A Robust Clue to the Nature of the Progenitors}",
      journal = {\aj},
         year = 2002,
        month = mar,
       volume = {123},
       number = {3},
        pages = {1111-1148},
          doi = {10.1086/338893},
archivePrefix = {arXiv},
       eprint = {astro-ph/0010176},
 primaryClass = {astro-ph},
       adsurl = {https://ui.adsabs.harvard.edu/abs/2002AJ....123.1111B}
}

@ARTICLE{Nakar2003,
       author = {{Nakar}, Ehud and {Piran}, Tsvi and {Granot}, Jonathan},
        title = "{Variability in GRB afterglows and GRB 021004}",
      journal = {\na},
         year = 2003,
        month = jul,
       volume = {8},
       number = {5},
        pages = {495-505},
          doi = {10.1016/S1384-1076(03)00044-7},
archivePrefix = {arXiv},
       eprint = {astro-ph/0210631},
 primaryClass = {astro-ph},
       adsurl = {https://ui.adsabs.harvard.edu/abs/2003NewA....8..495N}
}

@ARTICLE{Lazzati2002,
       author = {{Lazzati}, D. and {Rossi}, E. and {Covino}, S. and {Ghisellini}, G. and {Malesani}, D.},
        title = "{The afterglow of GRB 021004: Surfing on density waves}",
      journal = {\aap},
         year = 2002,
        month = dec,
       volume = {396},
        pages = {L5-L9},
          doi = {10.1051/0004-6361:20021618},
archivePrefix = {arXiv},
       eprint = {astro-ph/0210333},
 primaryClass = {astro-ph},
       adsurl = {https://ui.adsabs.harvard.edu/abs/2002A&A...396L...5L}
}

@ARTICLE{Abbott2017,
       author = {{Abbott}, B.~P. and {Abbott}, R. and {Abbott}, T.~D. and {Acernese}, F. and {Ackley}, K. and {Adams}, C. and {Adams}, T. and {Addesso}, P. and {Adhikari}, R.~X. and {Adya}, V.~B. and {Affeldt}, C. and {Afrough}, M. and {Agarwal}, B. and {Agathos}, M. and {Agatsuma}, K. and {Aggarwal}, N. and {Aguiar}, O.~D. and {Aiello}, L. and {Ain}, A. and {Ajith}, P. and {Allen}, B. and {Allen}, G. and {Allocca}, A. and {Aloy}, M.~A. and {Altin}, P.~A. and {Amato}, A. and {Ananyeva}, A. and {Anderson}, S.~B. and {Anderson}, W.~G. and {Angelova}, S.~V. and {Antier}, S. and {Appert}, S. and {Arai}, K. and {Araya}, M.~C. and {Areeda}, J.~S. and {Arnaud}, N. and {Arun}, K.~G. and {Ascenzi}, S. and {Ashton}, G. and {Ast}, M. and {Aston}, S.~M. and {Astone}, P. and {Atallah}, D.~V. and {Aufmuth}, P. and {Aulbert}, C. and {AultONeal}, K. and {Austin}, C. and {Avila-Alvarez}, A. and {Babak}, S. and {Bacon}, P. and {Bader}, M.~K.~M. and {Bae}, S. and {Baker}, P.~T. and {Baldaccini}, F. and {Ballardin}, G. and {Ballmer}, S.~W. and {Banagiri}, S. and {Barayoga}, J.~C. and {Barclay}, S.~E. and {Barish}, B.~C. and {Barker}, D. and {Barkett}, K. and {Barone}, F. and {Barr}, B. and {Barsotti}, L. and {Barsuglia}, M. and {Barta}, D. and {Bartlett}, J. and {Bartos}, I. and {Bassiri}, R. and {Basti}, A. and {Batch}, J.~C. and {Bawaj}, M. and {Bayley}, J.~C. and {Bazzan}, M. and {B{\'e}csy}, B. and {Beer}, C. and {Bejger}, M. and {Belahcene}, I. and {Bell}, A.~S. and {Berger}, B.~K. and {Bergmann}, G. and {Bero}, J.~J. and {Berry}, C.~P.~L. and {Bersanetti}, D. and {Bertolini}, A. and {Betzwieser}, J. and {Bhagwat}, S. and {Bhandare}, R. and {Bilenko}, I.~A. and {Billingsley}, G. and {Billman}, C.~R. and {Birch}, J. and {Birney}, R. and {Birnholtz}, O. and {Biscans}, S. and {Biscoveanu}, S. and {Bisht}, A. and {Bitossi}, M. and {Biwer}, C. and {Bizouard}, M.~A. and {Blackburn}, J.~K. and {Blackman}, J. and {Blair}, C.~D. and {Blair}, D.~G. and {Blair}, R.~M. and {Bloemen}, S. and {Bock}, O. and {Bode}, N. and {Boer}, M. and {Bogaert}, G. and {Bohe}, A. and {Bondu}, F. and {Bonilla}, E. and {Bonnand}, R. and {Boom}, B.~A. and {Bork}, R. and {Boschi}, V. and {Bose}, S. and {Bossie}, K. and {Bouffanais}, Y. and {Bozzi}, A. and {Bradaschia}, C. and {Brady}, P.~R. and {Branchesi}, M. and {Brau}, J.~E. and {Briant}, T. and {Brillet}, A. and {Brinkmann}, M. and {Brisson}, V. and {Brockill}, P. and {Broida}, J.~E. and {Brooks}, A.~F. and {Brown}, D.~A. and {Brown}, D.~D. and {Brunett}, S. and {Buchanan}, C.~C. and {Buikema}, A. and {Bulik}, T. and {Bulten}, H.~J. and {Buonanno}, A. and {Buskulic}, D. and {Buy}, C. and {Byer}, R.~L. and {Cabero}, M. and {Cadonati}, L. and {Cagnoli}, G. and {Cahillane}, C. and {Calder{\'o}n Bustillo}, J. and {Callister}, T.~A. and {Calloni}, E. and {Camp}, J.~B. and {Canepa}, M. and {Canizares}, P. and {Cannon}, K.~C. and {Cao}, H. and {Cao}, J. and {Capano}, C.~D. and {Capocasa}, E. and {Carbognani}, F. and {Caride}, S. and {Carney}, M.~F. and {Casanueva Diaz}, J. and {Casentini}, C. and {Caudill}, S. and {Cavagli{\`a}}, M. and {Cavalier}, F. and {Cavalieri}, R. and {Cella}, G. and {Cepeda}, C.~B. and {Cerd{\'a}-Dur{\'a}n}, P. and {Cerretani}, G. and {Cesarini}, E. and {Chamberlin}, S.~J. and {Chan}, M. and {Chao}, S. and {Charlton}, P. and {Chase}, E. and {Chassande-Mottin}, E. and {Chatterjee}, D. and {Chatziioannou}, K. and {Cheeseboro}, B.~D. and {Chen}, H.~Y. and {Chen}, X. and {Chen}, Y. and {Cheng}, H. -P. and {Chia}, H. and {Chincarini}, A. and {Chiummo}, A. and {Chmiel}, T. and {Cho}, H.~S. and {Cho}, M. and {Chow}, J.~H. and {Christensen}, N. and {Chu}, Q. and {Chua}, A.~J.~K. and {Chua}, S. and {Chung}, A.~K.~W. and {Chung}, S. and {Ciani}, G.},
        title = "{Gravitational Waves and Gamma-Rays from a Binary Neutron Star Merger: GW170817 and GRB 170817A}",
      journal = {\apjl},
         year = 2017,
        month = oct,
       volume = {848},
       number = {2},
          eid = {L13},
        pages = {L13},
          doi = {10.3847/2041-8213/aa920c},
archivePrefix = {arXiv},
       eprint = {1710.05834},
 primaryClass = {astro-ph.HE},
       adsurl = {https://ui.adsabs.harvard.edu/abs/2017ApJ...848L..13A}
}

@ARTICLE{Yang2015,
       author = {{Yang}, Bin and {Jin}, Zhi-Ping and {Li}, Xiang and {Covino}, Stefano and {Zheng}, Xian-Zhong and {Hotokezaka}, Kenta and {Fan}, Yi-Zhong and {Piran}, Tsvi and {Wei}, Da-Ming},
        title = "{A possible macronova in the late afterglow of the long-short burst GRB 060614}",
      journal = {Nature Communications},
         year = 2015,
        month = jun,
       volume = {6},
          eid = {7323},
        pages = {7323},
          doi = {10.1038/ncomms8323},
archivePrefix = {arXiv},
       eprint = {1503.07761},
 primaryClass = {astro-ph.HE},
       adsurl = {https://ui.adsabs.harvard.edu/abs/2015NatCo...6.7323Y}
}

@ARTICLE{Schroeder2025,
       author = {{Schroeder}, Genevieve and {Fong}, Wen-fai and {Kilpatrick}, Charles D. and {Rouco Escorial}, Alicia and {Laskar}, Tanmoy and {Nugent}, Anya E. and {Rastinejad}, Jillian and {Alexander}, Kate D. and {Berger}, Edo and {Brink}, Thomas G. and {Chornock}, Ryan and {de Bom}, Clecio R. and {Dong}, Yuxin and {Eftekhari}, Tarraneh and {Filippenko}, Alexei V. and {Fuentes-Carvajal}, Celeste and {Jacobson-Gal{\'a}n}, Wynn V. and {Malkan}, Matthew and {Margutti}, Raffaella and {Pearson}, Jeniveve and {Rhodes}, Lauren and {Salinas}, Ricardo and {Sand}, David J. and {Santana-Silva}, Luidhy and {Santos}, Andre and {Sears}, Huei and {Shrestha}, Manisha and {Smith}, Nathan and {Webb}, Wayne and {de Wet}, Simon and {Yang}, Yi},
        title = "{The Long-lived Broadband Afterglow of Short Gamma-Ray Burst 231117A and the Growing Radio-detected Short Gamma-Ray Burst Population}",
      journal = {\apj},
         year = 2025,
        month = mar,
       volume = {982},
       number = {1},
          eid = {42},
        pages = {42},
          doi = {10.3847/1538-4357/ada9e5},
archivePrefix = {arXiv},
       eprint = {2407.13822},
 primaryClass = {astro-ph.HE},
       adsurl = {https://ui.adsabs.harvard.edu/abs/2025ApJ...982...42S}
}

@ARTICLE{Yuan2025,
       author = {{Yuan}, Weimin and {Dai}, Lixin and {Feng}, Hua and {Jin}, Chichuan and {Jonker}, Peter and {Kuulkers}, Erik and {Liu}, Yuan and {Nandra}, Kirpal and {O'Brien}, Paul and {Piro}, Luigi and {Rau}, Arne and {Rea}, Nanda and {Sanders}, Jeremy and {Tao}, Lian and {Wang}, Junfeng and {Wu}, Xuefeng and {Zhang}, Bing and {Zhang}, Shuangnan and {Ai}, Shunke and {Buchner}, Johannes and {Bulbul}, Esra and {Chen}, Hechao and {Chen}, Minghua and {Chen}, Yong and {Chen}, Yu-Peng and {Coleiro}, Alexis and {Coti Zelati}, Francesco and {Dai}, Zigao and {Fan}, Xilong and {Fan}, Zhou and {Friedrich}, Susanne and {Gao}, He and {Ge}, Chong and {Ge}, Mingyu and {Geng}, Jinjun and {Ghirlanda}, Giancarlo and {Gianfagna}, Giulia and {Gou}, Lijun and {Guillot}, S{\'e}bastien and {Hou}, Xian and {Hu}, Jingwei and {Huang}, Yongfeng and {Ji}, Long and {Jia}, Shumei and {Komossa}, S. and {Kong}, Albert K.~H. and {Lan}, Lin and {Li}, An and {Li}, Ang and {Li}, Chengkui and {Li}, Dongyue and {Li}, Jian and {Li}, Zhaosheng and {Ling}, Zhixing and {Liu}, Ang and {Liu}, Jinzhong and {Liu}, Liangduan and {Liu}, Zhu and {Luo}, Jiawei and {Ma}, Ruican and {Maggi}, Pierre and {Maitra}, Chandreyee and {Marino}, Alessio and {Ng}, Stephen Chi-Yung and {Pan}, Haiwu and {Rukdee}, Surangkhana and {Soria}, Roberto and {Sun}, Hui and {Tam}, Pak-Hin Thomas and {Thakur}, Aishwarya Linesh and {Tian}, Hui and {Troja}, Eleonora and {Wang}, Wei and {Wang}, Xiangyu and {Wang}, Yanan and {Wei}, Junjie and {Wen}, Sixiang and {Wu}, Jianfeng and {Wu}, Ting and {Xiao}, Di and {Xu}, Dong and {Xu}, Renxin and {Xu}, Yanjun and {Xu}, Yu and {Yang}, Haonan and {You}, Bei and {Yu}, Heng and {Yu}, Yunwei and {Zhang}, Binbin and {Zhang}, Chen and {Zhang}, Guobao and {Zhang}, Liang and {Zhang}, Wenda and {Zhang}, Yu and {Zhou}, Ping and {Zou}, Zecheng},
        title = "{Science objectives of the Einstein Probe mission}",
      journal = {Science China Physics, Mechanics, and Astronomy},
         year = 2025,
        month = mar,
       volume = {68},
       number = {3},
          eid = {239501},
        pages = {239501},
          doi = {10.1007/s11433-024-2600-3},
archivePrefix = {arXiv},
       eprint = {2501.07362},
 primaryClass = {astro-ph.HE},
       adsurl = {https://ui.adsabs.harvard.edu/abs/2025SCPMA..6839501Y}
}

@ARTICLE{Gendre2013,
       author = {{Gendre}, B. and {Stratta}, G. and {Atteia}, J.~L. and {Basa}, S. and {Bo{\"e}r}, M. and {Coward}, D.~M. and {Cutini}, S. and {D'Elia}, V. and {Howell}, E.~J. and {Klotz}, A. and {Piro}, L.},
        title = "{The Ultra-long Gamma-Ray Burst 111209A: The Collapse of a Blue Supergiant?}",
      journal = {\apj},
         year = 2013,
        month = mar,
       volume = {766},
       number = {1},
          eid = {30},
        pages = {30},
          doi = {10.1088/0004-637X/766/1/30},
archivePrefix = {arXiv},
       eprint = {1212.2392},
 primaryClass = {astro-ph.HE},
       adsurl = {https://ui.adsabs.harvard.edu/abs/2013ApJ...766...30G}
}

@ARTICLE{Kann2019,
       author = {{Kann}, D.~A. and {Schady}, P. and {Olivares E.}, F. and {Klose}, S. and {Rossi}, A. and {Perley}, D.~A. and {Kr{\"u}hler}, T. and {Greiner}, J. and {Nicuesa Guelbenzu}, A. and {Elliott}, J. and {Knust}, F. and {Filgas}, R. and {Pian}, E. and {Mazzali}, P. and {Fynbo}, J.~P.~U. and {Leloudas}, G. and {Afonso}, P.~M.~J. and {Delvaux}, C. and {Graham}, J.~F. and {Rau}, A. and {Schmidl}, S. and {Schulze}, S. and {Tanga}, M. and {Updike}, A.~C. and {Varela}, K.},
        title = "{Highly luminous supernovae associated with gamma-ray bursts. I. GRB 111209A/SN 2011kl in the context of stripped-envelope and superluminous supernovae}",
      journal = {\aap},
         year = 2019,
        month = apr,
       volume = {624},
          eid = {A143},
        pages = {A143},
          doi = {10.1051/0004-6361/201629162},
archivePrefix = {arXiv},
       eprint = {1606.06791},
 primaryClass = {astro-ph.HE},
       adsurl = {https://ui.adsabs.harvard.edu/abs/2019A&A...624A.143K}
}

@ARTICLE{Nardini2014,
       author = {{Nardini}, M. and {Elliott}, J. and {Filgas}, R. and {Schady}, P. and {Greiner}, J. and {Kr{\"u}hler}, T. and {Klose}, S. and {Afonso}, P. and {Kann}, D.~A. and {Nicuesa Guelbenzu}, A. and {Olivares E.}, F. and {Rau}, A. and {Rossi}, A. and {Sudilovsky}, V. and {Schmidl}, S.},
        title = "{Afterglow rebrightenings as a signature of a long-lasting central engine activity?. The emblematic case of GRB 100814A}",
      journal = {\aap},
         year = 2014,
        month = feb,
       volume = {562},
          eid = {A29},
        pages = {A29},
          doi = {10.1051/0004-6361/201321525},
archivePrefix = {arXiv},
       eprint = {1312.1335},
 primaryClass = {astro-ph.HE},
       adsurl = {https://ui.adsabs.harvard.edu/abs/2014A&A...562A..29N}
}

@ARTICLE{vanDalen2025,
       author = {{van Dalen}, Joyce N.~D. and {Levan}, Andrew J. and {Jonker}, Peter G. and {Malesani}, Daniele Bj{\o}rn and {Izzo}, Luca and {Sarin}, Nikhil and {Quirola-V{\'a}squez}, Jonathan and {Mata S{\'a}nchez}, Daniel and {de Ugarte Postigo}, Antonio and {van Hoof}, Agnes P.~C. and {Torres}, Manuel A.~P. and {Schulze}, Steve and {Littlefair}, Stuart P. and {Chrimes}, Ashley and {Ravasio}, Maria E. and {Bauer}, Franz E. and {Martin-Carrillo}, Antonio and {Fraser}, Morgan and {van der Horst}, Alexander J. and {Jakobsson}, Pall and {O'Brien}, Paul and {De Pasquale}, Massimiliano and {Pugliese}, Giovanna and {Sollerman}, Jesper and {Tanvir}, Nial R. and {Zafar}, Tayyaba and {Anderson}, Joseph P. and {Galbany}, Llu{\'\i}s and {Gal-Yam}, Avishay and {Gromadzki}, Mariusz and {M{\"u}ller-Bravo}, Tom{\'a}s E. and {Ragosta}, Fabio and {Terwel}, Jacco H.},
        title = "{The Einstein Probe Transient EP240414a: Linking Fast X-Ray Transients, Gamma-Ray Bursts, and Luminous Fast Blue Optical Transients}",
      journal = {\apjl},
         year = 2025,
        month = apr,
       volume = {982},
       number = {2},
          eid = {L47},
        pages = {L47},
          doi = {10.3847/2041-8213/adbc7e},
archivePrefix = {arXiv},
       eprint = {2409.19056},
 primaryClass = {astro-ph.HE},
       adsurl = {https://ui.adsabs.harvard.edu/abs/2025ApJ...982L..47V}
}

@ARTICLE{Gianfagna2025,
       author = {{Gianfagna}, Giulia and {Piro}, Luigi and {Bruni}, Gabriele and {Thakur}, Aishwarya Linesh and {Van Eerten}, Hendrik and {Caballero-Garc{\'\i}a}, Maria D. and {Castro-Tirado}, Alberto and {Chen}, Yong and {Cheng}, Ye-hao and {Gritsevich}, Maria and {Guziy}, Sergiy and {He}, Han and {Hu}, You-Dong and {Jia}, Shumei and {Ling}, Zhixing and {Maiorano}, Elisabetta and {Paladino}, Rosita and {Pandey}, Shashi B. and {Tripodi}, Roberta and {Rossi}, Andrea and {S{\'a}nchez-Ram{\'\i}rez}, Rub{\'e}n and {Yang}, Shuaikang and {Yuan}, Jianghui and {Yuan}, Weimin and {Zhang}, Chen},
        title = "{The soft X-ray transient EP241021A: A cosmic explosion with a complex off-axis jet and cocoon from a massive progenitor}",
      journal = {\aap},
         year = 2025,
        month = nov,
       volume = {703},
          eid = {A92},
        pages = {A92},
          doi = {10.1051/0004-6361/202555450},
archivePrefix = {arXiv},
       eprint = {2505.05444},
 primaryClass = {astro-ph.HE},
       adsurl = {https://ui.adsabs.harvard.edu/abs/2025A&A...703A..92G}
}

@ARTICLE{Zhang2007,
       author = {{Zhang}, Bing},
        title = "{Gamma-Ray Bursts in the Swift Era}",
      journal = {\cjaa},
         year = 2007,
        month = feb,
       volume = {7},
       number = {1},
        pages = {1-50},
          doi = {10.1088/1009-9271/7/1/01},
archivePrefix = {arXiv},
       eprint = {astro-ph/0701520},
 primaryClass = {astro-ph},
       adsurl = {https://ui.adsabs.harvard.edu/abs/2007ChJAA...7....1Z}
}

@ARTICLE{Dey2019,
       author = {{Dey}, Arjun and {Schlegel}, David J. and {Lang}, Dustin and {Blum}, Robert and {Burleigh}, Kaylan and {Fan}, Xiaohui and {Findlay}, Joseph R. and {Finkbeiner}, Doug and {Herrera}, David and {Juneau}, St{\'e}phanie and {Landriau}, Martin and {Levi}, Michael and {McGreer}, Ian and {Meisner}, Aaron and {Myers}, Adam D. and {Moustakas}, John and {Nugent}, Peter and {Patej}, Anna and {Schlafly}, Edward F. and {Walker}, Alistair R. and {Valdes}, Francisco and {Weaver}, Benjamin A. and {Y{\`e}che}, Christophe and {Zou}, Hu and {Zhou}, Xu and {Abareshi}, Behzad and {Abbott}, T.~M.~C. and {Abolfathi}, Bela and {Aguilera}, C. and {Alam}, Shadab and {Allen}, Lori and {Alvarez}, A. and {Annis}, James and {Ansarinejad}, Behzad and {Aubert}, Marie and {Beechert}, Jacqueline and {Bell}, Eric F. and {BenZvi}, Segev Y. and {Beutler}, Florian and {Bielby}, Richard M. and {Bolton}, Adam S. and {Brice{\~n}o}, C{\'e}sar and {Buckley-Geer}, Elizabeth J. and {Butler}, Karen and {Calamida}, Annalisa and {Carlberg}, Raymond G. and {Carter}, Paul and {Casas}, Ricard and {Castander}, Francisco J. and {Choi}, Yumi and {Comparat}, Johan and {Cukanovaite}, Elena and {Delubac}, Timoth{\'e}e and {DeVries}, Kaitlin and {Dey}, Sharmila and {Dhungana}, Govinda and {Dickinson}, Mark and {Ding}, Zhejie and {Donaldson}, John B. and {Duan}, Yutong and {Duckworth}, Christopher J. and {Eftekharzadeh}, Sarah and {Eisenstein}, Daniel J. and {Etourneau}, Thomas and {Fagrelius}, Parker A. and {Farihi}, Jay and {Fitzpatrick}, Mike and {Font-Ribera}, Andreu and {Fulmer}, Leah and {G{\"a}nsicke}, Boris T. and {Gaztanaga}, Enrique and {George}, Koshy and {Gerdes}, David W. and {Gontcho}, Satya Gontcho A. and {Gorgoni}, Claudio and {Green}, Gregory and {Guy}, Julien and {Harmer}, Diane and {Hernandez}, M. and {Honscheid}, Klaus and {Huang}, Lijuan Wendy and {James}, David J. and {Jannuzi}, Buell T. and {Jiang}, Linhua and {Joyce}, Richard and {Karcher}, Armin and {Karkar}, Sonia and {Kehoe}, Robert and {Kneib}, Jean-Paul and {Kueter-Young}, Andrea and {Lan}, Ting-Wen and {Lauer}, Tod R. and {Le Guillou}, Laurent and {Le Van Suu}, Auguste and {Lee}, Jae Hyeon and {Lesser}, Michael and {Perreault Levasseur}, Laurence and {Li}, Ting S. and {Mann}, Justin L. and {Marshall}, Robert and {Mart{\'\i}nez-V{\'a}zquez}, C.~E. and {Martini}, Paul and {du Mas des Bourboux}, H{\'e}lion and {McManus}, Sean and {Meier}, Tobias Gabriel and {M{\'e}nard}, Brice and {Metcalfe}, Nigel and {Mu{\~n}oz-Guti{\'e}rrez}, Andrea and {Najita}, Joan and {Napier}, Kevin and {Narayan}, Gautham and {Newman}, Jeffrey A. and {Nie}, Jundan and {Nord}, Brian and {Norman}, Dara J. and {Olsen}, Knut A.~G. and {Paat}, Anthony and {Palanque-Delabrouille}, Nathalie and {Peng}, Xiyan and {Poppett}, Claire L. and {Poremba}, Megan R. and {Prakash}, Abhishek and {Rabinowitz}, David and {Raichoor}, Anand and {Rezaie}, Mehdi and {Robertson}, A.~N. and {Roe}, Natalie A. and {Ross}, Ashley J. and {Ross}, Nicholas P. and {Rudnick}, Gregory and {Safonova}, Sasha and {Saha}, Abhijit and {S{\'a}nchez}, F. Javier and {Savary}, Elodie and {Schweiker}, Heidi and {Scott}, Adam and {Seo}, Hee-Jong and {Shan}, Huanyuan and {Silva}, David R. and {Slepian}, Zachary and {Soto}, Christian and {Sprayberry}, David and {Staten}, Ryan and {Stillman}, Coley M. and {Stupak}, Robert J. and {Summers}, David L. and {Sien Tie}, Suk and {Tirado}, H. and {Vargas-Maga{\~n}a}, Mariana and {Vivas}, A. Katherina and {Wechsler}, Risa H. and {Williams}, Doug and {Yang}, Jinyi and {Yang}, Qian and {Yapici}, Tolga and {Zaritsky}, Dennis and {Zenteno}, A. and {Zhang}, Kai and {Zhang}, Tianmeng and {Zhou}, Rongpu and {Zhou}, Zhimin},
        title = "{Overview of the DESI Legacy Imaging Surveys}",
      journal = {\aj},
         year = 2019,
        month = may,
       volume = {157},
       number = {5},
          eid = {168},
        pages = {168},
          doi = {10.3847/1538-3881/ab089d},
archivePrefix = {arXiv},
       eprint = {1804.08657},
 primaryClass = {astro-ph.IM},
       adsurl = {https://ui.adsabs.harvard.edu/abs/2019AJ....157..168D}
}

@ARTICLE{Magnier2020,
       author = {{Magnier}, Eugene. A. and {Schlafly}, Edward. F. and {Finkbeiner}, Douglas P. and {Tonry}, J.~L. and {Goldman}, B. and {R{\"o}ser}, S. and {Schilbach}, E. and {Casertano}, S. and {Chambers}, K.~C. and {Flewelling}, H.~A. and {Huber}, M.~E. and {Price}, P.~A. and {Sweeney}, W.~E. and {Waters}, C.~Z. and {Denneau}, L. and {Draper}, P.~W. and {Hodapp}, K.~W. and {Jedicke}, R. and {Kaiser}, N. and {Kudritzki}, R. -P. and {Metcalfe}, N. and {Stubbs}, C.~W. and {Wainscoat}, R.~J.},
        title = "{Pan-STARRS Photometric and Astrometric Calibration}",
      journal = {\apjs},
         year = 2020,
        month = nov,
       volume = {251},
       number = {1},
          eid = {6},
        pages = {6},
          doi = {10.3847/1538-4365/abb82a},
archivePrefix = {arXiv},
       eprint = {1612.05242},
 primaryClass = {astro-ph.IM},
       adsurl = {https://ui.adsabs.harvard.edu/abs/2020ApJS..251....6M}
}

@INPROCEEDINGS{Basa2022,
       author = {{Basa}, St{\'e}phane and {Lee}, William H. and {Dolon}, Fran{\c{c}}ois and {Watson}, Alan M. and {Floriot}, Johan and {Atteia}, Jean-Luc and {Butler}, Nathaniel R. and {Dornic}, Damien and {Lombardo}, Simona and {Ronayette}, Samuel and {Ageron}, Michel and {Agneray}, Fran{\c{c}}ois and {{\'A}ngeles}, Fernando and {Bautista}, Ludovik and {Benamar-Aissa}, Hafid and {Blanpain}, Cyril and {Boulade}, Olivier and {Boy}, J{\'e}r{\'e}mie and {Buat}, Veronique and {Cadena}, Edgar and {Cuevas}, Salvador and {Farah}, Alejandro and {Figueroa}, Liliana and {Fuentes}, Jorge and {Ga{\"\i}ti}, Carole and {Gallais}, Pascal and {Kajfasz}, Eric and {Langarica}, Rosal{\'\i}a. and {Langlois}, Arthur and {Larrieu}, Marie and {Le Van Suu}, Auguste and {Lecubin}, Julien and {L{\'o}pez {\'A}ngeles}, Eduardo and {Lugo}, Erica and {Malgoyre}, Adrien and {Mathon}, Romain and {Moreau}, Chrystel and {Nouvel-De-La-Fl{\`e}che}, Alix and {Ochoa}, Jos{\'e} Luis and {Pedrayes-L{\'o}pez}, Maria and {Ramon}, Pascale and {Ru{\'\i}z-D{\'\i}az-Soto}, Jaime and {Tinoco}, Silvio and {Valentin}, Herv{\'e}},
        title = "{COLIBRI, a wide-field 1.3 m robotic telescope dedicated to the transient sky}",
    booktitle = {Ground-based and Airborne Telescopes IX},
         year = 2022,
       editor = {{Marshall}, Heather K. and {Spyromilio}, Jason and {Usuda}, Tomonori},
       series = {Society of Photo-Optical Instrumentation Engineers (SPIE) Conference Series},
       volume = {12182},
        month = aug,
          eid = {121821S},
        pages = {121821S},
          doi = {10.1117/12.2627139},
       adsurl = {https://ui.adsabs.harvard.edu/abs/2022SPIE12182E..1SB}
}

@ARTICLE{Yang2024,
       author = {{Yang}, Yu-Han and {Troja}, Eleonora and {O'Connor}, Brendan and {Fryer}, Chris L. and {Im}, Myungshin and {Durbak}, Joe and {Paek}, Gregory S.~H. and {Ricci}, Roberto and {Bom}, Cl{\'e}cio R. and {Gillanders}, James H. and {Castro-Tirado}, Alberto J. and {Peng}, Zong-Kai and {Dichiara}, Simone and {Ryan}, Geoffrey and {van Eerten}, Hendrik and {Dai}, Zi-Gao and {Chang}, Seo-Won and {Choi}, Hyeonho and {De}, Kishalay and {Hu}, Youdong and {Kilpatrick}, Charles D. and {Kutyrev}, Alexander and {Jeong}, Mankeun and {Lee}, Chung-Uk and {Makler}, Martin and {Navarete}, Felipe and {P{\'e}rez-Garc{\'\i}a}, Ignacio},
        title = "{A lanthanide-rich kilonova in the aftermath of a long gamma-ray burst}",
      journal = {\nat},
         year = 2024,
        month = feb,
       volume = {626},
       number = {8000},
        pages = {742-745},
          doi = {10.1038/s41586-023-06979-5},
archivePrefix = {arXiv},
       eprint = {2308.00638},
 primaryClass = {astro-ph.HE},
       adsurl = {https://ui.adsabs.harvard.edu/abs/2024Natur.626..742Y}
}

@ARTICLE{Levan2024,
       author = {{Levan}, Andrew J. and {Gompertz}, Benjamin P. and {Salafia}, Om Sharan and {Bulla}, Mattia and {Burns}, Eric and {Hotokezaka}, Kenta and {Izzo}, Luca and {Lamb}, Gavin P. and {Malesani}, Daniele B. and {Oates}, Samantha R. and {Ravasio}, Maria Edvige and {Rouco Escorial}, Alicia and {Schneider}, Benjamin and {Sarin}, Nikhil and {Schulze}, Steve and {Tanvir}, Nial R. and {Ackley}, Kendall and {Anderson}, Gemma and {Brammer}, Gabriel B. and {Christensen}, Lise and {Dhillon}, Vikram S. and {Evans}, Phil A. and {Fausnaugh}, Michael and {Fong}, Wen-fai and {Fruchter}, Andrew S. and {Fryer}, Chris and {Fynbo}, Johan P.~U. and {Gaspari}, Nicola and {Heintz}, Kasper E. and {Hjorth}, Jens and {Kennea}, Jamie A. and {Kennedy}, Mark R. and {Laskar}, Tanmoy and {Leloudas}, Giorgos and {Mandel}, Ilya and {Martin-Carrillo}, Antonio and {Metzger}, Brian D. and {Nicholl}, Matt and {Nugent}, Anya and {Palmerio}, Jesse T. and {Pugliese}, Giovanna and {Rastinejad}, Jillian and {Rhodes}, Lauren and {Rossi}, Andrea and {Saccardi}, Andrea and {Smartt}, Stephen J. and {Stevance}, Heloise F. and {Tohuvavohu}, Aaron and {van der Horst}, Alexander and {Vergani}, Susanna D. and {Watson}, Darach and {Barclay}, Thomas and {Bhirombhakdi}, Kornpob and {Breedt}, Elm{\'e} and {Breeveld}, Alice A. and {Brown}, Alexander J. and {Campana}, Sergio and {Chrimes}, Ashley A. and {D'Avanzo}, Paolo and {D'Elia}, Valerio and {De Pasquale}, Massimiliano and {Dyer}, Martin J. and {Galloway}, Duncan K. and {Garbutt}, James A. and {Green}, Matthew J. and {Hartmann}, Dieter H. and {Jakobsson}, P{\'a}ll and {Kerry}, Paul and {Kouveliotou}, Chryssa and {Langeroodi}, Danial and {Le Floc'h}, Emeric and {Leung}, James K. and {Littlefair}, Stuart P. and {Munday}, James and {O'Brien}, Paul and {Parsons}, Steven G. and {Pelisoli}, Ingrid and {Sahman}, David I. and {Salvaterra}, Ruben and {Sbarufatti}, Boris and {Steeghs}, Danny and {Tagliaferri}, Gianpiero and {Th{\"o}ne}, Christina C. and {de Ugarte Postigo}, Antonio and {Kann}, David Alexander},
        title = "{Heavy-element production in a compact object merger observed by JWST}",
      journal = {\nat},
         year = 2024,
        month = feb,
       volume = {626},
       number = {8000},
        pages = {737-741},
          doi = {10.1038/s41586-023-06759-1},
archivePrefix = {arXiv},
       eprint = {2307.02098},
 primaryClass = {astro-ph.HE},
       adsurl = {https://ui.adsabs.harvard.edu/abs/2024Natur.626..737L}
}

@ARTICLE{Troja2022,
       author = {{Troja}, E. and {Fryer}, C.~L. and {O'Connor}, B. and {Ryan}, G. and {Dichiara}, S. and {Kumar}, A. and {Ito}, N. and {Gupta}, R. and {Wollaeger}, R.~T. and {Norris}, J.~P. and {Kawai}, N. and {Butler}, N.~R. and {Aryan}, A. and {Misra}, K. and {Hosokawa}, R. and {Murata}, K.~L. and {Niwano}, M. and {Pandey}, S.~B. and {Kutyrev}, A. and {van Eerten}, H.~J. and {Chase}, E.~A. and {Hu}, Y. -D. and {Caballero-Garcia}, M.~D. and {Castro-Tirado}, A.~J.},
        title = "{A nearby long gamma-ray burst from a merger of compact objects}",
      journal = {\nat},
         year = 2022,
        month = dec,
       volume = {612},
       number = {7939},
        pages = {228-231},
          doi = {10.1038/s41586-022-05327-3},
archivePrefix = {arXiv},
       eprint = {2209.03363},
 primaryClass = {astro-ph.HE},
       adsurl = {https://ui.adsabs.harvard.edu/abs/2022Natur.612..228T}
}

@ARTICLE{Rastinejad2022,
       author = {{Rastinejad}, Jillian C. and {Gompertz}, Benjamin P. and {Levan}, Andrew J. and {Fong}, Wen-fai and {Nicholl}, Matt and {Lamb}, Gavin P. and {Malesani}, Daniele B. and {Nugent}, Anya E. and {Oates}, Samantha R. and {Tanvir}, Nial R. and {de Ugarte Postigo}, Antonio and {Kilpatrick}, Charles D. and {Moore}, Christopher J. and {Metzger}, Brian D. and {Ravasio}, Maria Edvige and {Rossi}, Andrea and {Schroeder}, Genevieve and {Jencson}, Jacob and {Sand}, David J. and {Smith}, Nathan and {Ag{\"u}{\'\i} Fern{\'a}ndez}, Jos{\'e} Feliciano and {Berger}, Edo and {Blanchard}, Peter K. and {Chornock}, Ryan and {Cobb}, Bethany E. and {De Pasquale}, Massimiliano and {Fynbo}, Johan P.~U. and {Izzo}, Luca and {Kann}, D. Alexander and {Laskar}, Tanmoy and {Marini}, Ester and {Paterson}, Kerry and {Escorial}, Alicia Rouco and {Sears}, Huei M. and {Th{\"o}ne}, Christina C.},
        title = "{A kilonova following a long-duration gamma-ray burst at 350 Mpc}",
      journal = {\nat},
         year = 2022,
        month = dec,
       volume = {612},
       number = {7939},
        pages = {223-227},
          doi = {10.1038/s41586-022-05390-w},
archivePrefix = {arXiv},
       eprint = {2204.10864},
 primaryClass = {astro-ph.HE},
       adsurl = {https://ui.adsabs.harvard.edu/abs/2022Natur.612..223R}
}

@ARTICLE{Ahumada2021,
       author = {{Ahumada}, Tom{\'a}s and {Singer}, Leo P. and {Anand}, Shreya and {Coughlin}, Michael W. and {Kasliwal}, Mansi M. and {Ryan}, Geoffrey and {Andreoni}, Igor and {Cenko}, S. Bradley and {Fremling}, Christoffer and {Kumar}, Harsh and {Pang}, Peter T.~H. and {Burns}, Eric and {Cunningham}, Virginia and {Dichiara}, Simone and {Dietrich}, Tim and {Svinkin}, Dmitry S. and {Almualla}, Mouza and {Castro-Tirado}, Alberto J. and {De}, Kishalay and {Dunwoody}, Rachel and {Gatkine}, Pradip and {Hammerstein}, Erica and {Iyyani}, Shabnam and {Mangan}, Joseph and {Perley}, Dan and {Purkayastha}, Sonalika and {Bellm}, Eric and {Bhalerao}, Varun and {Bolin}, Bryce and {Bulla}, Mattia and {Cannella}, Christopher and {Chandra}, Poonam and {Duev}, Dmitry A. and {Frederiks}, Dmitry and {Gal-Yam}, Avishay and {Graham}, Matthew and {Ho}, Anna Y.~Q. and {Hurley}, Kevin and {Karambelkar}, Viraj and {Kool}, Erik C. and {Kulkarni}, S.~R. and {Mahabal}, Ashish and {Masci}, Frank and {McBreen}, Sheila and {Pandey}, Shashi B. and {Reusch}, Simeon and {Ridnaia}, Anna and {Rosnet}, Philippe and {Rusholme}, Benjamin and {Carracedo}, Ana Sagu{\'e}s and {Smith}, Roger and {Soumagnac}, Maayane and {Stein}, Robert and {Troja}, Eleonora and {Tsvetkova}, Anastasia and {Walters}, Richard and {Valeev}, Azamat F.},
        title = "{Discovery and confirmation of the shortest gamma-ray burst from a collapsar}",
      journal = {Nature Astronomy},
         year = 2021,
        month = jul,
       volume = {5},
        pages = {917-927},
          doi = {10.1038/s41550-021-01428-7},
archivePrefix = {arXiv},
       eprint = {2105.05067},
 primaryClass = {astro-ph.HE},
       adsurl = {https://ui.adsabs.harvard.edu/abs/2021NatAs...5..917A}
}

@ARTICLE{Metzger2011,
       author = {{Metzger}, B.~D. and {Giannios}, D. and {Thompson}, T.~A. and {Bucciantini}, N. and {Quataert}, E.},
        title = "{The protomagnetar model for gamma-ray bursts}",
      journal = {\mnras},
         year = 2011,
        month = may,
       volume = {413},
       number = {3},
        pages = {2031-2056},
          doi = {10.1111/j.1365-2966.2011.18280.x},
archivePrefix = {arXiv},
       eprint = {1012.0001},
 primaryClass = {astro-ph.HE},
       adsurl = {https://ui.adsabs.harvard.edu/abs/2011MNRAS.413.2031M}
}

@ARTICLE{Meszaros1997,
       author = {{M{\'e}sz{\'a}ros}, P. and {Rees}, M.~J.},
        title = "{Optical and Long-Wavelength Afterglow from Gamma-Ray Bursts}",
      journal = {\apj},
         year = 1997,
        month = feb,
       volume = {476},
       number = {1},
        pages = {232-237},
          doi = {10.1086/303625},
archivePrefix = {arXiv},
       eprint = {astro-ph/9606043},
 primaryClass = {astro-ph},
       adsurl = {https://ui.adsabs.harvard.edu/abs/1997ApJ...476..232M}
}

@ARTICLE{Hjorth2003,
       author = {{Hjorth}, Jens and {Sollerman}, Jesper and {M{\o}ller}, Palle and {Fynbo}, Johan P.~U. and {Woosley}, Stan E. and {Kouveliotou}, Chryssa and {Tanvir}, Nial R. and {Greiner}, Jochen and {Andersen}, Michael I. and {Castro-Tirado}, Alberto J. and {Castro Cer{\'o}n}, Jos{\'e} Mar{\'\i}a and {Fruchter}, Andrew S. and {Gorosabel}, Javier and {Jakobsson}, P{\'a}ll and {Kaper}, Lex and {Klose}, Sylvio and {Masetti}, Nicola and {Pedersen}, Holger and {Pedersen}, Kristian and {Pian}, Elena and {Palazzi}, Eliana and {Rhoads}, James E. and {Rol}, Evert and {van den Heuvel}, Edward P.~J. and {Vreeswijk}, Paul M. and {Watson}, Darach and {Wijers}, Ralph A.~M.~J.},
        title = "{A very energetic supernova associated with the {\ensuremath{\gamma}}-ray burst of 29 March 2003}",
      journal = {\nat},
         year = 2003,
        month = jun,
       volume = {423},
       number = {6942},
        pages = {847-850},
          doi = {10.1038/nature01750},
archivePrefix = {arXiv},
       eprint = {astro-ph/0306347},
 primaryClass = {astro-ph},
       adsurl = {https://ui.adsabs.harvard.edu/abs/2003Natur.423..847H}
}

@ARTICLE{Woosley1993,
       author = {{Woosley}, S.~E.},
        title = "{Gamma-Ray Bursts from Stellar Mass Accretion Disks around Black Holes}",
      journal = {\apj},
         year = 1993,
        month = mar,
       volume = {405},
        pages = {273},
          doi = {10.1086/172359},
       adsurl = {https://ui.adsabs.harvard.edu/abs/1993ApJ...405..273W}
}

@ARTICLE{Li2012,
       author = {{Li}, Liang and {Liang}, En-Wei and {Tang}, Qing-Wen and {Chen}, Jie-Min and {Xi}, Shao-Qiang and {L{\"u}}, Hou-Jun and {Gao}, He and {Zhang}, Bing and {Zhang}, Jin and {Yi}, Shuang-Xi and {Lu}, Rui-Jing and {L{\"u}}, Lian-Zhong and {Wei}, Jian-Yan},
        title = "{A Comprehensive Study of Gamma-Ray Burst Optical Emission. I. Flares and Early Shallow-decay Component}",
      journal = {\apj},
         year = 2012,
        month = oct,
       volume = {758},
       number = {1},
          eid = {27},
        pages = {27},
          doi = {10.1088/0004-637X/758/1/27},
archivePrefix = {arXiv},
       eprint = {1203.2332},
 primaryClass = {astro-ph.HE},
       adsurl = {https://ui.adsabs.harvard.edu/abs/2012ApJ...758...27L}
}

@ARTICLE{Berger2014,
       author = {{Berger}, Edo},
        title = "{Short-Duration Gamma-Ray Bursts}",
      journal = {\araa},
         year = 2014,
        month = aug,
       volume = {52},
        pages = {43-105},
          doi = {10.1146/annurev-astro-081913-035926},
archivePrefix = {arXiv},
       eprint = {1311.2603},
 primaryClass = {astro-ph.HE},
       adsurl = {https://ui.adsabs.harvard.edu/abs/2014ARA&A..52...43B}
}

@ARTICLE{Becerra2019a,
       author = {{Becerra}, R.~L. and {Watson}, A.~M. and {Fraija}, N. and {Butler}, N.~R. and {Lee}, W.~H. and {Troja}, E. and {Rom{\'a}n-Z{\'u}{\~n}iga}, C.~G. and {Kutyrev}, A.~S. and {{\'A}lvarez Nu{\~n}ez}, L.~C. and {{\'A}ngeles}, F. and {Chapa}, O. and {Cuevas}, S. and {Farah}, A.~S. and {Fuentes-Fern{\'a}ndez}, J. and {Figueroa}, L. and {Langarica}, R. and {Quir{\'o}s}, F. and {Ru{\'\i}z-D{\'\i}az-Soto}, J. and {Tejada}, C.~G. and {Tinoco}, S.~J.},
        title = "{Late Central-engine Activity in GRB 180205A}",
      journal = {\apj},
         year = 2019,
        month = feb,
       volume = {872},
       number = {2},
          eid = {118},
        pages = {118},
          doi = {10.3847/1538-4357/ab0026},
archivePrefix = {arXiv},
       eprint = {1901.06051},
 primaryClass = {astro-ph.HE},
       adsurl = {https://ui.adsabs.harvard.edu/abs/2019ApJ...872..118B}
}

@ARTICLE{Granot2002,
       author = {{Granot}, Jonathan and {Sari}, Re'em},
        title = "{The Shape of Spectral Breaks in Gamma-Ray Burst Afterglows}",
      journal = {\apj},
         year = 2002,
        month = apr,
       volume = {568},
       number = {2},
        pages = {820-829},
          doi = {10.1086/338966},
archivePrefix = {arXiv},
       eprint = {astro-ph/0108027},
 primaryClass = {astro-ph},
       adsurl = {https://ui.adsabs.harvard.edu/abs/2002ApJ...568..820G}
}

@ARTICLE{Sari1998,
       author = {{Sari}, Re'em and {Piran}, Tsvi and {Narayan}, Ramesh},
        title = "{Spectra and Light Curves of Gamma-Ray Burst Afterglows}",
      journal = {\apjl},
         year = 1998,
        month = apr,
       volume = {497},
       number = {1},
        pages = {L17-L20},
          doi = {10.1086/311269},
archivePrefix = {arXiv},
       eprint = {astro-ph/9712005},
 primaryClass = {astro-ph},
       adsurl = {https://ui.adsabs.harvard.edu/abs/1998ApJ...497L..17S}
}

@ARTICLE{Evans2009,
       author = {{Evans}, P.~A. and {Beardmore}, A.~P. and {Page}, K.~L. and {Osborne}, J.~P. and {O'Brien}, P.~T. and {Willingale}, R. and {Starling}, R.~L.~C. and {Burrows}, D.~N. and {Godet}, O. and {Vetere}, L. and {Racusin}, J. and {Goad}, M.~R. and {Wiersema}, K. and {Angelini}, L. and {Capalbi}, M. and {Chincarini}, G. and {Gehrels}, N. and {Kennea}, J.~A. and {Margutti}, R. and {Morris}, D.~C. and {Mountford}, C.~J. and {Pagani}, C. and {Perri}, M. and {Romano}, P. and {Tanvir}, N.},
        title = "{Methods and results of an automatic analysis of a complete sample of Swift-XRT observations of GRBs}",
      journal = {\mnras},
         year = 2009,
        month = aug,
       volume = {397},
       number = {3},
        pages = {1177-1201},
          doi = {10.1111/j.1365-2966.2009.14913.x},
archivePrefix = {arXiv},
       eprint = {0812.3662},
 primaryClass = {astro-ph},
       adsurl = {https://ui.adsabs.harvard.edu/abs/2009MNRAS.397.1177E}
}

@ARTICLE{Becerra2019b,
       author = {{Becerra}, R.~L. and {Dichiara}, S. and {Watson}, A.~M. and {Troja}, E. and {Fraija}, N. and {Klotz}, A. and {Butler}, N.~R. and {Lee}, W.~H. and {Veres}, P. and {Turpin}, D. and {Bloom}, J.~S. and {Boer}, M. and {Gonz{\'a}lez}, J.~J. and {Kutyrev}, A.~S. and {Prochaska}, J.~X. and {Ramirez-Ruiz}, E. and {Richer}, M.~G.},
        title = "{Reverse Shock Emission Revealed in Early Photometry in the Candidate Short GRB 180418A}",
      journal = {\apj},
         year = 2019,
        month = aug,
       volume = {881},
       number = {1},
          eid = {12},
        pages = {12},
          doi = {10.3847/1538-4357/ab275b},
archivePrefix = {arXiv},
       eprint = {1904.05987},
 primaryClass = {astro-ph.HE},
       adsurl = {https://ui.adsabs.harvard.edu/abs/2019ApJ...881...12B}
}

@ARTICLE{Becerra2023,
       author = {{Becerra}, R.~L. and {Troja}, E. and {Watson}, A.~M. and {O'Connor}, B. and {Veres}, P. and {Dichiara}, S. and {Butler}, N.~R. and {De Colle}, F. and {Sakamoto}, T. and {L{\'o}pez}, K.~O.~C. and {Aoki}, K. and {Fraija}, N. and {Im}, M. and {Kutyrev}, A.~S. and {Lee}, W.~H. and {Paek}, G.~S.~H. and {Pereyra}, M. and {Ravi}, S. and {Urata}, Y.},
        title = "{Deciphering the unusual stellar progenitor of GRB 210704A}",
      journal = {\mnras},
         year = 2023,
        month = jul,
       volume = {522},
       number = {4},
        pages = {5204-5216},
          doi = {10.1093/mnras/stad1372},
archivePrefix = {arXiv},
       eprint = {2303.06909},
 primaryClass = {astro-ph.HE},
       adsurl = {https://ui.adsabs.harvard.edu/abs/2023MNRAS.522.5204B}
}

@ARTICLE{Pereyra2022,
       author = {{Pereyra}, M. and {Fraija}, N. and {Watson}, A.~M. and {Becerra}, R.~L. and {Butler}, N.~R. and {De Colle}, F. and {Troja}, E. and {Dichiara}, S. and {Fraire-Bonilla}, E. and {Lee}, W.~H. and {Ramirez-Ruiz}, E. and {Bloom}, J.~S. and {Prochaska}, J.~X. and {Kutyrev}, A.~S. and {Gonz{\'a}lez}, J.~J. and {Richer}, M.~G.},
        title = "{GRB 191016A: The onset of the forward shock and evidence of late energy injection}",
      journal = {\mnras},
         year = 2022,
        month = apr,
       volume = {511},
       number = {4},
        pages = {6205-6217},
          doi = {10.1093/mnras/stac389},
archivePrefix = {arXiv},
       eprint = {2111.15168},
 primaryClass = {astro-ph.HE},
       adsurl = {https://ui.adsabs.harvard.edu/abs/2022MNRAS.511.6205P}
}

@ARTICLE{Becerra2017,
       author = {{Becerra}, R.~L. and {Watson}, A.~M. and {Lee}, W.~H. and {Fraija}, N. and {Butler}, N.~R. and {Bloom}, J.~S. and {Capone}, J.~I. and {Cucchiara}, A. and {de Diego}, J.~A. and {Fox}, O.~D. and {Gehrels}, N. and {Georgiev}, L.~N. and {Gonz{\'a}lez}, J.~J. and {Kutyrev}, A.~S. and {Littlejohns}, O.~M. and {Prochaska}, J.~X. and {Ramirez-Ruiz}, E. and {Richer}, M.~G. and {Rom{\'a}n-Z{\'u}{\~n}iga}, C.~G. and {Toy}, V.~L. and {Troja}, E.},
        title = "{Photometric Observations of Supernova 2013cq Associated with GRB 130427A}",
      journal = {\apj},
         year = 2017,
        month = mar,
       volume = {837},
       number = {2},
          eid = {116},
        pages = {116},
          doi = {10.3847/1538-4357/aa610f},
archivePrefix = {arXiv},
       eprint = {1702.04762},
 primaryClass = {astro-ph.HE},
       adsurl = {https://ui.adsabs.harvard.edu/abs/2017ApJ...837..116B}
}

@ARTICLE{Lee2007,
       author = {{Lee}, William H. and {Ramirez-Ruiz}, Enrico},
        title = "{The progenitors of short gamma-ray bursts}",
      journal = {New Journal of Physics},
         year = 2007,
        month = jan,
       volume = {9},
       number = {1},
        pages = {17},
          doi = {10.1088/1367-2630/9/1/017},
archivePrefix = {arXiv},
       eprint = {astro-ph/0701874},
 primaryClass = {astro-ph},
       adsurl = {https://ui.adsabs.harvard.edu/abs/2007NJPh....9...17L}
}

@ARTICLE{Kumar2015,
       author = {{Kumar}, Pawan and {Zhang}, Bing},
        title = "{The physics of gamma-ray bursts \& relativistic jets}",
      journal = {\physrep},
         year = 2015,
        month = feb,
       volume = {561},
        pages = {1-109},
          doi = {10.1016/j.physrep.2014.09.008},
archivePrefix = {arXiv},
       eprint = {1410.0679},
 primaryClass = {astro-ph.HE},
       adsurl = {https://ui.adsabs.harvard.edu/abs/2015PhR...561....1K}
}

@ARTICLE{Kouveliotou1993,
       author = {{Kouveliotou}, Chryssa and {Meegan}, Charles A. and {Fishman}, Gerald J. and {Bhat}, Narayana P. and {Briggs}, Michael S. and {Koshut}, Thomas M. and {Paciesas}, William S. and {Pendleton}, Geoffrey N.},
        title = "{Identification of Two Classes of Gamma-Ray Bursts}",
      journal = {\apjl},
         year = 1993,
        month = aug,
       volume = {413},
        pages = {L101},
          doi = {10.1086/186969},
       adsurl = {https://ui.adsabs.harvard.edu/abs/1993ApJ...413L.101K}
}

@ARTICLE{Bertin1996,
       author = {{Bertin}, E. and {Arnouts}, S.},
        title = "{SExtractor: Software for source extraction.}",
      journal = {\aaps},
         year = 1996,
        month = jun,
       volume = {117},
        pages = {393-404},
          doi = {10.1051/aas:1996164},
       adsurl = {https://ui.adsabs.harvard.edu/abs/1996A&AS..117..393B}
}

@ARTICLE{Atteia2017,
       author = {{Atteia}, J. -L. and {Heussaff}, V. and {Dezalay}, J. -P. and {Klotz}, A. and {Turpin}, D. and {Tsvetkova}, A.~E. and {Frederiks}, D.~D. and {Zolnierowski}, Y. and {Daigne}, F. and {Mochkovitch}, R.},
        title = "{The Maximum Isotropic Energy of Gamma-ray Bursts}",
      journal = {\apj},
         year = 2017,
        month = mar,
       volume = {837},
       number = {2},
          eid = {119},
        pages = {119},
          doi = {10.3847/1538-4357/aa5ffa},
archivePrefix = {arXiv},
       eprint = {1702.02961},
 primaryClass = {astro-ph.HE},
       adsurl = {https://ui.adsabs.harvard.edu/abs/2017ApJ...837..119A}
}

@ARTICLE{Paczynski1986,
       author = {{Paczynski}, B.},
        title = "{Gamma-ray bursters at cosmological distances}",
      journal = {\apjl},
         year = 1986,
        month = sep,
       volume = {308},
        pages = {L43-L46},
          doi = {10.1086/184740},
       adsurl = {https://ui.adsabs.harvard.edu/abs/1986ApJ...308L..43P}
}

@INPROCEEDINGS{Paczynski1991,
       author = {{Paczy{\'n}ski}, Bohdan},
        title = "{Extragalactic scenarios for gamma-ray bursts}",
    booktitle = {Gamma-ray Bursts},
         year = 1991,
       series = {American Institute of Physics Conference Series},
       volume = {265},
        month = sep,
    publisher = {AIP},
        pages = {144-148},
          doi = {10.1063/1.42815},
       adsurl = {https://ui.adsabs.harvard.edu/abs/1991AIPC..265..144P}
}

@ARTICLE{39471,
       author = {{Palmer}, D.~M. and {Barthelmy}, S.~D. and {Caputo}, R. and {Gupta}, R. and {Krimm}, H.~A. and {Laha}, S. and {Lien}, A.~Y. and {Markwardt}, C.~B. and {Moss}, M.~J. and {Parsotan}, T. and {Sadaula}, D. and {Sakamoto}, T.},
        title = "{GRB 250221A: Swift-BAT refined analysis}",
      journal = {GRB Coordinates Network},
         year = 2025,
        month = feb,
       volume = {39471},
        pages = {1},
       adsurl = {https://ui.adsabs.harvard.edu/abs/2025GCN.39471....1P}
}

@ARTICLE{Greiner2015,
       author = {{Greiner}, Jochen and {Mazzali}, Paolo A. and {Kann}, D. Alexander and {Kr{\"u}hler}, Thomas and {Pian}, Elena and {Prentice}, Simon and {Olivares E.}, Felipe and {Rossi}, Andrea and {Klose}, Sylvio and {Taubenberger}, Stefan and {Knust}, Fabian and {Afonso}, Paulo M.~J. and {Ashall}, Chris and {Bolmer}, Jan and {Delvaux}, Corentin and {Diehl}, Roland and {Elliott}, Jonathan and {Filgas}, Robert and {Fynbo}, Johan P.~U. and {Graham}, John F. and {Guelbenzu}, Ana Nicuesa and {Kobayashi}, Shiho and {Leloudas}, Giorgos and {Savaglio}, Sandra and {Schady}, Patricia and {Schmidl}, Sebastian and {Schweyer}, Tassilo and {Sudilovsky}, Vladimir and {Tanga}, Mohit and {Updike}, Adria C. and {van Eerten}, Hendrik and {Varela}, Karla},
        title = "{A very luminous magnetar-powered supernova associated with an ultra-long {\ensuremath{\gamma}}-ray burst}",
      journal = {\nat},
         year = 2015,
        month = jul,
       volume = {523},
       number = {7559},
        pages = {189-192},
          doi = {10.1038/nature14579},
archivePrefix = {arXiv},
       eprint = {1509.03279},
 primaryClass = {astro-ph.HE},
       adsurl = {https://ui.adsabs.harvard.edu/abs/2015Natur.523..189G}
}

@ARTICLE{39396,
       author = {{Caputo}, R. and {Gronwall}, C. and {Gupta}, R. and {Page}, K.~L. and {Palmer}, D.~M. and {Parsotan}, T.~M. and {Siegel}, M.~H. and {Neil Gehrels Swift Observatory Team}},
        title = "{GRB 250221A: Swift detection of a burst with an optical counterpart}",
      journal = {GRB Coordinates Network},
         year = 2025,
        month = feb,
       volume = {39396},
        pages = {1},
       adsurl = {https://ui.adsabs.harvard.edu/abs/2025GCN.39396....1C}
}

@INPROCEEDINGS{Langarica2024,
       author = {{Langarica}, Rosal{\'\i}a. and {Watson}, Alan M. and {{\'A}lvarez N{\'u}{\~n}ez}, Luis Carlos and {Angeles}, Fernando and {Atteia}, Jean-Luc and {Basa}, St{\'e}phane and {Cuevas}, Salvador and {Dolon}, Fran{\c{c}}ois and {Farah}, Alejandro and {Dornic}, Damien and {Floriot}, Johan and {Fuentes-Fern{\'a}ndez}, Jorge and {Langlois}, Arthur and {Lee}, William H. and {Lombardo}, Simona and {Pereyra}, Margarita and {Ru{\'\i}z D{\'\i}az-Soto}, Jaime and {Ronayette}, Samuel and {Tinoco}, Silvio and {Valent{\'\i}n}, Herv{\'e}},
        title = "{The DDRAGO wide-field imager for the COLIBR{\'I} telescope}",
    booktitle = {Ground-based and Airborne Instrumentation for Astronomy X},
         year = 2024,
       editor = {{Bryant}, Julia J. and {Motohara}, Kentaro and {Vernet}, Jo{\"e}l. R.~D.},
       series = {Society of Photo-Optical Instrumentation Engineers (SPIE) Conference Series},
       volume = {13096},
        month = jul,
          eid = {130963D},
        pages = {130963D},
          doi = {10.1117/12.3020545},
       adsurl = {https://ui.adsabs.harvard.edu/abs/2024SPIE13096E..3DL}
}

@ARTICLE{Galama1999,
       author = {{Galama}, T.~J. and {Vreeswijk}, P.~M. and {van Paradijs}, J. and {Kouveliotou}, C. and {Augusteijn}, T. and {Patat}, F. and {Heise}, J. and {in 't Zand}, J. and {Groot}, P.~J. and {Wijers}, R.~A.~M.~J. and {Pian}, E. and {Palazzi}, E. and {Frontera}, F. and {Masetti}, N.},
        title = "{On the possible association of SN 1998bw and GRB 980425}",
      journal = {\aaps},
         year = 1999,
        month = sep,
       volume = {138},
        pages = {465-466},
          doi = {10.1051/aas:1999311},
       adsurl = {https://ui.adsabs.harvard.edu/abs/1999A&AS..138..465G}
}

@ARTICLE{39397,
       author = {{Watson}, Alan M. and {Angulo}, Camila and {Lee}, William H. and {Moreno M{\'e}ndez}, Enrique and {Basa}, St{\'e}phane and {Akl}, Dalya and {Antier}, Sarah and {Atteia}, Jean-Luc and {Becerra}, Rosa L. and {Butler}, Nathaniel R. and {Dornic}, Damien and {Ducoin}, Jean-Gr{\'e}goire and {Fortin}, Francis and {Globus}, No{\'e}mie and {Ocelotl L{\'o}pez}, Kin and {L{\'o}pez-C{\'a}mara}, Diego and {Magnani}, Francesco and {Pereyra}, Margarita and {Avo Rakotondrainibe}, Ny and {Schneider}, Benjamin and {de Ugarte Postigo}, Antonio},
        title = "{GRB 250221A: COLIBR{\'I}/DDRAGO Confirmation of a Bright Optical Counterpart}",
      journal = {GRB Coordinates Network},
         year = 2025,
        month = feb,
       volume = {39397},
        pages = {1},
       adsurl = {https://ui.adsabs.harvard.edu/abs/2025GCN.39397....1W}
}

@ARTICLE{Anderson+25,
       author = {{Anderson}, G.~E. and {Lamb}, G.~P. and {Gompertz}, B.~P. and {Rhodes}, L. and {Martin-Carrillo}, A. and {van der Horst}, A.~J. and {Rowlinson}, A. and {Bell}, M.~E. and {Chen}, T.-W. and {Fausey}, H.~M. and {Ferro}, M. and {Hancock}, P.~J. and {Oates}, S.~R. and {Schulze}, S. and {Starling}, R.~L.~C. and {Yang}, S. and {Ackley}, K. and {Anderson}, J.~P. and {Andersson}, A. and {Ag{\"u}{\'\i} Fern{\'a}ndez}, J.~F. and {Brivio}, R. and {Burns}, E. and {Chambers}, K.~C. and {de Boer}, T. and {D'Elia}, V. and {De Pasquale}, M. and {de Ugarte Postigo}, A. and {Dimple} and {Fender}, R. and {Fulton}, M.~D. and {Gao}, H. and {Gillanders}, J.~H. and {Green}, D.~A. and {Gromadzki}, M. and {Gulati}, A. and {Hartmann}, D.~H. and {Huber}, M.~E. and {Klingler}, N.~J. and {Kuin}, N.~P.~M. and {Leung}, J.~K. and {Levan}, A.~J. and {Lin}, C.-C. and {Magnier}, E. and {Malesani}, D.~B. and {Minguez}, P. and {Mooley}, K.~P. and {Mukherjee}, T. and {Nicholl}, M. and {O'Brien}, P.~T. and {Pugliese}, G. and {Rossi}, A. and {Ryder}, S.~D. and {Sbarufatti}, B. and {Schneider}, B. and {Sch{\"u}ssler}, F. and {Smartt}, S.~J. and {Smith}, K.~W. and {Srivastav}, S. and {Steeghs}, D. and {Tanvir}, N.~R. and {Thoene}, C.~C. and {Vergani}, S.~D. and {Wainscoat}, R.~J. and {Wang}, Z.-N. and {Wijers}, R.~A.~M.~J. and {Williams-Baldwin}, D. and {Worssam}, I. and {Zafar}, T.},
        title = "{The Radio Flare and Multiwavelength Afterglow of the Short GRB 231117A: Energy Injection from a Violent Shell Collision}",
      journal = {\apj},
         year = 2025,
        month = nov,
       volume = {994},
       number = {1},
          eid = {5},
        pages = {5},
          doi = {10.3847/1538-4357/adfed7},
archivePrefix = {arXiv},
       eprint = {2508.14650},
 primaryClass = {astro-ph.HE},
       adsurl = {https://ui.adsabs.harvard.edu/abs/2025ApJ...994....5A}
}

@ARTICLE{Beniamini+16,
       author = {{Beniamini}, Paz and {Nava}, Lara and {Piran}, Tsvi},
        title = "{A revised analysis of gamma-ray bursts' prompt efficiencies}",
      journal = {\mnras},
         year = 2016,
        month = sep,
       volume = {461},
       number = {1},
        pages = {51-59},
          doi = {10.1093/mnras/stw1331},
archivePrefix = {arXiv},
       eprint = {1606.00311},
 primaryClass = {astro-ph.HE},
       adsurl = {https://ui.adsabs.harvard.edu/abs/2016MNRAS.461...51B}
}

@ARTICLE{Blandford-Mckee-76,
       author = {{Blandford}, R.~D. and {McKee}, C.~F.},
        title = "{Fluid dynamics of relativistic blast waves}",
      journal = {Physics of Fluids},
         year = 1976,
        month = aug,
       volume = {19},
        pages = {1130-1138},
          doi = {10.1063/1.861619},
       adsurl = {https://ui.adsabs.harvard.edu/abs/1976PhFl...19.1130B}
}

@ARTICLE{Bromberg-Tchekhovskoy-16,
       author = {{Bromberg}, Omer and {Tchekhovskoy}, Alexander},
        title = "{Relativistic MHD simulations of core-collapse GRB jets: 3D instabilities and magnetic dissipation}",
      journal = {\mnras},
         year = 2016,
        month = feb,
       volume = {456},
       number = {2},
        pages = {1739-1760},
          doi = {10.1093/mnras/stv2591},
archivePrefix = {arXiv},
       eprint = {1508.02721},
 primaryClass = {astro-ph.HE},
       adsurl = {https://ui.adsabs.harvard.edu/abs/2016MNRAS.456.1739B}
}

@ARTICLE{Chiang-Dermer-99,
       author = {{Chiang}, James and {Dermer}, Charles D.},
        title = "{Synchrotron and Synchrotron Self-Compton Emission and the Blast-Wave Model of Gamma-Ray Bursts}",
      journal = {\apj},
         year = 1999,
        month = feb,
       volume = {512},
       number = {2},
        pages = {699-710},
          doi = {10.1086/306789},
archivePrefix = {arXiv},
       eprint = {astro-ph/9803339},
 primaryClass = {astro-ph},
       adsurl = {https://ui.adsabs.harvard.edu/abs/1999ApJ...512..699C}
}

@ARTICLE{Dai-Lu-98a,
       author = {{Dai}, Z.~G. and {Lu}, T.},
        title = "{{\ensuremath{\gamma}}-Ray Bursts and Afterglows from Rotating Strange Stars and Neutron Stars}",
      journal = {\prl},
         year = 1998,
        month = nov,
       volume = {81},
       number = {20},
        pages = {4301-4304},
          doi = {10.1103/PhysRevLett.81.4301},
archivePrefix = {arXiv},
       eprint = {astro-ph/9810332},
 primaryClass = {astro-ph},
       adsurl = {https://ui.adsabs.harvard.edu/abs/1998PhRvL..81.4301D}
}

@ARTICLE{Dai-Lu-98b,
       author = {{Dai}, Z.~G. and {Lu}, T.},
        title = "{Gamma-ray burst afterglows and evolution of postburst fireballs with energy injection from strongly magnetic millisecond pulsars}",
      journal = {\aap},
         year = 1998,
        month = may,
       volume = {333},
        pages = {L87-L90},
          doi = {10.48550/arXiv.astro-ph/9810402},
archivePrefix = {arXiv},
       eprint = {astro-ph/9810402},
 primaryClass = {astro-ph},
       adsurl = {https://ui.adsabs.harvard.edu/abs/1998A&A...333L..87D}
}

@ARTICLE{de-Ugarte+07,
       author = {{de Ugarte Postigo}, A. and {Fatkhullin}, T.~A. and {J{\'o}hannesson}, G. and {Gorosabel}, J. and {Sokolov}, V.~V. and {Castro-Tirado}, A.~J. and {Balega}, Yu. Yu. and {Spiridonova}, O.~I. and {Jel{\'\i}nek}, M. and {Guziy}, S. and {P{\'e}rez-Ram{\'\i}rez}, D. and {Hjorth}, J. and {Laursen}, P. and {Bersier}, D. and {Pandey}, S.~B. and {Bremer}, M. and {Monfardini}, A. and {Huang}, K.~Y. and {Urata}, Y. and {Ip}, W.~H. and {Tamagawa}, T. and {Kinoshita}, D. and {Mizuno}, T. and {Arai}, Y. and {Yamagishi}, H. and {Soyano}, T. and {Usui}, F. and {Tashiro}, M. and {Abe}, K. and {Onda}, K. and {Aslan}, Z. and {Khamitov}, I. and {Ozisik}, T. and {Kiziloglu}, U. and {Bikmaev}, I. and {Sakhibullin}, N. and {Burenin}, R. and {Pavlinsky}, M. and {Sunyaev}, R. and {Bhattacharya}, D. and {Kamble}, A.~P. and {Ishwara Chandra}, C.~H. and {Trushkin}, S.~A.},
        title = "{Extensive multiband study of the X-ray rich GRB 050408. A likely off-axis event with an intense energy injection}",
      journal = {\aap},
         year = 2007,
        month = feb,
       volume = {462},
       number = {3},
        pages = {L57-L60},
          doi = {10.1051/0004-6361:20066660},
archivePrefix = {arXiv},
       eprint = {astro-ph/0612545},
 primaryClass = {astro-ph},
       adsurl = {https://ui.adsabs.harvard.edu/abs/2007A&A...462L..57D}
}

@ARTICLE{deWet+23,
       author = {{de Wet}, S. and {Laskar}, T. and {Groot}, P.~J. and {Cavallaro}, F. and {Nicuesa Guelbenzu}, A. and {Chastain}, S. and {Izzo}, L. and {Levan}, A. and {Malesani}, D.~B. and {Monageng}, I.~M. and {van der Horst}, A.~J. and {Zheng}, W. and {Bloemen}, S. and {Filippenko}, A.~V. and {Kann}, D.~A. and {Klose}, S. and {Pieterse}, D.~L.~A. and {Rau}, A. and {Vreeswijk}, P.~M. and {Woudt}, P. and {Zhu}, Z. -P.},
        title = "{The triple-peaked afterglow of GRB 210731A from X-ray to radio frequencies}",
      journal = {\aap},
         year = 2023,
        month = mar,
       volume = {671},
          eid = {A116},
        pages = {A116},
          doi = {10.1051/0004-6361/202244917},
archivePrefix = {arXiv},
       eprint = {2301.11985},
 primaryClass = {astro-ph.HE},
       adsurl = {https://ui.adsabs.harvard.edu/abs/2023A&A...671A.116D}
}

@ARTICLE{deWet+24,
       author = {{de Wet}, Simon and {Laskar}, Tanmoy and {Groot}, Paul J. and {Barniol Duran}, Rodolfo and {Berger}, Edo and {Bhandari}, Shivani and {Eftekhari}, Tarraneh and {Guidorzi}, Cristiano and {Kobayashi}, Shiho and {Perley}, Daniel A. and {Sari}, Re'em and {Schroeder}, Genevieve},
        title = "{A Millimeter Rebrightening in GRB 210702A}",
      journal = {\apj},
         year = 2024,
        month = oct,
       volume = {974},
       number = {2},
          eid = {279},
        pages = {279},
          doi = {10.3847/1538-4357/ad77bb},
archivePrefix = {arXiv},
       eprint = {2408.14641},
 primaryClass = {astro-ph.HE},
       adsurl = {https://ui.adsabs.harvard.edu/abs/2024ApJ...974..279D}
}

@ARTICLE{Eichler-Waxman-05,
       author = {{Eichler}, David and {Waxman}, Eli},
        title = "{The Efficiency of Electron Acceleration in Collisionless Shocks and Gamma-Ray Burst Energetics}",
      journal = {\apj},
         year = 2005,
        month = jul,
       volume = {627},
       number = {2},
        pages = {861-867},
          doi = {10.1086/430596},
archivePrefix = {arXiv},
       eprint = {astro-ph/0502070},
 primaryClass = {astro-ph},
       adsurl = {https://ui.adsabs.harvard.edu/abs/2005ApJ...627..861E}
}

@ARTICLE{Fan-Xu-06,
       author = {{Fan}, Yi-Zhong and {Xu}, Dong},
        title = "{The X-ray afterglow flat segment in short GRB 051221A: Energy injection from a millisecond magnetar?}",
      journal = {\mnras},
         year = 2006,
        month = oct,
       volume = {372},
       number = {1},
        pages = {L19-L22},
          doi = {10.1111/j.1745-3933.2006.00217.x},
archivePrefix = {arXiv},
       eprint = {astro-ph/0605445},
 primaryClass = {astro-ph},
       adsurl = {https://ui.adsabs.harvard.edu/abs/2006MNRAS.372L..19F}
}

@ARTICLE{Fong-Berger-13,
       author = {{Fong}, W. and {Berger}, E.},
        title = "{The Locations of Short Gamma-Ray Bursts as Evidence for Compact Object Binary Progenitors}",
      journal = {\apj},
         year = 2013,
        month = oct,
       volume = {776},
       number = {1},
          eid = {18},
        pages = {18},
          doi = {10.1088/0004-637X/776/1/18},
archivePrefix = {arXiv},
       eprint = {1307.0819},
 primaryClass = {astro-ph.HE},
       adsurl = {https://ui.adsabs.harvard.edu/abs/2013ApJ...776...18F}
}

@ARTICLE{Fong+15,
       author = {{Fong}, W. and {Berger}, E. and {Margutti}, R. and {Zauderer}, B.~A.},
        title = "{A Decade of Short-duration Gamma-Ray Burst Broadband Afterglows: Energetics, Circumburst Densities, and Jet Opening Angles}",
      journal = {\apj},
         year = 2015,
        month = dec,
       volume = {815},
       number = {2},
          eid = {102},
        pages = {102},
          doi = {10.1088/0004-637X/815/2/102},
archivePrefix = {arXiv},
       eprint = {1509.02922},
 primaryClass = {astro-ph.HE},
       adsurl = {https://ui.adsabs.harvard.edu/abs/2015ApJ...815..102F}
}

@ARTICLE{Gill-Granot-18,
       author = {{Gill}, Ramandeep and {Granot}, Jonathan},
        title = "{Afterglow imaging and polarization of misaligned structured GRB jets and cocoons: breaking the degeneracy in GRB 170817A}",
      journal = {\mnras},
         year = 2018,
        month = aug,
       volume = {478},
       number = {3},
        pages = {4128-4141},
          doi = {10.1093/mnras/sty1214},
archivePrefix = {arXiv},
       eprint = {1803.05892},
 primaryClass = {astro-ph.HE},
       adsurl = {https://ui.adsabs.harvard.edu/abs/2018MNRAS.478.4128G}
}

@ARTICLE{Gill+19,
       author = {{Gill}, Ramandeep and {Nathanail}, Antonios and {Rezzolla}, Luciano},
        title = "{When Did the Remnant of GW170817 Collapse to a Black Hole?}",
      journal = {\apj},
         year = 2019,
        month = may,
       volume = {876},
       number = {2},
          eid = {139},
        pages = {139},
          doi = {10.3847/1538-4357/ab16da},
archivePrefix = {arXiv},
       eprint = {1901.04138},
 primaryClass = {astro-ph.HE},
       adsurl = {https://ui.adsabs.harvard.edu/abs/2019ApJ...876..139G}
}

@ARTICLE{Gill-Granot-23,
       author = {{Gill}, Ramandeep and {Granot}, Jonathan},
        title = "{GRB 221009A afterglow from a shallow angular structured jet}",
      journal = {\mnras},
         year = 2023,
        month = sep,
       volume = {524},
       number = {1},
        pages = {L78-L83},
          doi = {10.1093/mnrasl/slad075},
archivePrefix = {arXiv},
       eprint = {2304.14331},
 primaryClass = {astro-ph.HE},
       adsurl = {https://ui.adsabs.harvard.edu/abs/2023MNRAS.524L..78G}
}

@ARTICLE{Granot+03,
       author = {{Granot}, Jonathan and {Nakar}, Ehud and {Piran}, Tsvi},
        title = "{Astrophysics: refreshed shocks from a {\ensuremath{\gamma}}-ray burst}",
      journal = {\nat},
         year = 2003,
        month = nov,
       volume = {426},
       number = {6963},
        pages = {138-139},
          doi = {10.1038/426138a},
archivePrefix = {arXiv},
       eprint = {astro-ph/0304563},
 primaryClass = {astro-ph},
       adsurl = {https://ui.adsabs.harvard.edu/abs/2003Natur.426..138G}
}

@ARTICLE{Hascoet+12,
       author = {{Hasco{\"e}t}, R. and {Daigne}, F. and {Mochkovitch}, R.},
        title = "{The origin of the late rebrightening in GRB 080503}",
      journal = {\aap},
         year = 2012,
        month = may,
       volume = {541},
          eid = {A88},
        pages = {A88},
          doi = {10.1051/0004-6361/201118722},
archivePrefix = {arXiv},
       eprint = {1204.0941},
 primaryClass = {astro-ph.HE},
       adsurl = {https://ui.adsabs.harvard.edu/abs/2012A&A...541A..88H}
}

@ARTICLE{Huang+99,
       author = {{Huang}, Y.~F. and {Dai}, Z.~G. and {Lu}, T.},
        title = "{A generic dynamical model of gamma-ray burst remnants}",
      journal = {\mnras},
         year = 1999,
        month = oct,
       volume = {309},
       number = {2},
        pages = {513-516},
          doi = {10.1046/j.1365-8711.1999.02887.x},
archivePrefix = {arXiv},
       eprint = {astro-ph/9906370},
 primaryClass = {astro-ph},
       adsurl = {https://ui.adsabs.harvard.edu/abs/1999MNRAS.309..513H}
}

@ARTICLE{Kobayashi-00,
       author = {{Kobayashi}, Shiho},
        title = "{Light Curves of Gamma-Ray Burst Optical Flashes}",
      journal = {\apj},
         year = 2000,
        month = dec,
       volume = {545},
       number = {2},
        pages = {807-812},
          doi = {10.1086/317869},
archivePrefix = {arXiv},
       eprint = {astro-ph/0009319},
 primaryClass = {astro-ph},
       adsurl = {https://ui.adsabs.harvard.edu/abs/2000ApJ...545..807K}
}

%%%%%%%%%%%%%%%%% APPENDICES %%%%%%%%%%%%%%%%%%%%%
\appendix
%%%%%%%%%%%%%%%%%%%%%%%%%%%%%%%%%%%%%%%%%%%%%%%%%%
\section{Refreshed Shock from Two Shell Collision}
\label{sec:ref-shock}
%%%%%%%%%%%%%%%%%%%%%%%%%%%%%%%%%%%%%%%%%%%%%%%%%%
We consider two spherical shells, where the first ejected shell has initial bulk-LF $\Gamma_{(3),0}$ and the second has $\Gamma_{(6),0}$; the shells here are labelled according to Figure\,\ref{fig:shells} where the parenthesized number refers to the region number. The two shells travel the same radial distance $R_{\rm coll}$ from the central engine, where the collision occurs at the lab-frame time $t_{\rm coll}$. The lab-frame time to reach a given radius $R$ is determined by the shell velocity, $\beta(R)=\sqrt{1-\Gamma^{-2}(R)}$. Here we approximate the radial evolution of the bulk-LF of the first shell as a broken power-law, with $\Gamma_{(3)} = \Gamma_{(3),0}$ for $R<R_{\rm dec}$, when the shell has not reached the deceleration radius, and $(\Gamma_{(3)}/\Gamma_{(3),0})^2 = \hat R^{-m}$ for $R>R_{\rm dec}$, where we define $\hat R=R/R_{\rm dec}$ and $m=3-k$. The lab-frame time taken by the first shell to arrive at a given radius $\hat R>1$ is $t = R_{\rm dec}/c\beta_{(3),0} + (R_{\rm dec}/c)\int_1^{\hat R}d\hat R/\beta(\hat R)$, where the first term is the time at which the shell decelerates. The arrival time of photons from different parts of the shell when the shell is at radius $R$ is determined by the EATS equation, such that $T_z\equiv T/(1+z) = t(R) - (R/c)\mu$, where $t(R)$ is the lab-frame time and $\mu = \cos\theta$ with $\theta$ being the angle measured from the line-of-sight (LOS) to the emitting material. The on-axis arrival time of photons, i.e. along the LOS, when $\Gamma^2\propto R^{-m}$, is given by $T_z = R/2(1+m)\Gamma^2c$ for $\Gamma\gg1$. Since the time of the re-brightening in the observer frame, $T_{\rm bright}\gg T_{\rm dec}$, it also means that the corresponding collision radius $\hat R_{\rm coll}\gg(1+m)^{1/(1+m)}=\sqrt{2}$ for $m=3-k$ using $k=0$. Therefore, the lab-frame time at which the collision occurs is also much greater than the time at which the shell decelerates, i.e $t_{\rm coll}\gg R_{\rm dec}/c\beta_{(3),0}$. In this case, the first shell arrives at $R_{\rm coll}$ at lab-frame time $t_{\rm coll} = [1 + \hat R_{\rm coll}^m/2(1+m)\Gamma_{(3),0}^2]R_{\rm coll}/c$. The second shell, which does not suffer any deceleration, reaches $R_{\rm coll}$ in a lab-frame time $t_{\rm coll} = t_{\rm em} + (1+1/2\Gamma_{(6),0}^2)R_{\rm coll}/c$, where $t_{\rm em}$ is the engine quiescent time after which it emits the second shell. By comparing the arrival times of the two shells at $R=R_{\rm coll}$, we get
\begin{equation}
    t_{\rm em} \approx \left[1 - \frac{(1+m)\Gamma_{(3)}^2}{\Gamma_{(6),0}^2}\right]T_{\rm bright}\,,
\end{equation}
which further yields
\begin{equation}\label{eq:Gamma_0-ratio}
    \left(\frac{\Gamma_{(3),0}}{\Gamma_{(6),0}}\right)^2 = \left[1-(1+z)\frac{t_{\rm em}}{T_{\rm bright}}\right]
    \frac{[(1+m)\hat T_{\rm bright}]^{m\over1+m}}{(1+m)}\,,
\end{equation}
where $\hat T_{\rm bright}=T_{\rm bright}/T_{\rm dec}$. In general, if $t_{\rm em}\ll T_{\rm bright}/(1+z)$ then it fixes the ratio of the bulk LFs of the two shells,
\begin{equation}
    \left(\frac{\Gamma_{(3),0}}{\Gamma_{(6),0}}\right)^2 = \frac{[(1+m)\hat T_{\rm bright}]^{m\over1+m}}{(1+m)}\,.
\end{equation}
Since $t_{\rm em}/T_{\rm bright}<T_{90}/T_{\rm bright}\ll1$ is guaranteed in a binary merger scenario, it forces $\Gamma_{(3)}/\Gamma_{(6),0}\approx(1+m)^{-1/2}=1/2$ for $m=3$ \citep{Kumar-Piran-00}. This also implies that the delay between the emission time of the shells can be approximated as negligible, with $t_{\rm em}\approx0$. The arrival time of radiation from the collision radius is given by $\hat T_{\rm bright} = \hat R_{\rm coll}^{1+m}/(1+m)$ when $\Gamma_{(3)}\gg1$, where
\begin{equation}
    \hat R_{\rm coll} \equiv \frac{R_{\rm coll}}{R_{\rm dec}} = \left[(1+m)\left(\frac{\Gamma_{(3),0}}{\Gamma_{(6),0}}\right)^2\right]^{1/m}\,.
\end{equation}
In this case it is required that $\Gamma_{(6),0}<\Gamma_{(3),0}$ for $\hat R_{\rm coll}>1$. 

The relative LF between the two shells at the time of collision is $\Gamma_{\rm rel}\simeq(\Gamma_{(3)}/\Gamma_{(6),0} + \Gamma_{(6),0}/\Gamma_{(3)})/2 = 5/4$ for $\Gamma_{(3)},\,\Gamma_{(6),0}\gg1$. In order to produce a strong shock, $\Gamma_{\rm rel}$ must significantly exceed the LF corresponding to the local sound speed ($c_s$) in the first shell. Since the collision occurs after the passage of the reverse shock in the first shell, due to its interaction with the external medium, it can be assumed to be relativistically hot. In that case, the sound speed is $\beta_s=c_s/c=1/\sqrt{3}$, with the corresponding LF $\Gamma_s=\sqrt{3/2}\simeq1.22\lesssim\Gamma_{\rm rel}$.

\begin{figure}
    \centering
    \includegraphics[width=0.48\textwidth]{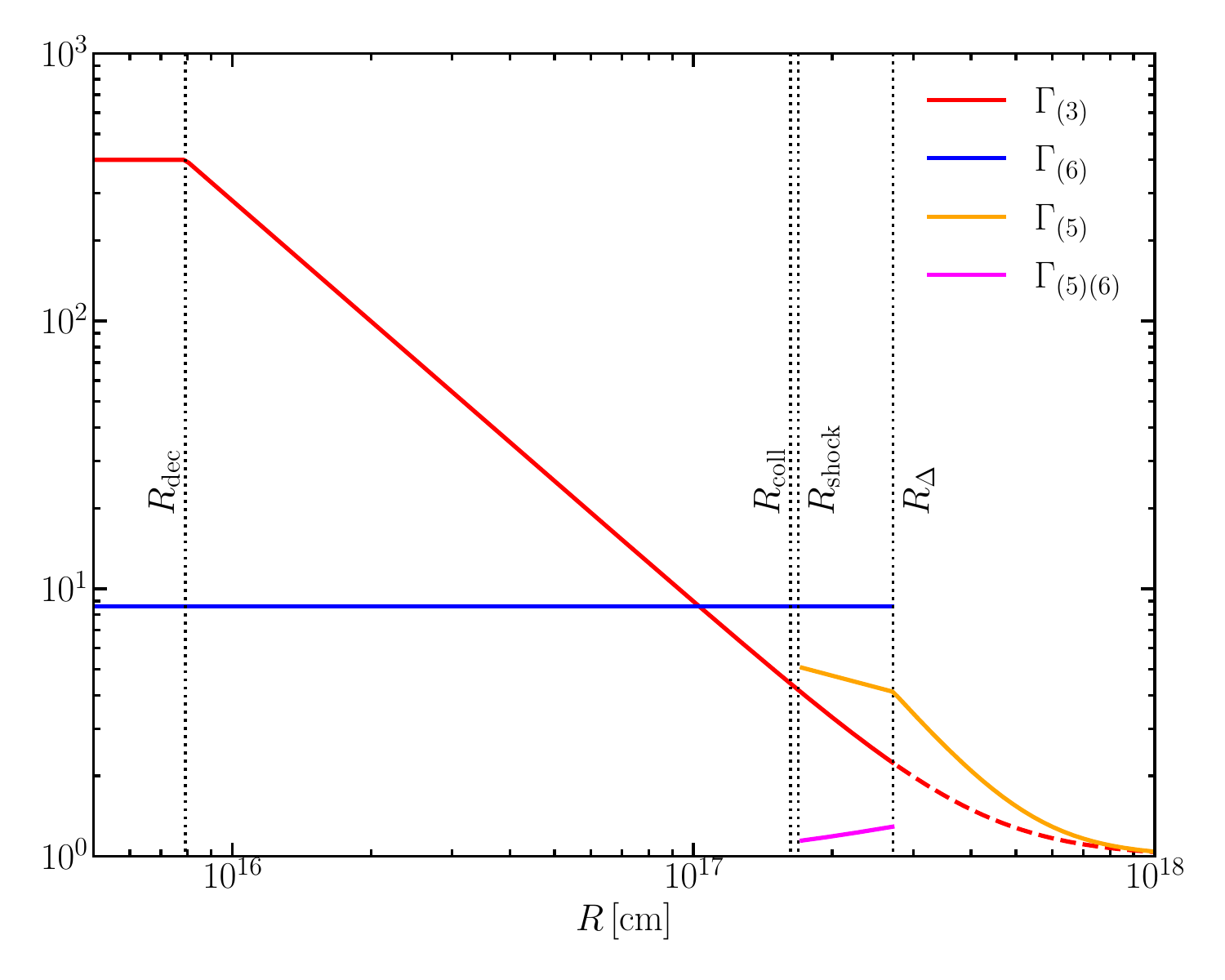}
    \caption{
    Shell dynamics with $E_{(6),\rm k,iso}/E_{(3),\rm k,iso} = 24$ and $\Gamma_{(3),0}/\Gamma_{(6),0}=46.5$. All other parameters are the same as the best-fit parameters shown in the legend of Figure\,\ref{fig:lc-fit}. Different important radii are also shown, where the first shell decelerates at $R_{\rm dec}$ and the second shell catches up from behind and merges with the first at $R_{\rm coll}$. Two shocks, in addition to the external forward shock by the first shell, form at $R_{\rm shock}$, from which a reverse shock crosses the second shell at $R_\Delta$. The dynamics of the first decelerating shell are altered (dashed line) at $R>R_\Delta$ due to injection of energy by the second shell. 
    }
    \label{fig:shell-dynamics}
\end{figure}

Next, we calculate the emission that arises from the two shell collision and which 
produces the re-brightening. As the two shells collide, a forward shock is launched into the first shell and a reverse shock into the second shell. In total, there are six different regions with different dynamical evolution as illustrated in Figure~\ref{fig:shells} \citep{Zhang-Meszaros-02}: (1) unshocked ISM, 
(2) ISM shocked by the forward shock from the first shell, (3) material from the first shell shocked by the early passage ($T_{\times 1}\lesssim T_{\rm dec}$) of the reverse shock, (4) relativistically hot material in the first shell now shocked by the forward shock from the second shell, (5) material of the second shell now shocked by the reverse shock from the collision, and (6) unshocked material of the second shell. All physical quantities in what follows refer to these regions with the corresponding number. To relate the properties of the upstream flow with that of the downstream across the three shocks, we use the shock jump conditions. For a strong shock going into a cold medium, i.e. $\rho_uc^2\gg e_u + P_u$, where $\rho_u$, $e_u$, and $P_u$ are the proper mass density, internal energy density, and pressure of the upstream medium (with subscript `u'), the shock-jump conditions yield \citep{Blandford-Mckee-76}
\begin{equation}\label{eq:shock-jump-BM76}
    \frac{e_d}{n_d} = (\Gamma_{ud}-1)m_pc^2\,,
    \quad
    \frac{n_d}{n_u} = \frac{\hat\gamma\Gamma_{ud}+1}{\hat\gamma-1}=4\Gamma_{ud}\,,
    \quad
    \hat\gamma = \frac{4\Gamma_{ud}+1}{3\Gamma_{ud}}
\end{equation}
where $\Gamma_{ud}$ is the relative LF between the upstream and downstream (with subscript `d') 
media and $\hat\gamma$ is the adiabatic index of the downstream medium with $\hat\gamma=4/3\,(5/3)$ 
for a relativistic (non-relativistic) fluid. The above shock-jump condition applies to the forward 
shock between regions (1) and (2) and the reverse shock between regions (5) and (6). The shock 
between regions (3) and (4) is propagating into a relativistically hot region, i.e. $e_u+P_u\gg\rho_uc^2$ 
and $P_u = (\hat\gamma-1)e_u=e_u/3$ for $\hat\gamma=4/3$. Therefore, it admits different shock-jump 
conditions \citep{Kumar-Piran-00,Zhang-Meszaros-02}
\begin{equation}\label{eq:shock-jump-hot}
    \Gamma_{ud}^2 = \frac{(3e_d+e_u)(e_d+3e_u)}{16e_de_u}\,,
    \quad\quad
    \left(\frac{n_d}{n_u}\right)^2 = \frac{(3e_d+e_u)}{e_d+3e_u}\frac{e_d}{e_u}\,.
\end{equation}

Since the emission from the reverse shock makes the dominant contribution to the total flux until the shock crosses the second shell, we first derive the flux scaling relations for that. When the collision occurs and the second shell decelerates, it is important to know if the reverse shock is 
propagating into a radially expanding shell (thin shell regime) or that with a fixed radial width (thick shell regime; \citealt{Sari-Piran-95}). If the initial radial width of the second shell is $\Delta_{(6),0} = c\delta T/(1+z)$, where $\delta T$ is the activity time of the central engine, the shell starts to spread at the radius $R_s = \Delta_{(6),0}'\Gamma_{(6),0}=\Delta_{(6),0}\Gamma_{(6),0}^2$. For the shell to be thick at the time of collision, i.e. $\hat R_s>\hat R_{\rm coll}$, its initial LF must exceed that of the first shell,

\begin{eqnarray}
    \frac{\Gamma_{(6),0}}{\Gamma_{(3),0}} &&> \sqrt{\frac{2T_{\rm dec}}{\delta T}}[(1+m)\hat T_{\rm bright}]^{1\over2(1+m)} \\
    &&\simeq 16.8\,T_{\rm bright,5}^{1/8}\,T_{\rm dec,1}^{3/8}\,\delta T^{-1/2}\nonumber\,,
\end{eqnarray}
which contradicts the earlier condition of $\Gamma_{(2),0}<\Gamma_{(1),0}$. Therefore, the second shell must start to spread before the collision, in which case the comoving number density of the shell is $n_{(6)}=M_{(6)}\Gamma_{(6),0}/4\pi R^3m_p\propto R^{-3}$, where $M_{(6)} = E_{(6),\rm k,iso}/\Gamma_{(6),0}c^2$ is the shell mass and $E_{(6),\rm k,iso}$ is its kinetic energy.

From the shock-jump condition in Equation\,(\ref{eq:shock-jump-BM76}), we find $n_{(5)}=4\Gamma_{(5)(6)}n_{(6)}$ and 
$e_{(5)}=4(\Gamma_{(5)(6)}-1)\Gamma_{(5)(6)}n_{(6)}m_pc^2$. Regions (4) and (5) are separated by a contact discontinuity, where the shocked fluids in the two regions are in pressure equilibrium, i.e. $P_{(5)}=P_{(4)}$, and also move with the same velocity, i.e. $\Gamma_{(5)}=\Gamma_{(4)}$, which implies that $\Gamma_{(5)(6)}=\Gamma_{(4)(6)}$. In order to calculate $\Gamma_{(5)(6)}$ we need to know $\Gamma_{(4)}$, where the latter is obtained from the shock-jump condition in Equation\,(\ref{eq:shock-jump-hot}) that yields, $\Gamma_{(4)(3)}^2=(3e_{(4)}/e_{(3)}+1)(e_{(4)}/e_{(3)}+3)/16(e_{(4)}/e_{(3)})$. Here, $e_{(4)} = e_{(5)}$ due to both being relativistically hot and in pressure equilibrium. We assume that region (3) is still relativistically hot after RS passage, so that $P_{(3)}\approx P_{(2)}$ and $e_{(3)}\approx e_{(2)}$, and also $\Gamma_{(3)}\approx\Gamma_{(2)}$. In that case, the jump condition in Equation\,(\ref{eq:shock-jump-BM76}) yields $e_{(4)}/e_{(3)}\approx e_{(5)}/e_{(2)}\approx(\Gamma_{(5)(6)}-1)\Gamma_{(5)(6)}n_{(6)}/\Gamma_{(2)}^2n_{(1)}$.  

We solve the coupled dynamics of the different regions and show the LFs of the relevant regions in the left column of Figure\,\ref{fig:shell-dynamics} for the two different solutions obtained from the light curve fits in Figure\,\ref{fig:lc-fit}. Even though the two shells make contact at the radius $R_{\rm coll}$, the shocks do not form until the shells reach $R_{\rm shock}\geq R_{\rm coll}$ when the condition that $e_{(4)}>e_{(3)}$ is met. At $R>R_{\rm shock}$, the RS starts to propagate into the inner shell and extracts its kinetic energy. The differential number of baryons entering the shock can be calculated as $dN_{(6)} = 4\pi R^2 n_{(6)}\Gamma_{(6)} d\Delta_{(5)}$, where $d\Delta_{(5)} = (\beta_{(6)}-\beta_{(5)})(1 - \Gamma_{(6)}n_{(6)}/\Gamma_{(5)}n_{(5)})^{-1}dR$ is the infinitesimal width of the shocked region, as obtained from baryon number conservation. The total mass shocked by the RS is then $M_{(5)}=\int_{R_{\rm shock}}^{R_\Delta} m_p dN_{(6)}$, where $R_\Delta$ is the shock crossing radius, and it is obtained when $M_{(5)}=M_{(6)}$. In the cases shown in Figure\,\ref{fig:shell-dynamics}, the RS moves through the inner shell at a non-relativistic speed, with $\Gamma_{\rm RS}\sim\Gamma_{(5)(6)}$, and crosses the shell over $\Delta R/R_{\rm cross}\lesssim1.5 - 2$. After RS passage, it is expected that the LF of the now entirely shocked inner shell follows $\Gamma_{(5)}\propto R^{-g}$, where $3/2 \leq g \leq 7/2$ \citep{Kobayashi-Sari-00} and we assume $g=2$. Due to energy injection by the inner shell, the dynamics of the merged two-shell system are altered, which requires more careful modelling and are not considered here.

\section{Central Engine}
\label{App:CentralEngine}

In order to produce a GRB we need a central engine, whether the event is coming from a compact-objects merger, i.e., a low mass BH-NS, or a NS-NS \citep{1974LattimerSchramm, Metzger2011}--\citet[likely from a compact-objects merger, i.e., a low mass BH-NS, or a NS-NS, see, e.g.,][]{}, or from a Collapsar as was explained in Section~\ref{sec:introduction} \citep{Woosley1993}.
So far, neither our restricted observations nor our analytical models can tell which one is responsible for GRB~250221A.

We know from our analysis (see Section~\ref{sec:refreshed-shock}) that the central engine must deliver not less than 
$E_k > 6.3\times10^{51} (\theta_j/0.25)^2$ erg for a GRB in $G2$; but this makes the dubious assumption that efficiencies within the process are close to one, casting aside the equipartition theorem. Hence, we look for more robust engines which can deliver several tens of times these energies.

The most accepted and well studied central engines (CE) can be summarized as follows:
\begin{enumerate}

    \item $\nu\bar{\nu}$ annihilation from accretion onto a NS or BH \citep[see, e.g.][]{1999MacFadyenWoosley,Zalamea2011, globus2014}:
    When the mass-accretion rate close to the compact object (CO: BH or NS) is larger than a few hundredths of solar masses per second and the spin (of the CO) is large; or accretion rates around a few tenths of a solar mass per second and the spin is low; conditions are optimal for nuclear burning in the inner disk and, thus, production of large amounts of $\nu\bar{\nu}$. \citet{Zalamea2011} have shown that powers  $\dot{E}_{\nu\bar{\nu}}\gtrsim10^{52}$~erg s$^{-1}$ can be reached for accretion rates $\gtrsim1$~M$_\odot$ s$^{-1}$, and cannot exceed $\sim5\,\times 10^{52}$~erg s$^{-1}$.
    
    \item Millisecond magnetar spinning down \citep{Metzger2011}:
    Utilizing a large dipolar magnetic field to extract the rotational kinetic energy stored in a ms magnetar.  If the field is of order $B\simeq10^{15}$ G, the spin downtime is of order of seconds, and the energy is around  
    $E_{rot} = \frac{I\Omega^2}{2} \lesssim 10^{53} {\rm erg}$ for a 3-$M_\odot$, ms-spin, NS (although unstable, and above the TOV limit, such a large mass may be expected to briefly survive after a NS-NS merger, or after accreting during a core collapse event, when centrifugal force helps to achieve hydrostatic equilibrium).
    
    \item A Kerr BH with Blandford-Znajek engine \citep{1977BZ,2000Brown}:
    A Penrose process \citep{1971Penrose} where a Kerr BH, with an attached magnetic field finds itself surrounded by a plasma-built accretion disk, works as an electric generator and the average of the Poynting vector points in the direction of the rotational axis, delivering thus, a fraction of the Kerr's BH rotational kinetic energy into jets. Typical Blandford-Znajek (BZ) luminosities ($L_{BZ}$) are of order $L_{BZ} \simeq 10^{52} a_\star^2 \frac{\Omega(\omega_{BH}-\Omega)}{\omega_{BH}^2} \left(\frac{M_{BH}}{5 M_\odot}\right)^2 \left<\frac{B}{10^{15} {\rm G}}\right>^2$ erg s$^{-1}$, where $a_\star$ is the dimensionless Kerr (or spin) parameter of the BH, $\omega_{BH}$ is the angular velocity of the BH, and $\Omega$ is the angular velocity of the accretion disk.
    
    \item Schwarschild BH or Kerr BH with BP engine \citep{1982BlandfordPayne}:
    Again, an electric-generator like process, but not a Penrose process.
    This is magneto-rotational process that extracts rotational energy from the accretion disk. 
\end{enumerate}

A two-peaked afterglow may be explained by the collision of two shells at a radius of $R_\mathrm{coll} \simeq 10^{17}$~cm.
The simplest model assumes they both move with similar initial Lorentz factors, but are launched about two seconds apart from each other. The first shell sweeps the ISM away, gradually slowing down as it accumulates material on its head.  The second shell, having a clear path, maintains a constant speed until it finally catches up with the first one, producing the rebrightening we observe.

Producing two peaks, instead of one, or three, allows for placing constraints on the nature of the central engine.
A single-episode engine can be produced by both (CO merger or a Collapsar) and they may use any one of the four central engines (or a combination of them) described above.
Instead, a three peaked process may involve, as pointed out in \citet{2015MM} for GRB 110709B, an initial CE powered by a ms magnetar that powers the first peak; the magnetar collapses onto a BH as it loses its centrifugal support due to the powering of the jet. The collapse releases the binding energy of the BH, sending the blast wave down the jet-produced funnels as they get collimated (this allowed for the prediction and discovery of the intermediate peak in GRB 110709B in that work).  Finally, after substantial mass and angular momentum accretion, the BH turns from a Schwarschild into a Kerr BH; this, along with the surrounding disk, allows for the turning on of a BZ central engine which powers the third and last peak.
If only two peaks are produced, this implies that the conditions for the third, BZ, engine were not fulfilled.  This is most easily explained if the CE is not produced by a Collapsar but, instead, by a NS-NS merger with a $B\gtrsim10^{15}$ G field which produces a centrifugally-supported massive, ms magnetar powering up a two-second GRB; as it slows down, it collapses onto a Schwarschild BH, releasing a blast wave with energy $E_B \simeq {\mathcal G}\frac{3}{5}\left(\frac{M_\mathrm{BH}^2}{R_\mathrm{BH}} - \frac{M_\mathrm{NS}^2}{R_\mathrm{NS}}\right) \lesssim (20-15) \cdot 10^{53}  {\rm erg}$, which gets collimated by the infalling, accreting, material.  The infall, nonetheless, is too small to spin the BH up and power up a BZ engine.

Although other mechanisms exist to switch on and off a CE engine (e.g., interrupted accretion stages), allowing for the timing and strong energy differences is simple using the aforementioned mechanism.  
Two peaks in the emission coming from two colliding shells allow putting constraints on the nature of CE and type of event (Collapsar vs merger), favouring, thus, a SGRB from a NS-NS merger for GRB~250221A. However, if a single shell (with energy injection, density gradient, or other mechanism resulting from different accretion stages) is the preferred model for producing the rebrightening, then no constraints can be placed on either of them.

%%%%%%%%%%%%%%%%%%%%%%%%%%%%%%%%%%%%%%%%%%%%%%%%%%

% Don't change these lines
\bsp	% typesetting comment
\label{lastpage}
\end{document}